\documentclass[prc,twocolumn,showpacs,showkeys,floatfix]{revtex4}
\usepackage{graphicx}
\usepackage{subfigure}
\usepackage{color}
\usepackage{multirow}
\usepackage{setspace}
\usepackage{dcolumn}
\usepackage{amsmath}
\usepackage{float}
\usepackage{tablefootnote}
\usepackage{longtable}

\begin{document}
\title{Spin/Parity Dependent Level Density}

\author{R.B.~Firestone}
\affiliation{University of California, Department of Nuclear Engineering,  Berkeley, CA 94720, USA}

\date{\today}

\begin{abstract}

It is shown that the Constant Temperature (CT) model of nuclear level density is a direct consequence of a symmetrized Poisson distribution of nuclear level spacings.  The standard CT model describing the total level density is shown to be fatally flawed due to discontinuities at the Yrast energies, the onset of new $J^{\pi}$ sequences, that disrupt the exponential formula and cause the back shift parameter to become nonphysically negative.  A new CT-JPI level density model is proposed with a constant temperature and separate back shift parameters for each $J^{\pi}$ sequence.  The CT-JPI model is also constrained to reproduce the spin distribution predicted by Ericson's spin distribution function~\cite{Eric60} at the neutron separation energy.  A fitting procedure is described for determining the temperature $T$, back shifts $E_0(J^{\pi})$, and spin cutoff parameters $\sigma_c$ from nuclear structure and resonance data.  The CT-JPI model is demonstrated to successfully predict the level densities for a wide range of spins and parities for 46 nuclear with Z=7-92.  In variance with earlier predictions the spin cut-off parameters show no mass dependence and instead substantial variation at all mass regions.

\end{abstract}


\maketitle

\section{Introduction}

Nuclear level density is an important ingredient of statistical models used for nuclear reaction studies and nuclear astrophysics.  It is defined as the number of energy levels per unit energy at a given nuclear excitation energy.  Two simple statistical models are widely used to describe nuclear density.  The Back-Shifted Fermi Gas (BSFG) model was proposed by Hans Bethe~\cite{Bethe36,Bethe37} assuming that level density propagates as a Fermi gas.  The Constant Temperature (CT) model was proposed by Torleif Ericson~\cite{Eric60} assuming that level density propagates exponentially with a constant temperature.  Both models describe only total level density which is of limited applied value because reaction calculations require more detailed information about the spin and parity distributions.  The two models were unified by Gilbert and Cameron~\cite{Gilbert,Gilbert0} who proposed that the CT model applies at lower nuclear excitation energies while the BSFG model should be applied at higher excitation energies.  Gilbert and Cameron proposed a recipe for determining the parameters of each model based on fitting the experimental level energy sequence to the s-wave neutron capture level spacing, $D_0$, corrected to total level spacing at the neutron separation energy by a spin distribution function.

Both models require a back shift parameter, $E_0$, that defines the initial energy of the level density distribution.  This parameter is typically a very negative energy that is difficult to justify theoretically.  Normalization of the total level density to $D_0$ introduces another problem because the spin distribution function proposed by Ericson~\cite{Gilbert,Gilbert0} contains no parity information.  It is usually assumed that the level density of both parities are equal at the neutron separation energy.  That assumption was shown to be false by S.M. Grimes~\cite{Grimes88}.  Finally, these models cannot predict the level density as a function of spin and parity because the onset of each $J^{\pi}$ sequence begins at the Yrast energy, not $E_0$.  These deficiencies suggest that the level density parameters provided in the Reference Input Parameter Library (RIPL-3)~\cite{RIPL3} are not optimal.

In this paper I will show that the CT model is a consequence of the statistical level spacing distribution.  The CT model can be modified to the CT-JPI model by including individual back shifts for each $J^{\pi}$ sequence.  The level densities for all spins and parities can then be fit, with a single temperature, to reproduce the predictions of the spin distribution function.  A procedure for fitting the back shifts and temperature is demonstrated for a wide range of nuclei with Z=7-92.

\section{The CT-JPI model}

\begin{figure}[t]
\begin{center}
\includegraphics[width=8.5cm]{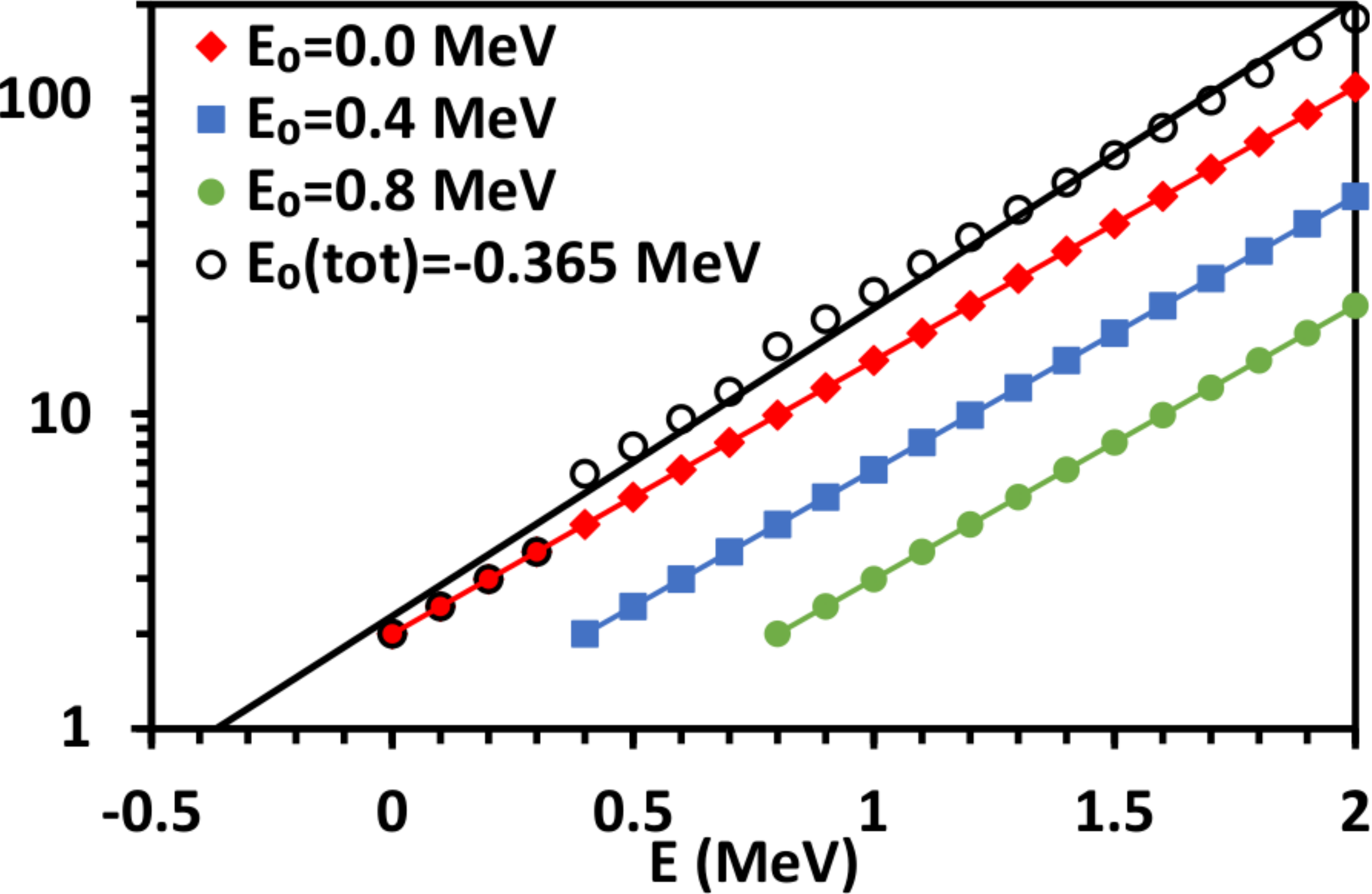}
\caption{\label{BS} Total level density (black circles) gives an effective back shift ${E_0=-}$0.365 MeV assuming separate $J^{\pi}$ level sequences with back shifts starting at 0.0, 0.4, and 0.8 MeV with a temperature, $T$=0.5 MeV.}
\end{center}
\end{figure}

The standard CT model defines the total level density, $\rho$(E), by Eq.~\ref{CT} where $N(E)$ is the level sequence number and $T$ is a constant temperature.
\begin{equation}\label{CT}
\rho(E) = N(E)/T = N(E_0)e^{(E-E_0)/T}
\end{equation}
The back shift, $E_0$, is defined as the excitation energy where $N(E_0)$=1.  This back shift is nearly always very negative and has little physical justification.  If we assume that the level densities for each $J^{\pi}$ also propagate exponentially then the cause of this negative back shift immediately becomes apparent as shown in Fig.~\ref{BS}.

The onset of new levels at each Yrast energy raises the apparent exponential that can be fit through the total level density leading to the negative back shift.  I will show that the exponential growth of level density is a fundamental consequence of the statistical level spacings and it can be applied to sequences of levels of each $J^{\pi}$ with individual back shifts and a constant temperature.  This modification of the CT model will be called the CT-JPI model.

\subsection{Derivation of the CT Model}

\begin{figure}[!h]
\centering
\includegraphics[width=8.5cm]{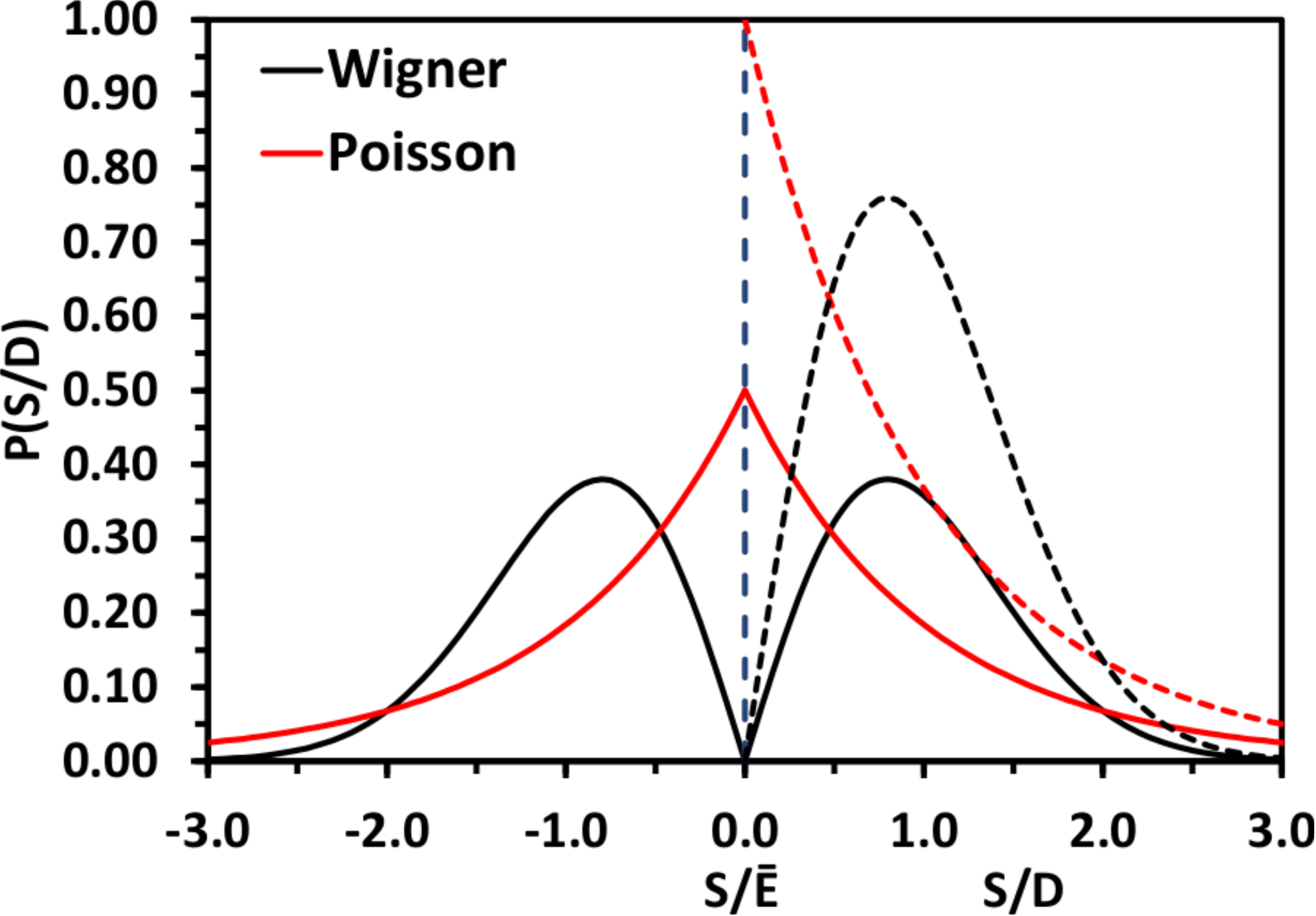}
\caption{Statistical distribution, $P(S/D)$, of level energy spacings, $S$, with respect to the expected spacing, $D=1/\rho(E)$ for Wigner (black), Poisson (red), and Normal (blue) distributions.}
\label{Wign}
\end{figure}

If an ensemble of nuclear levels were randomly spaced in energy then their average spacing, $\bar{E}$, would be constant.  Conventionally the average level energy spacing is expected to follow a Poisson distribution if the quantum system is characterized by regular motion or a Wigner distribution~\cite{Wigner51,Brody73} if the motion is chaotic as given by Eq.~\ref{Wigner} and shown in Fig.~\ref{Wign}

\begin{equation}
\begin{aligned}\label{Wigner}
& \textrm{Wigner:} &P\bigg(\frac{S}{D}\bigg) & =\frac{\pi}{2}\bigg(\frac{S}{D}\bigg)\textrm{exp}\bigg[-\frac{\pi}{4}\bigg(\frac{S}{D}\bigg)^2\bigg] \\
& \textrm{Poisson:} && =\textrm{exp}\bigg[-\bigg(\frac{S}{D}\bigg) \bigg]
\end{aligned}
\end{equation}
where $P(\frac{S}{D})$ is the probability of a given level spacing between level numbers $N$ and ${N+1}$, ${S=E(N+1)-E(N)}$, and the expected level spacing ${D=1/ \rho(E)}$.  Commonly it is assumed that the level spacing can be described by a combination of both distributions~\cite{Sigmac}.

There are several significant problems with the Poisson and Wigner distributions.  The mean spacing is $\bar{P}(\frac{S}{D})$=0.692 for levels with a Poisson distribution and $\bar{P}(\frac{S}{D})$=0.938 for levels with a Wigner distribution.  This implies that the average statistical level spacing is inconsistent with the actual level density.  Another problem with the statistical distributions is that they only consider the level spacing in the direction of increasing energy as is shown in Fig.~\ref{Wigner}.  The level spacing probability should be the same whether considering higher or lower energy levels.  Finally the Poisson and Wigner distributions only consider the spacings between pairs of adjacent levels and ignore the contributions of all other levels.  There is no a priori reason to assume that the level spacing distributions only apply to neighboring levels.

These problems are resolved if we assume that the level spacings follow a normal distribution, described by Eq.~\ref{Norm} and shown in Fig.~\ref{Wigner}, which describes the spacing of
\begin{equation}
\label{Norm}
P\bigg(\frac{S}{\bar{E}}\bigg) = \frac{1}{\sqrt{2\pi}}\textrm{exp}(-0.5\bigg(\frac{S}{\bar{E}}\bigg)^2
\end{equation}
both higher and lower energy nearby levels.  The level spacing distribution, $P(\frac{S}{\bar{E}})$, is with respect to the unperturbed level spacing, $\bar{E}$, and can be considered to predict the compression of level spacing due to quantum effects.  This distribution applies not only to adjacent levels but to all levels.  That creates a conflict because the levels must have fixed energies while the level spacing probabilities would predict different energies depending on how the order that the level separation probabilities were applied. In order to resolve this conflict a complete average level scheme can then be generated using the mean level separations to all levels predicted by the level spacing distributions.  Assuming that the unconstrained level separation of a set of levels is $\bar{E}$, the average energy of the first excited state, $E_2$, can have two possible average values,  ${E_2^1=E_1+\bar{E}\bar{P}(\frac{S}{\bar{E}})}$ for the upward transition or ${E_2^2=E_2-\bar{E}\bar{P}(\frac{S}{\bar{E}})}$ for the downward transition, where $E_1$=0.0 is the GS energy.  Both constraints cannot be achieved simultaneously so we can assume that the most likely energy of the first excited state is the average, ${E_2=(E_2^1+E_2^2)/2=0.5\bar{E}}$.  Similarly the average energy of the second excited state, $E_3$, has four constraints, ${E_3^1=(E_2+\bar{E})\bar{P}(\frac{S}{\bar{E}})}$, ${E_3^2=(E_2-\bar{E})[1-\bar{P}(\frac{S}{\bar{E}})]}$, ${E_3^3=E_2+\bar{E}\bar{P}(\frac{S}{\bar{E}})}$, and ${E_3^4=E_2+\bar{E}[1-\bar{P}(\frac{S}{\bar{E}})]}$.  Assuming that the energy of the second excited state is the average of all four constraints, ${E_3=(E_3^1+E_3^2+E_3^3+E_3^4)/4}$.  Similarly, the energies of higher lying levels can be calculated by averaging the energies of all combinations of level separation constraints to lower and higher energy levels.  These calculated level energies are the statistically most probable values although in actual nuclei they would fluctuate statistically about the expectation values.

\begin{figure}[!ht]
\begin{center}
\includegraphics[width=8.5cm]{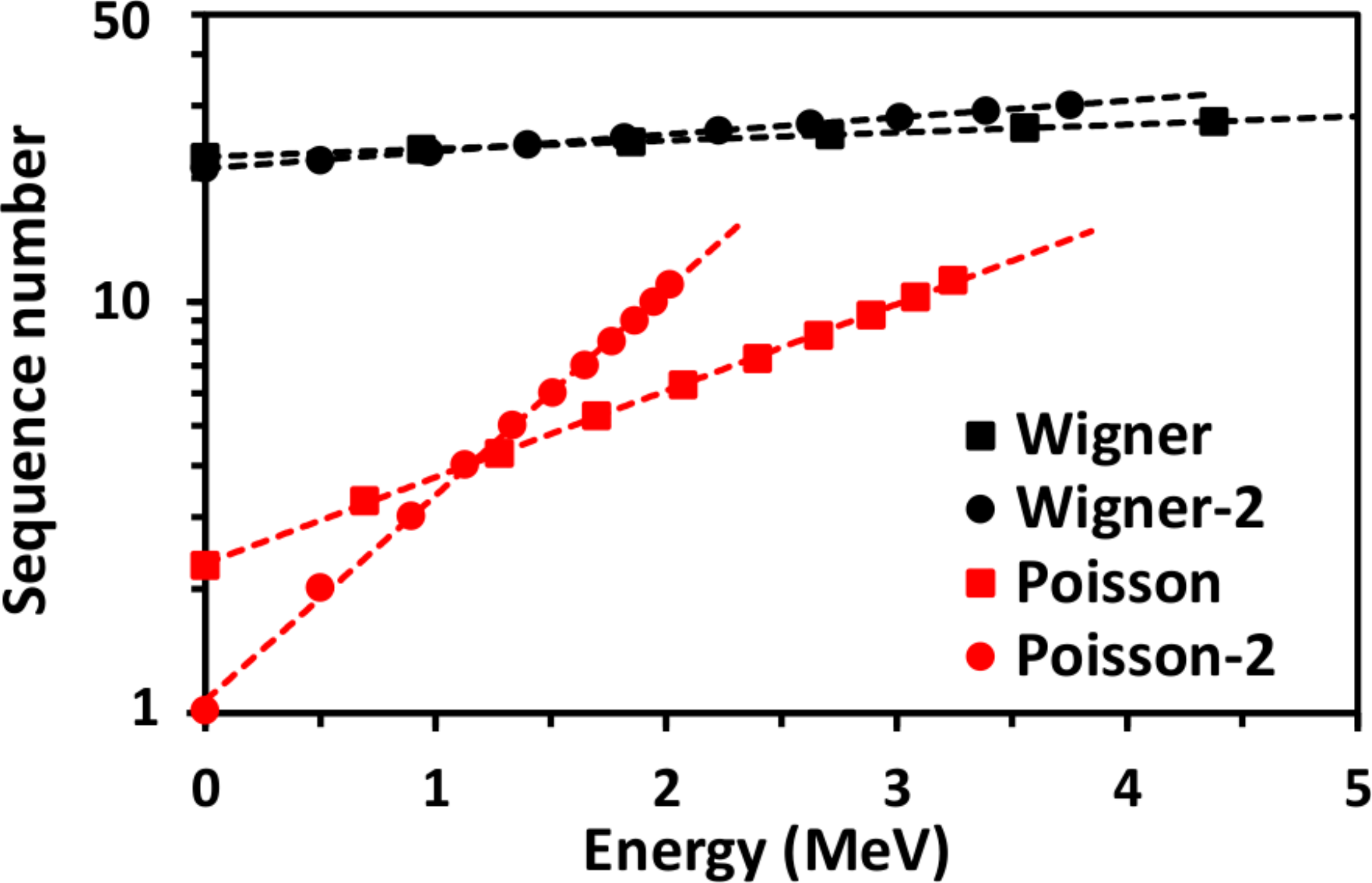}
\caption{\label{WigLS} Predicted level energies, $E$, and sequence numbers, $N(E)$, for $\bar{E}$=1 MeV assuming the Wigner~\cite{Wigner51} (black), Poisson (red), and Normal (Blue) distributions.}
\end{center}
\end{figure}

Average level energies and sequence numbers calculated with the Wigner, Poisson, and normal distributions for a sequence of 10 levels assuming $\bar{E}$=1 MeV are shown in Fig.~\ref{WigLS}.
The calculated level energies for each statistical distribution are given in Table~\ref{WigLev}.  In each case the level energies, $E$, and sequence numbers, $N(E)$, could be fitted to the CT model exponential, ${N(E)=N(E_0)\textrm{exp}[(E-E_0)/T]}$ where $T$ is the temperature and $E_0$=0.0 is the GS energy.

The Wigner distribution gives a fitted GS level sequence number N($E_0$)=22.5 and $T$=22.20 MeV and the Poisson distribution gives N($E_0$)=2.28 and a temperature $T$=1.806 MeV, which are both inconsistent with the standard CT model and experimental data.  The normal Poisson distribution gives N($E_0$)=1.00 and a temperature $T$=0.82 MeV that is completely consistent with the CT model and experimental data.  This analysis eliminates both the Wigner and Poisson descriptions of nuclear level spacing and establishes that the CT model as a natural consequence of a normal distribution of level energy spacings.

\begin{table}[!ht]
\tabcolsep=7pt
\caption{\label{WigLev} Average level energies, $E$, calculated for the standard Wigner and Poisson level spacing distributions and the symmetrized Wigner and Poisson distributions. The temperature, $T$,and GS level sequence number, N($E_0$), are fitted to the CT model ${N(E)=\textrm{exp}[(E-E_0)/T}$. }
\begin{tabular}{clll}
\toprule
\multicolumn{1}{c}{Level}&\multicolumn{1}{c}{E(Wigner)}&\multicolumn{1}{c}{E(Poissant)}&\multicolumn{1}{c}{E(Normal)}\\
&\multicolumn{3}{c}{(MeV)}\\
\colrule
1&0.00&0.00&0.00\\
2&0.94&0.69&0.50\\
3&1.85&1.28&0.89\\
4&2.71&1.70&1.12\\
5&3.56&2.07&1.32\\
6&4.38&2.40&1.48\\
7&5.17&2.66&1.61\\
8&5.95&2.89&1.72\\
9&6.70&3.08&1.81\\
10&7.42&3.24&1.88\\
\colrule
$T$ (MeV)&22.20&2.05&0.82\\
N($E_0$)&22.5&2.28&1.00\\
\botrule
\end{tabular}
\end{table}

\subsection{CT-JPI model formulation}

The standard CT model is insufficient because discontinuities in the level density at the Yrast energies distort the exponential curve and create physically meaningless back shift parameters.  This is not a problem for each $J^{\pi}$ level sequence which is continuous at least up to the next shell closure.  A new CT-JPI model is proposed with a constant temperature, $T$, and separate back shifts, $E_0 (J^{\pi})$, for each spin and parity.  The back shift energies will occur near the Yrast energy for each $J^{\pi}$ and represent the energy at which each $J^{\pi}$ sequence commences.  The temperature is defined in Eq.~\ref{CT} and is assumed to be constant for all $J^{\pi}$.  This assumption is largely based on experimental observation although it can be viewed as the result of a phase transition from superfluid to normal nuclear matter where $T$ is constant as the nucleus is heated~\cite{Moretto15,Zelev18}.

The CT-JPI model is also constrained by the spin distribution function~\cite{Eric60} where the spins associated with the excited degrees of freedom are assumed to be coupled at random as given in Eq.~\ref{CTSDF}
\begin{equation}\label{CTSDF}
f(J) = \frac{2J+1}{2\sigma_c^2} \textrm{exp}\bigg[-\frac{(J+1/2)^2}{2\sigma_c^2}\bigg]
\end{equation}
where $f(J)$ is the fraction of levels at the neutron separation energy, $S_n$, with spin $J$ and $\sigma_c$ is the spin cutoff factor.  T. von Egidy \textit{et al}~\cite{Sigmac} have approximated the spin cutoff factor as $\sigma_c = 0.98A^{0.29}$.  The CT-JPI model gives $f(J^{\pi})$ values for each parity which can be summed to give $f(J)$ for each spin.  The CT-JPI $f(J)$ values can be compared with values from the spin distribution function by varying the spin cutoff parameter.  The CT-JPI model is thus rigorously constrained by both an exponential decrease in level energy spacing and the expected spin distribution at the neutron separation energy.

With many more parameters in the CT-JPI model a new evaluation procedure is required.  Despite the greater complexity of this model, the determination of level densities for all spins and parities will offer greater capabilities for performing nuclear transport and astrophysical calculations.

\section{CT-JPI model evaluation procedure}

The CT-JPI model requires the determination of the back shift for each $J^{\pi}$ sequence with a common temperature.  This can be accomplished if sufficient nuclear structure and resonance data are available for several spins and parities.  Both $T$ and $E_0 (J^{\pi})$ in Eq.~\ref{CT} can be fit to the experimental level and resonance energies for each $J^{\pi}$ by minimizing the average deviation of the data from the fitted integral level sequence numbers ${\overline{\Delta N(J^{\pi})} = \overline{\lvert(N(E_N)_{fit}-N(E_N)_{exp}\rvert}}$.  This can be done using an iterative optimization procedure such as Excel Solver.  It is not necessary that a complete sequence of levels and resonances for each $J^{\pi}$ are known since they should all fall on the same exponential curve even if some are missing.  Assuming that the temperature is constant for all $J^{\pi}$ values in the CT-JPI model, several $J^{\pi}$ sequences can be fit simultaneously to obtain the best back shifts and temperature.  Resonance data are not required if sufficient nuclear structure data are available so more nuclei can be analyzed than with the standard CT model.

Spurious local minima will be found in the ${\overline{\Delta N(J^{\pi})}}$ fit.  The path to the deepest minima is guided by the assumption that the fractional abundances vary smoothly for each parity and the fitted spin fractions, ${f(J)=f(J^{\pi=+})+f(J^{\pi=-})}$, can be fit to the spin distribution function by varying the spin cutoff parameter, $\sigma_c$.  Nuclear structure and resonance data need not be available for all $J^{\pi}$ values.  If data for only one parity of a given spin exists, the back shift for the other parity is constrained by interpolation of the fractional abundances of other levels with the same parity and the fractional abundance for that spin predicted by the spin distribution function.  If no data for a given spin are available the back shifts for both parities are constrained by fractional abundance trends of both parities and the spin distribution function.

For actual nuclei Eq.~\ref{CT} is not exact because the CT-JPI model level energy spacings vary by a folded Normal distribution.  If the level energy spacings were random we would expect $\overline{\Delta N}(J^{\pi})$=0.25.  For the folded Normal distribution the statistically weighted value of $\overline{\Delta N}(J^{\pi})$=0.125 is expected.  Mistakes in spin/parity assignments and large level/resonance energy uncertainties will deteriorate this fit so great care must be taken in the data selection,  In the following sections the validity of the CT-JPI model will be demonstrated for a wide range of nuclei with Z=7-92.  Only levels with $J=1/2-11/2$, for odd A nuclei and $J=0-6$ for even A nuclei will be considered because the validity of spin distribution function is questionable at high spin.

\subsection{Evaluation of $^{235}$U using resonance and nuclear structure data}

\begin{figure}[!h]
\centering
\includegraphics[width=8.5cm]{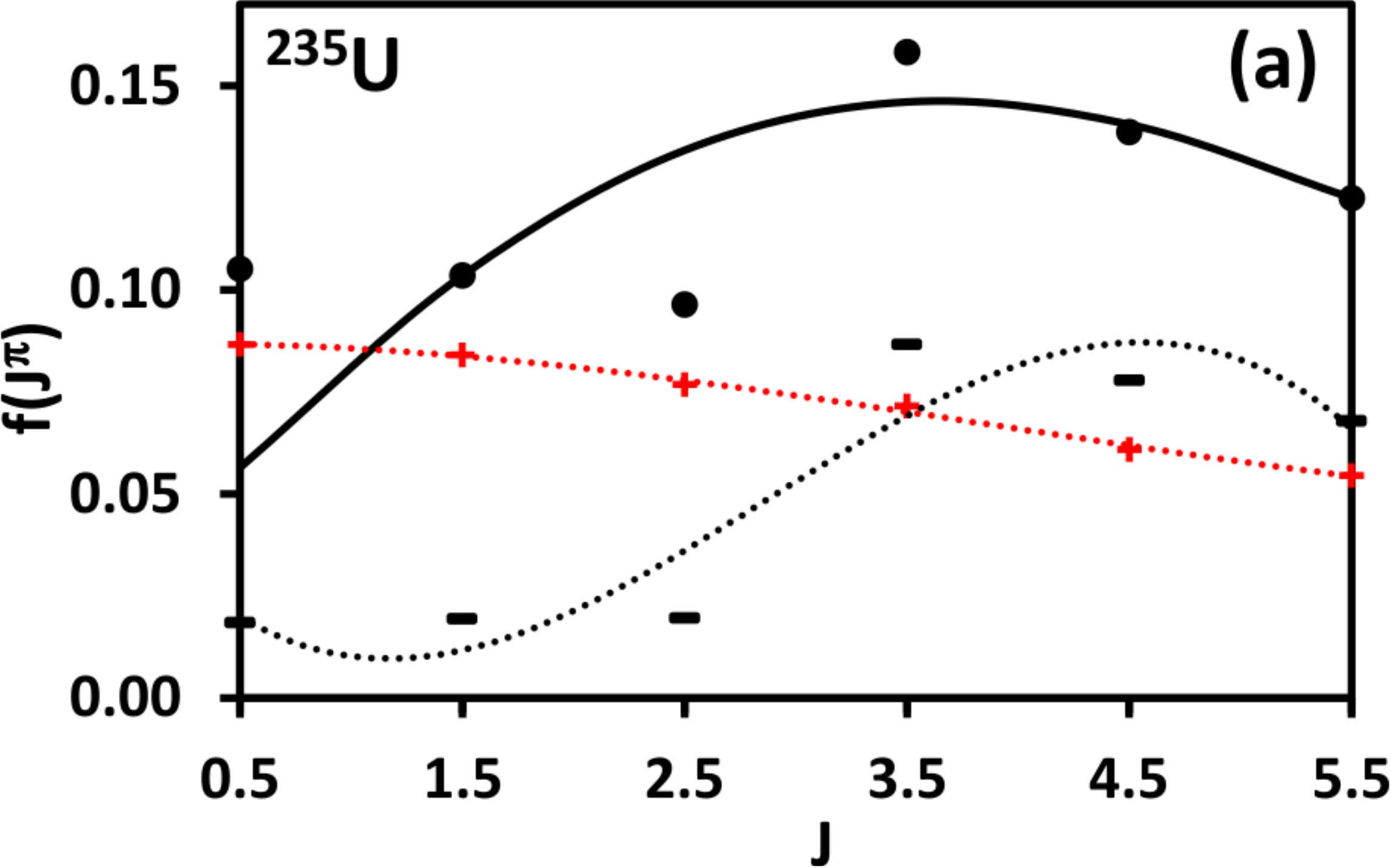}
\includegraphics[width=8.5cm]{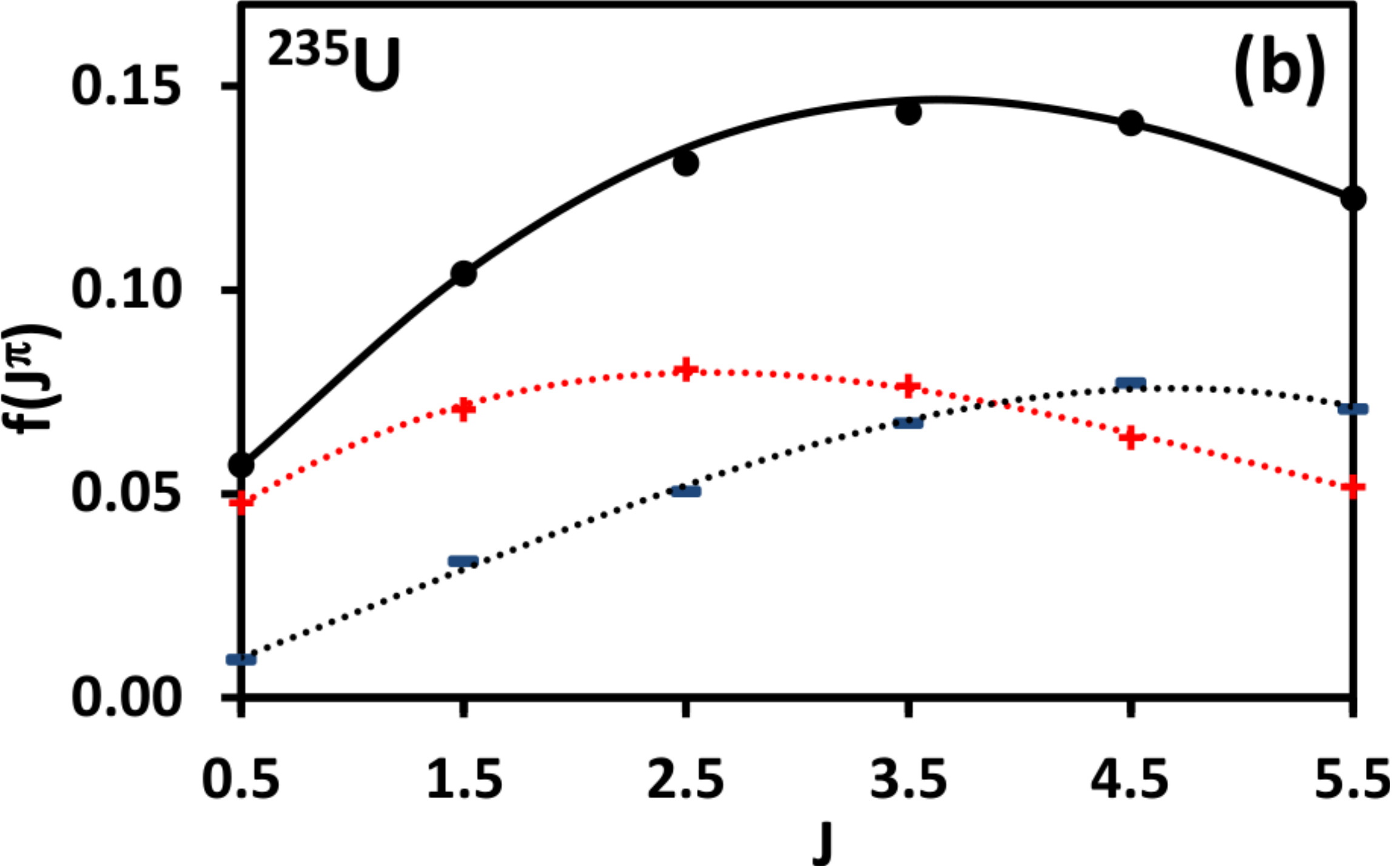}
\caption{First (a) and final (b) iterations of the $^{235}$U level density analysis.  Positive($\textbf{\textcolor{red}+}$), negative(\textbf{$-$}), and total($\bullet$) fractions of level density for each spin at the neutron separation energy are shown.  The solid black curve represents the fraction of each spin predicted by the spin distribution function.}
\label{235U}
\end{figure}

Both nuclear structure~\cite{ENSDF235} and s-wave resonance~\cite{Mugh} data can be used simultaneously to fit the CT-JPI parameters for ${J^{\pi}=1/2^+}$ levels in $^{235}$U.  Initially we can assume the Yrast energy, $E_0 (1/2^+)$=0.000076 MeV, for the ${J^{\pi}=1/2^+}$ back shift.  Then we can determine a first estimate of the temperature by minimizing $\overline{\Delta N(1/2^+)}$ for all $J^{\pi}=1/2^+$ states, giving $T$=0.426 MeV and ${\overline{\Delta N(1/2^+})}=0.181$.  With this temperature and assuming that back shifts correspond to the Yrast energies of the other levels with ${J=1/2-11/2}$ we can calculate a preliminary spin distribution at the neutron separation energy, $S_n$=5.29749 MeV~\cite{AME2013} which is plotted in Fig.~\ref{235U}a.  The positive parity states show a smooth distribution with increasing spin while the negative parity states have large oscillations.  The overall fit to all spins and parities gives $\overline{\Delta N}(J^{\pi})$=0.198 is poorer than desired.  An initial spin distribution function can be fit to the experimental data, as shown in Fig.~\ref{235U}a, with a spin cutoff parameter, $\sigma_c$=4.15, that agrees with the data to within 16\%.

\begin{figure}[!h]
\centering
\includegraphics[width=8cm]{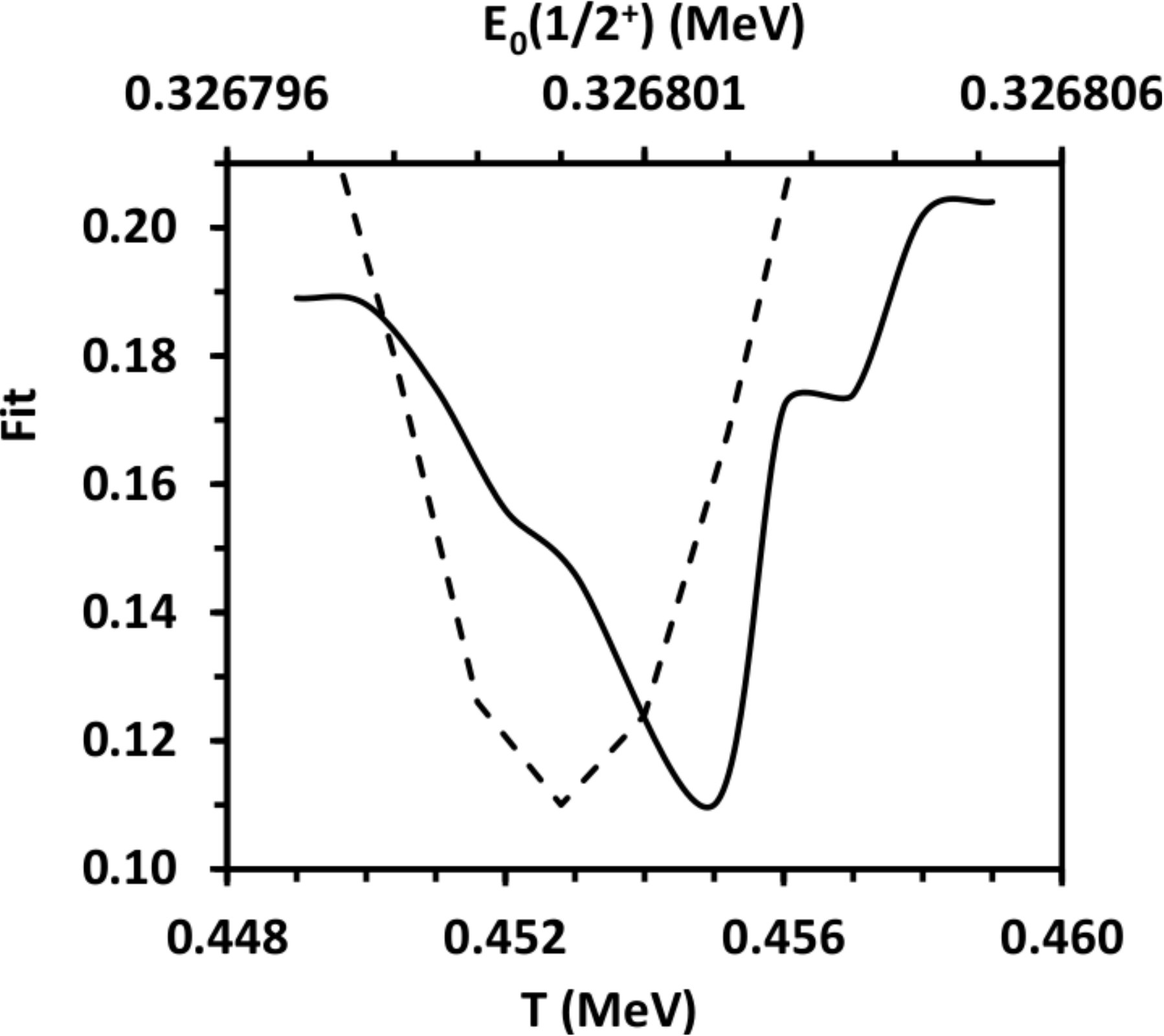}
\caption{Minimization of the fit to the $^{235}$U temperature (solid line) and $J^{\pi}=1/2^+$ back shift (dotted line). }
\label{Dfit}
\end{figure}

The next step is to search for values of $T$ and $E_0(1/2^+)$, consistent with the spin distribution function, that minimize the fit to $\overline{\Delta N}(1/2^+)$.  Using the  Excel Solver we find $T$=0.455 MeV and $E_0 (1/2^+)$=0.32680 MeV with $\overline{\Delta N}(1/2^+)$=0.110.  The fitted temperature gives a sharp minimum with a FWHM=4 keV, and the fit to $E_0(1/2^+)$ gives an extremely sharp minimum with a FWHM=3 eV, as shown in Fig.~\ref{Dfit}.
Other spurious $E_0(1/2^+)$ minima can be found but none are either so narrow or deep.  Using this temperature the remaining back shifts can be similarly solved giving $\overline{\Delta N}(J^{\pi})$=0.127, consistent with the expected value.  The fitted $E_0 (J^{\pi})$ values are compared with the $^{235}$U Yrast and Yrare energies for each spin in Table~\ref{U235}.  For positive parity states the fitted values occur between the Yrast and Yrare energy.  The same is true for negative parity states with $J=7/2-11/2$, but $E_0 (1/2^-)$ is well above the Yrare energy and $E_0 (3/2^-,5/2^-)$ is well below the Yrast energy.
\begin{table}[!ht]
\tabcolsep=8pt
\caption{\label{U235} Comparison of fitted $E_0 (J^{\pi})$ values with the $^{235}$U Yrast and Yrare energies.}
\begin{tabular}{cccc}
\toprule
$J^{\pi}$&E(Yrast)&E(Yrare)&E(fit)\\
&\multicolumn{3}{c}{(MeV)}\\
\colrule
$0^+$&0.000&0.9227&1.159\\
$1^+$&1.209&1.516&0.943\\
$2^+$&0.045&0.966&0.828\\
$3^+$&1.060&1.106&0.796\\
$4^+$&0.148&1.056&0.885\\
$5^+$&1.232&$-$&1.051\\
$6^+$&0.307&1.269&1.311\\
\colrule
$0^-$&$-$&$-$&1.373\\
$1^-$&0.680&0.931&0.764\\
$2^-$&0.950&1.129&0.650\\
$3^-$&0.732&0.998&0.714\\
$4^-$&1.028&$-$&0.824\\
$5^-$&0.827&$-$&1.011\\
$6^-$&1.151&$-$&1.243\\
\botrule
\end{tabular}
\end{table}

The new fitted spin distribution at the neutron separation energy is agrees with the spin distribution function, assuming $\sigma_c$=4.14, within 0.9\% as shown in Fig.~\ref{235U}b.  The positive and negative state abundances vary smoothly with increasing spin where the trend lines are arbitrarily fit the data to a third order polynomial.  The fitted spin cutoff parameter is lower than $\sigma_c$=4.77 calculated from von Egidy's \textit{et al}~\cite{Sigmac} formulation.   The fitted temperature is 8\% higher than $T$=0.420 MeV recommended in RIPL-3~\cite{RIPL3} and the average s-wave spacing at the neutron separation energy is ${D_0=1/\rho(S_n,1/2^+)}$=8.19 eV, which is significantly lower than $D_0$=12.0$\pm$0.8 eV from RIPL-3 consistent with the short comings of the standard CT model.

\subsection{Evaluation of $^{238}$U using nuclear structure data}

\begin{figure}[!h]
\centering
\includegraphics[width=8.5cm]{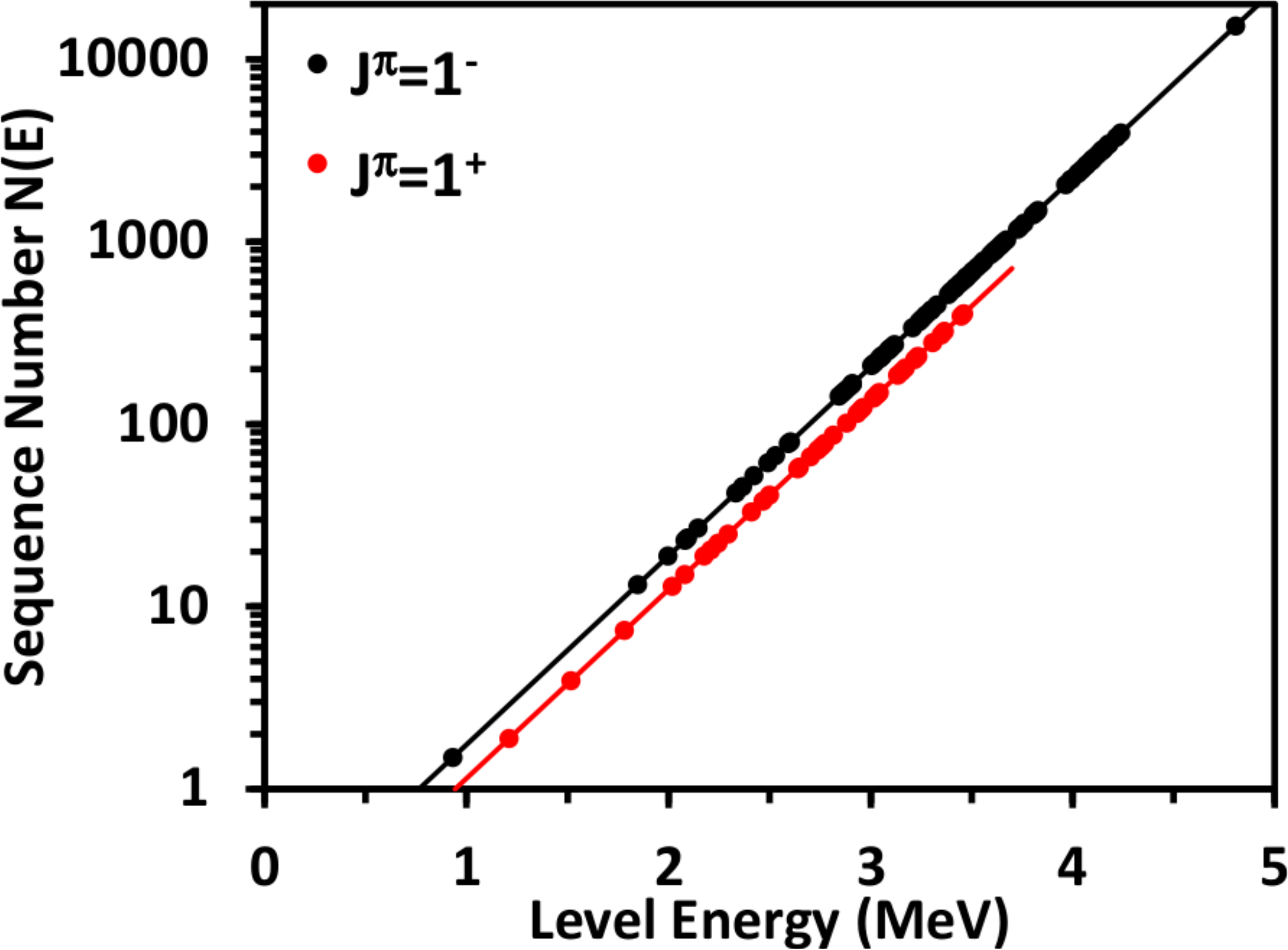}
\caption{Fitted exponential growth of $^{238}$U ${J^{\pi}=1/2^-,1/2^+}$ level sequence numbers with energy.  Although not all levels of each $J^{\pi}$ are known, an excellent fit (solid lines) can be obtained through all levels.}
\label{238UJPI}
\end{figure}

\begin{figure}[!h]
\centering
\includegraphics[width=8.5cm]{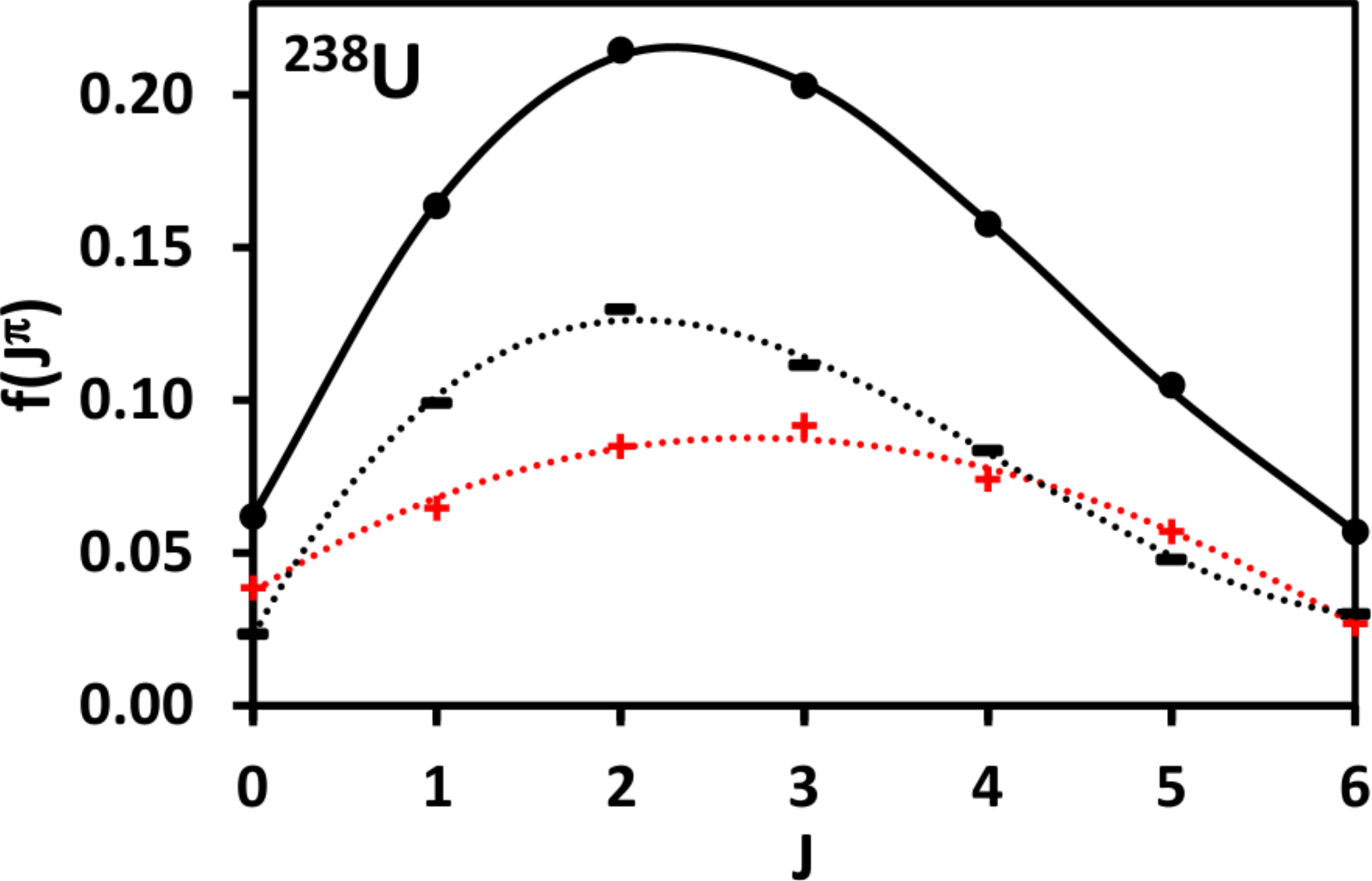}
\caption{Comparison of the positive parity ($\textbf{\textcolor{red}+}$), negative parity (\textbf{$-$}), and total spin ($\bullet$) fitted $^{238}$U spin/parity fractions with the spin distribution function calculation (solid black curve) at the neutron separation energy, $S_n$=6.1543 MeV~\cite{AME2013}, assuming $\sigma_c$=2.81.}
\label{238U}
\end{figure}

Although no resonance data are available for $^{238}$U considerable nuclear structure data are available~\cite{ENSDF238}.  There are 128 ${J^{\pi}=1^-,1^+}$ known levels.  The CT-JPI model sequence numbers for each $J^{\pi}$ series can be fit to an exponential as shown in Fig.~\ref{238UJPI}.  This fit gives a temperature $T$=0.420 MeV and
back shifts  $E_0(1^+)$=0.943 MeV, and $E_0(1^-)$=0.764 MeV with $\overline{\Delta N}(1^+,1^-)$=0.145.  The fitted temperature is higher than the RIPL-3 value, $T$=0.393 MeV.  The remaining back shifts for levels with $J^{\pi}=0^+,1,2,3,4^+,5^+,6^+$ can be fit, as described for $^{235}$U,  assuming the constant temperature giving $\overline{\Delta N}(J^{\pi})$=0.133, consistent with the expected minimization limit.  The fitted fractional abundances for levels with $J$=0-6 are compared with the values calculated with the spin distribution function, assuming $\sigma_c$=2.81, at the neutron separation energy, $S_n$=6.1543 MeV~\cite{AME2013}, in Fig.~\ref{238U} with an average uncertainty of 0.6\%.  This fitted spin cutoff parameter is much lower than the von Egidy \textit{et al} estimate~\cite{Sigmac}, $\sigma_c$=4.79.

The fractional abundances of levels with ${J^{\pi}=0^-,4^-,6^-}$, where little experimental data exist, were determined from the systematics of the fitted negative parity fractions and the difference between the total $J$ fraction from the spin distribution function and the fitted positive parity component.  For the $J$=5 states, where there is no data to fit for either parity, the fractional abundances for each parity were determined by interpolation of the abundance trends for each parity constrained by the total fractional abundance for $J$=5 from the spin distribution function.

The fitted $E_0 (J^{\pi})$ values are compared with the $^{238}$U Yrast and Yrare energies for each spin in Table~\ref{U238}.  Back shifts for levels with $J^{\pi}=2^-,3,4^-$ lie
\begin{table}[!ht]
\tabcolsep=8pt
\caption{\label{U238} Comparison of fitted $E_0 (J^{\pi})$ values with the $^{238}$U Yrast and Yrare energies.}
\begin{tabular}{cccc}
\toprule
$J^{\pi}$&E(Yrast)&E(Yrare)&E(fit)\\
\colrule
$0^+$&0.000&0.9227&1.159\\
$1^+$&1.209&1.516&0.943\\
$2^+$&0.045&0.966&0.828\\
$3^+$&1.060&1.106&0.796\\
$4^+$&0.148&1.056&0.885\\
$5^+$&1.232&$-$&1.051\\
$6^+$&0.307&1.269&1.311\\
\colrule
$0^-$&$-$&$-$&1.373\\
$1^-$&0.680&0.931&0.764\\
$2^-$&0.950&1.129&0.650\\
$3^-$&0.732&0.998&0.714\\
$4^-$&1.028&$-$&0.824\\
$5^-$&0.827&$-$&1.011\\
$6^-$&1.151&$-$&1.243\\
\botrule
\end{tabular}
\end{table}
below the Yrast energy and back shifts for levels with $J^{\pi}=0^+,6^+$ lie above the Yrare energy.  Although several fitted back shifts lie outside the Yrast-Yrare energy region, the folded Normal distribution predicts that only $\approx$58\% of back shifts will lie within the Yrast-Yrare energy range.

\section{Evaluation of CT-JPI model level densities for nuclei with Z=7-92}

The CT-JPI model evaluation procedure has been applied to 46 nuclei with Z=7-92.  In these evaluations the difference between the fitted and observed level sequence numbers, $\overline{\Delta N}(J^{\pi})$, were minimized, as discussed above, by varying the temperature, $T$, and back shifts, $E_0(J^{\pi})$ for each spin and parity.  The fits were also constrained to minimize the difference between the fitted CT-JPI model fractional spin distribution at the neutron separation energy, $S_n$, and the calculated distribution from the spin distribution function, defined in Eq.~\ref{CTSDF}, by varying the spin cutoff parameter, $\sigma_c$.  In all cases minimization process could simultaneously give ${\overline{\Delta N}(J^{\pi})\approx}$1.25, as expected from the level spacing distribution, and consistency to within $\lesssim1\%$ with the fractional abundances calculated with the spin distribution function.

\subsection{Uranium}

The uranium isotopes are of considerable applied importance so any deficiencies of the standard CT model should be of great concern.  The CT-JPI model has been applied to the isotopes $^{233-239}$U where sufficient nuclear structure and resonance data are available to provide reasonable fits.  The fitted $E_0(J^{\pi})$ back shifts are shown in Table~\ref{Udata} and the corresponding neutron separation energies, $S_n$, temperatures, $T$, spin cutoff parameters, $\sigma_c$, resonance spacings, $D_0$ and $D_1$, and quality of fit, $\overline{\Delta N}(J^{\pi})$, are shown in Table~\ref{UCTfit}.  The average fit for all isotopes is $\overline{\Delta N}(J^{\pi})$=0.132(9) in excellent agreement with the expected value from the Wigner+Poisson distribution.  The temperatures and resonance spacings that were fitted to the CT-JPI model are compared with the values from RIPL-3~\cite{RIPL3} in Table~\ref{UCTfit}.  The fitted temperatures range from  7-22\% higher than the RIPL-3 values~\cite{RIPL3}.  The fitted level spacings, $D_0$ and $D_1$, vary widely from the RIPL-3 values, sometimes by a factor of $\approx$3.  The average spin cutoff parameter, $\sigma_c$=3.6(4), is smaller that the value predicted by T. von Egidy \textit{et al}~\cite{Egidy88}, $\sigma_c$=4.8.

\begin{table*}[!ht]
\tabcolsep=5pt
\caption{\label{Udata} Back shifts, $E_0(J^{\pi})$, derived from the CT-JPI model, for $^{233-239}$U. }
\begin{tabular}{cdddddddd}
\toprule
&\multicolumn{2}{c}{$\hspace{0.5cm}E_0 (J^{\pi}) ^{234}U$}&\multicolumn{2}{c}{$\hspace{0.5cm}E_0 (J^{\pi}) ^{236}U$}&\multicolumn{2}{c}{$\hspace{0.5cm}E_0 (J^{\pi}) ^{238}U$}&\\
J&\multicolumn{1}{c}{$\hspace{0.5cm}\pi=+$}&\multicolumn{1}{c}{$\hspace{0.5cm}\pi=-$}&\multicolumn{1}{c}{$\hspace{0.5cm}\pi=+$}&\multicolumn{1}{c}{$\hspace{0.5cm}\pi=-$}&
\multicolumn{1}{c}{$\hspace{0.5cm}\pi=+$}&\multicolumn{1}{c}{$\hspace{0.5cm}\pi=-$}&\\
\colrule
0&2.000&(1.455)&\hspace{0.5cm}2.074&(1.213)&\hspace{0.5cm}1.159&(1.371)&\hspace{0.5cm}&\\
1&1.294&1.106&\hspace{0.5cm}0.961&1.105&\hspace{0.5cm}0.943&0.764&\hspace{0.5cm}&\\
2&1.092&0.970&\hspace{0.5cm}(0.782)&1.009&\hspace{0.5cm}0.828&0.650&\hspace{0.5cm}&\\
3&0.935&0.999&\hspace{0.5cm}0.739&0.994&\hspace{0.5cm}0.796&0.714&\hspace{0.5cm}&\\
4&0.919&1.048&\hspace{0.5cm}0.750&1.117&\hspace{0.5cm}0.885&(0.825)&\hspace{0.5cm}&\\
5&1.000&1.094&\hspace{0.5cm}(0.821)&1.282&\hspace{0.5cm}(1.073)&(0.999)&\hspace{0.5cm}&\\
6&1.122&1.162&\hspace{0.5cm}(0.963)&1.472&\hspace{0.5cm}1.311&(1.265)&\hspace{0.5cm}&\\
\toprule
&\multicolumn{2}{c}{$\hspace{0.5cm}E_0 (J^{\pi}) ^{233}U$}&\multicolumn{2}{c}{$\hspace{0.5cm}E_0 (J^{\pi}) ^{235}U$}&\multicolumn{2}{c}{$\hspace{0.5cm}E_0 (J^{\pi}) ^{237}U$}&\multicolumn{2}{c}{$\hspace{0.5cm}E_0 (^{239}U)$}\\
J&\multicolumn{1}{c}{$\hspace{0.5cm}\pi=+$}&\multicolumn{1}{c}{$\hspace{0.5cm}\pi=-$}&\multicolumn{1}{c}{$\hspace{0.5cm}\pi=+$}&\multicolumn{1}{c}{$\hspace{0.5cm}\pi=-$}&
\multicolumn{1}{c}{$\hspace{0.5cm}\pi=+$}&\multicolumn{1}{c}{$\hspace{0.5cm}\pi=-$}&\multicolumn{1}{c}{$\hspace{0.5cm}\pi=+$}&\multicolumn{1}{c}{$\hspace{0.5cm}\pi=-$}\\
\colrule
1/2&0.364&(0.676)&\hspace{0.5cm}0.327&1.073&\hspace{0.5cm}0.402&0.769&\hspace{0.5cm}0.665&0.740\\
3/2&0.090&0.393&\hspace{0.5cm}0.149&0.490&\hspace{0.5cm}0.203&0.470&\hspace{0.5cm}0.291&0.759\\
5/2&0.038&0.233&\hspace{0.5cm}0.089&0.301&\hspace{0.5cm}0.095&(0.426&\hspace{0.5cm}0.219&0.558\\
7/2&0.044&0.195&\hspace{0.5cm}0.113&0.171&\hspace{0.5cm}0.138&(0.353)&\hspace{0.5cm}0.194&(0.549)\\
9/2&0.094&0.293&\hspace{0.5cm}0.195&0.109&\hspace{0.5cm}0.201&(0.420)&\hspace{0.5cm}0.242&(0.555)\\
11/2&0.235&0.400&\hspace{0.5cm}0.291&0.148&\hspace{0.5cm}0.295&(0.577)&\hspace{0.5cm}(0.432)&(0.484)\\
\botrule
\end{tabular}
\end{table*}
\begin{table*}[t]
\tabcolsep=4pt
\caption{\label{UCTfit} Neutron separation energies~\cite{AME2013}, $S_n$, spin cutoff parameters, $\sigma_c$, temperatures, $T$, resonance spacings, $D_0$ and $D_1$, and minimization uncertainty fitted to the CT-JPI model and compared with the values from RIPL-3~\cite{RIPL3} for $^{233-239}$U. }
\begin{tabular}{lddddddd}
\toprule
&\multicolumn{1}{c}{$^{233}$U}&\multicolumn{1}{c}{$^{234}$U}&\multicolumn{1}{c}{$^{235}$U}&\multicolumn{1}{c}{$^{236}$U}&\multicolumn{1}{c}{$^{237}$U}&
\multicolumn{1}{c}{$^{238}$U}&\multicolumn{1}{c}{$^{239}$U}\\
\colrule
$S_n$ (MeV)&5.7621&6.8447&5.29749&6.54545&5.1258&6.1543&4.80638\\
$\sigma_c$&3.53&3.82&4.13&3.38&3.45&2.81&3.83\\
T $_{CT-JPI}$(MeV) &0.470&0.444&0.455&0.428&0.416&0.420&0.419\\
T $_{RIPL-3}$(MeV) &0.411&0.408&0.420&0.394&0.386&0.393&0.344\\
$D_0$(CT-JPI)(eV)&4.83&1.04&8.19&0.569&4.81&1.07&21.5\\
$D_0$(RIPL-3)(eV)&4.6(7)&0.52(2)&11.2(8)&0.45(3)&14.0(10)&3.5(8)&20.3(6)\\
$D_1$(CT-JPI)(eV)&4.17&0.225&9.19&0.148&3.81&0.437&13.1\\
$D_1$(RIPL-3)(eV)&$-$&$-$&$-$&$-$&$-$&$-$&7.7(30)\\
$\Delta$N&0.141&0.137&0.127&0.129&0.116&0.133&0.139\\
\botrule
\end{tabular}
\end{table*}

\begin{figure*}[!ht]
  \centering
    \includegraphics[width=8cm]{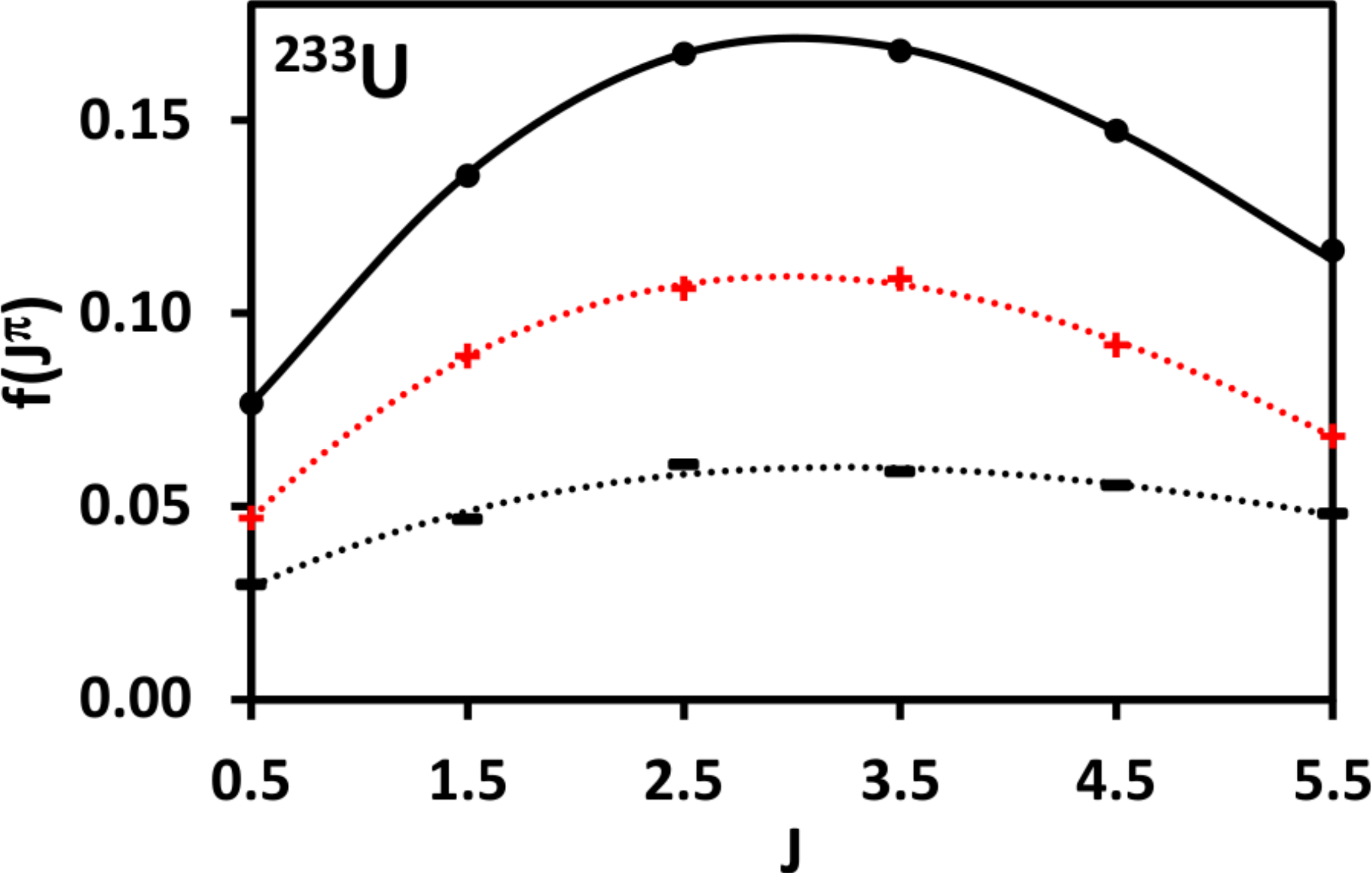}
    \includegraphics[width=8cm]{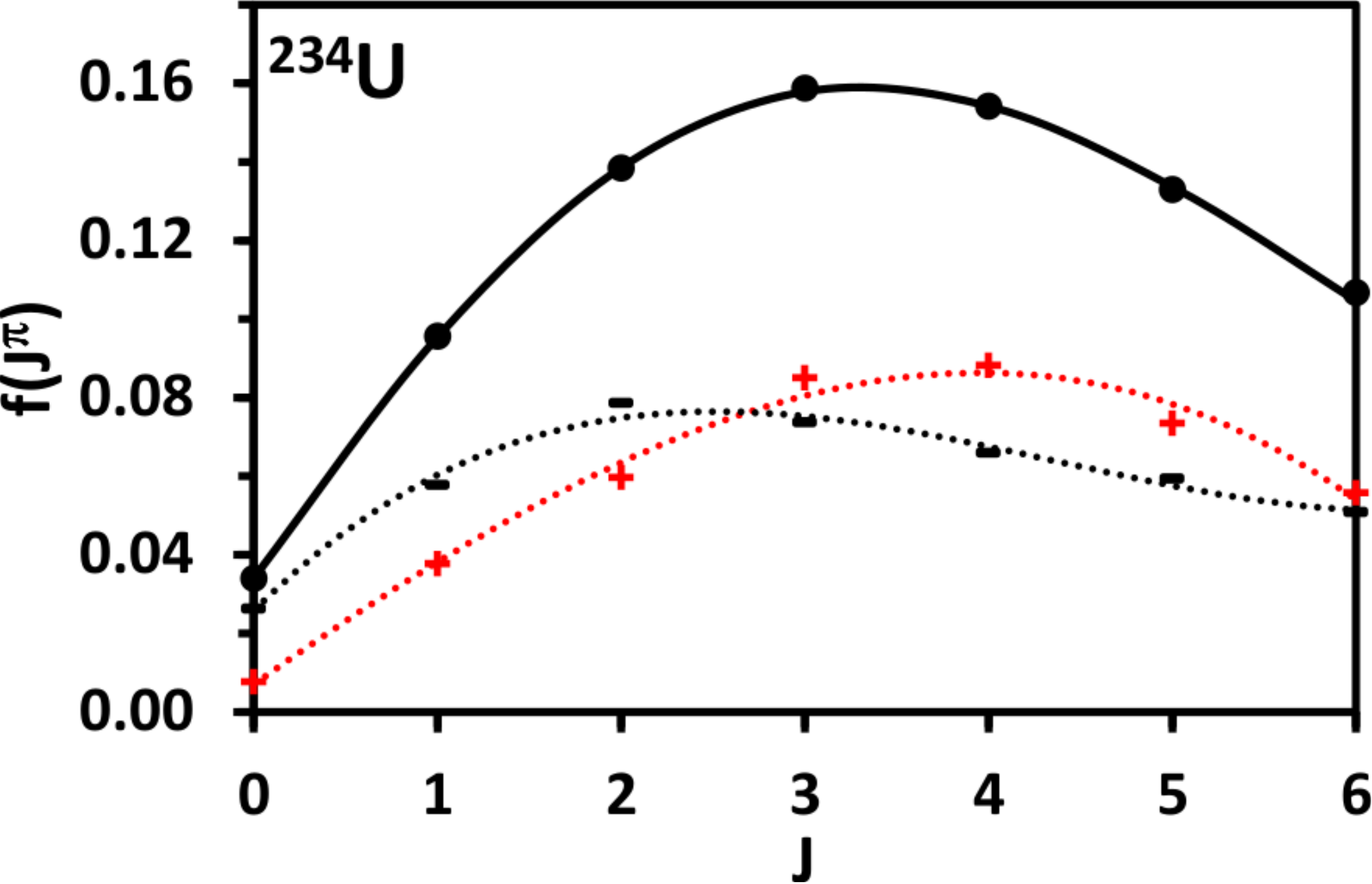}
    \includegraphics[width=8cm]{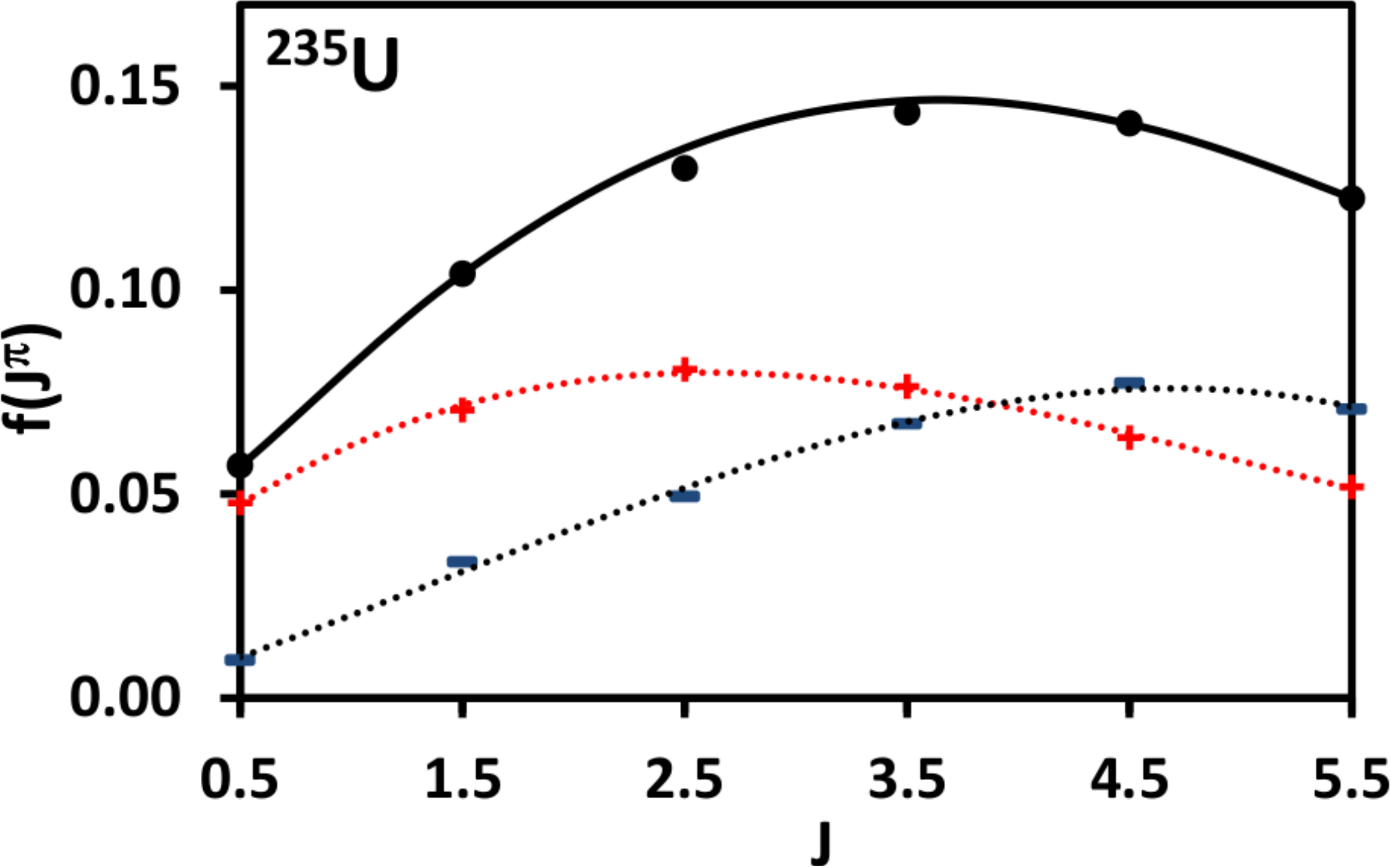}
    \includegraphics[width=8cm]{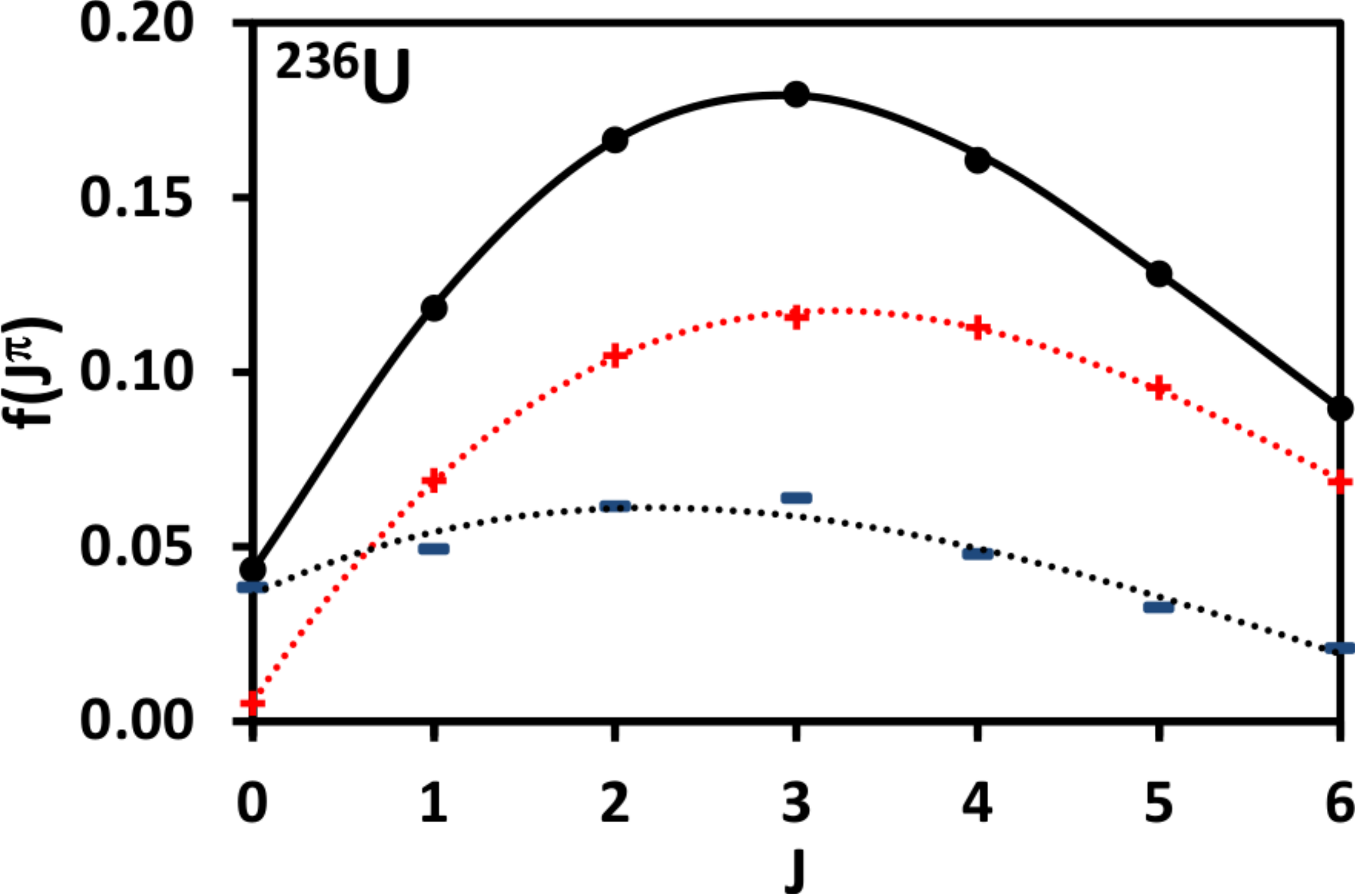}
    \includegraphics[width=8cm]{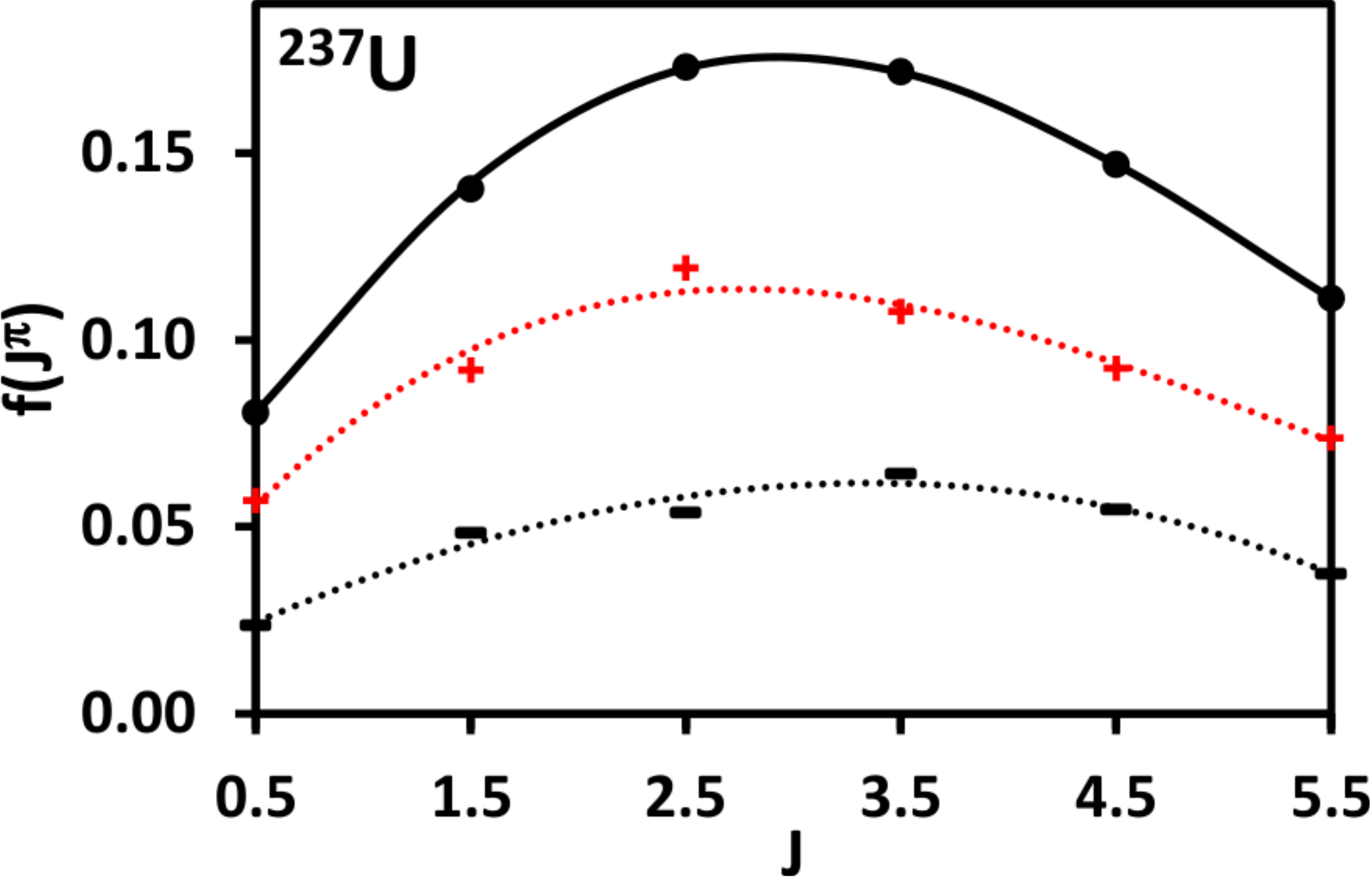}
    \includegraphics[width=8cm]{238U-1.pdf}
    \includegraphics[width=8cm]{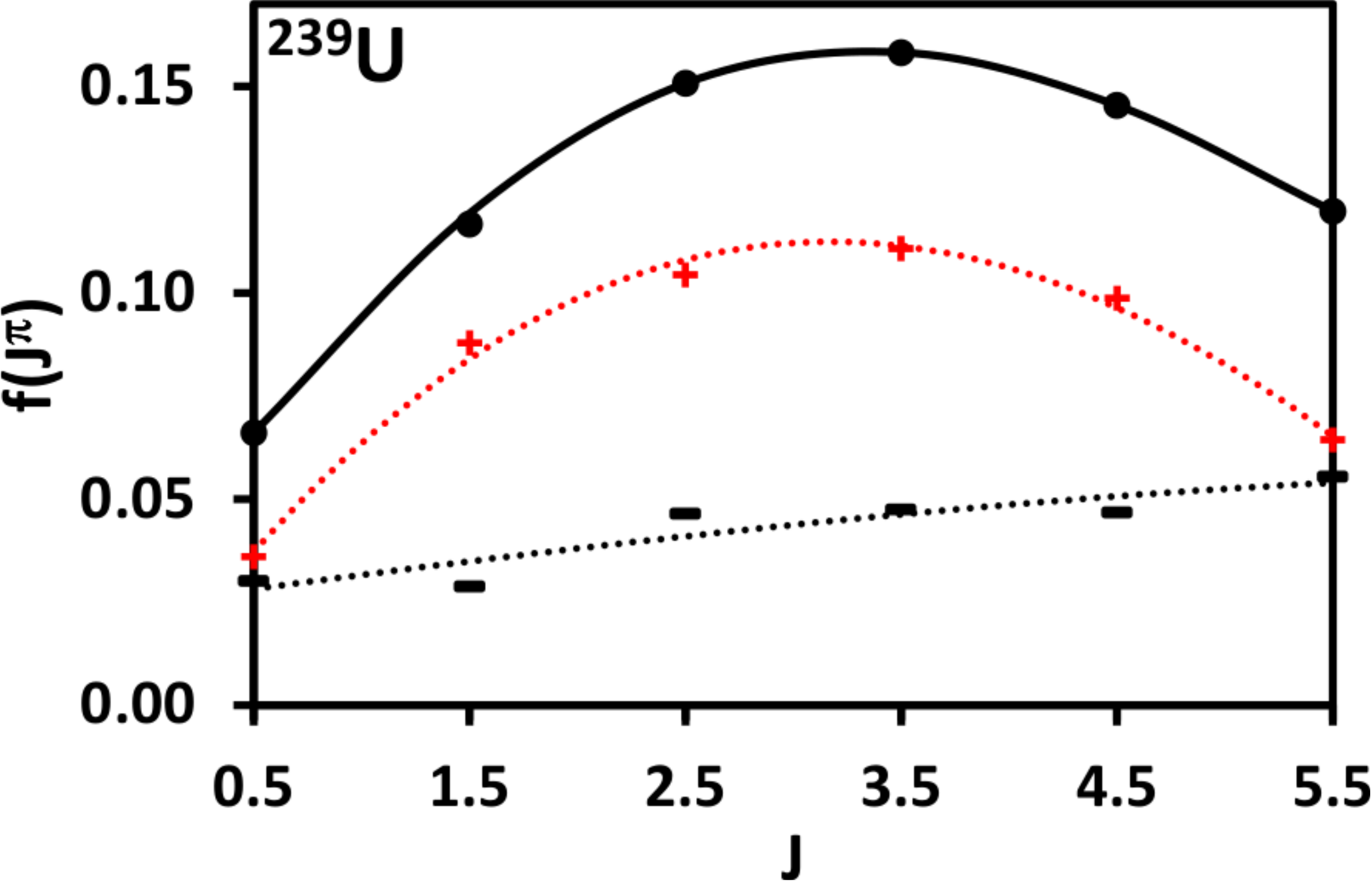}
  \caption{Fit of the CT-JPI model level densities ($\bullet$) to the spin distribution function for $^{233-239}$U (solid black lines).  The dotted lines show third order polynomial fits to the positive ($\textbf{\textcolor{red}+}$) and negative (\textbf{-}) parity $J^{\pi}$ fractions.}
  \label{Ufig}
\end{figure*}

The fitted positive parity, negative parity and total spin fractions at $S_n$ are plotted in Fig.~\ref{Ufig} where they are compared with the total spin fractions calculated with the spin distribution function with the spin cutoff parameters, $\sigma_c$, in Table~\ref{UCTfit}.  The fitted and calculated spin distributions differ by $\lesssim$1\%.  The distributions of both positive and negative parity spins are seen to vary smoothly and were fit with a third order polynomial to guide the eye in Fig.~\ref{Udata}.

For $^{233}$U the temperature and back shift, $E_0(1/2^+)$, were fit to the ${J^{\pi}=1/2^+}$ Yrast energy, 0.398 MeV, and the first 10 ${J=1/2^+}$ resonance energies to give $T$=0.440 MeV and $E_0 (1/2^+)$=0.399 MeV with $\overline{\Delta N}(1/2^+)$=0.137.  The back shifts, $E_0 (J^{\pi})$, for all levels with ${J^{\pi}=1/2^+,3/2^-,5/2,7/2,9/2,11/2}$ were then fit by the  minimization procedure, assuming a constant temperature, giving $\overline{\Delta N}(J^{\pi})$=0.135 averaged over 53 levels.  The average deviation of the fitted spin distribution from the fitted value is 0.5\%.

The temperature and back shifts for $^{234}$U were fit to the 10 known ${J^{\pi}=2^+,3^+}$ levels and 35 ${J^{\pi}=2^+,3^+}$ resonances giving $T$=0.444 MeV, $E_0 (2^+)$=1.092 MeV, and $E_0 (3^+)$=0.935 MeV with $\overline{\Delta N}(2^+,3^+)$=0.131.  The $E_0 (J^{\pi})$ back shifts for all levels with ${J^{\pi}=0^+,2,3,4,5,6}$ were fit assuming the constant temperature with $\overline{\Delta N}(J^{\pi})$=0.137 averaged over 111 levels.  The average deviation of the fitted spin distribution from the calculated value is 0.5\%.

For $^{235}$U the temperature and ${J=1/2^+}$ back shift were fit to the 2nd and 3rd known ${J^{\pi}=1/2^+}$ levels and the first 20 ${J^{\pi}=1/2^+}$ resonance energies to give $T$=0.455 MeV and ${E_0 (1/2^+)}$=0.327 MeV with $\overline{\Delta N}(1/2^+)$=0.110.  It is often found that a better fit is obtained by ignoring the Yrast energy when the fitted back shift is significantly above the Yrast value.  The $E_0 (J^{\pi})$ back shifts for all levels with ${J=1/2-11/2}$ were fit, assuming the constant temperature, with $\overline{\Delta N}(J^{\pi})$=0.127 averaged over 83 levels.  The average deviation of the fitted spin distribution from the calculated value is 0.9\%.

The temperature and back shifts for $^{236}$U were fit to the 9 known ${J^{\pi}=3^-,4^-}$ levels and 15 ${J^{\pi}=3^-,4^-}$ resonances giving $T$=0.428 MeV, $E_0 (3^-)$=0.994 MeV, and $E_0 (4^-)$=1.117 MeV with $\overline{\Delta N}(3^-,4^-)$=0.110.  The $E_0 (J^{\pi})$ back shifts for levels with ${J^{\pi}=0^+,1,2^-,3,4,5^+,6^-}$, were then fit, assuming the constant temperature, with $\overline{\Delta N}(J^{\pi})$=0.129 averaged over 56 levels.  The average deviation of the fitted spin distribution from the calculated value is 0.3\%.

For $^{237}$U the temperature and ${J^{\pi}=1/2^+}$ back shift were fit to the 2 known ${J^{\pi}=1/2^+}$ levels and the first 10 ${J=1/2^+}$ resonances energies to give $T$=0.416 MeV and $E_0 (1/2^+)$=0.402 MeV with $\overline{\Delta N}(1/2^+)$=0.114.  The $E_0 (J^{\pi})$ back shifts for levels with ${J^{\pi}=1/2,3/2,5/2^+,7/2^+,9/2^+,11/2^+}$ were then fit assuming a constant temperature with $\overline{\Delta N}(J^{\pi})$=0.116 averaged over 42 levels.  The average deviation of the fitted spin distribution from the calculated value is 0.3\%.

\begin{table*}[!ht]
\tabcolsep=5pt
\caption{\label{Pbdata} Back shifts, $E_0(J^{\pi})$, derived from the CT-JPI model, for $^{204-209}$Pb. }
\begin{tabular}{cdddddd}
\toprule
&\multicolumn{2}{c}{$\hspace{0.5cm}E_0 (J^{\pi}) ^{204}Pb$}&\multicolumn{2}{c}{$\hspace{0.5cm}E_0 (J^{\pi}) ^{206}Pb$}&\multicolumn{2}{c}{$\hspace{0.5cm}E_0 (J^{\pi}) ^{208}Pb$}\\
J&\multicolumn{1}{c}{$\hspace{0.5cm}\pi=+$}&\multicolumn{1}{c}{$\hspace{0.5cm}\pi=-$}&\multicolumn{1}{c}{$\hspace{0.5cm}\pi=+$}&\multicolumn{1}{c}{$\hspace{0.5cm}\pi=-$}&
\multicolumn{1}{c}{$\hspace{0.5cm}\pi=+$}&\multicolumn{1}{c}{$\hspace{0.5cm}\pi=-$}\\
\colrule
0&2.050&(4.136)&\hspace{0.5cm}2.575&(3.951)&\hspace{0.5cm}5.247&6.368\\
1&(1.320)&(3.472)&\hspace{0.5cm}1.800&3.751&\hspace{0.5cm}4.602&5.110\\
2&1.165&(2.426)&\hspace{0.5cm}1.615&3.502&\hspace{0.5cm}4.507&4.462\\
3&1.329&(1.743)&\hspace{0.5cm}1.708&(2.884)&\hspace{0.5cm}(4.496)&4.185\\
4&1.658&(1.529)&\hspace{0.5cm}1.923&2.866&\hspace{0.5cm}4.324&4.207\\
5&2.214&1.508&\hspace{0.5cm}(2.381)&2.969&\hspace{0.5cm}(4.395)&4.170\\
6&2.837&1.689&\hspace{0.5cm}(2.838)&3.113&\hspace{0.5cm}4.402&4.240\\
\toprule
&\multicolumn{2}{c}{$\hspace{0.5cm}E_0 (J^{\pi}) ^{205}Pb$}&\multicolumn{2}{c}{$\hspace{0.5cm}E_0 (J^{\pi}) ^{207}Pb$}&\multicolumn{2}{c}{$\hspace{0.5cm}E_0 (J^{\pi}) ^{209}Pb$}\\
J&\multicolumn{1}{c}{$\hspace{0.5cm}\pi=+$}&\multicolumn{1}{c}{$\hspace{0.5cm}\pi=-$}&\multicolumn{1}{c}{$\hspace{0.5cm}\pi=+$}&\multicolumn{1}{c}{$\hspace{0.5cm}\pi=-$}&
\multicolumn{1}{c}{$\hspace{0.5cm}\pi=+$}&\multicolumn{1}{c}{$\hspace{0.5cm}\pi=-$}\\
\colrule
1/2&3.213&0.769&\hspace{0.5cm}4.526&3.383&\hspace{0.5cm}3.383&2.395\\
3/2&(1.457)&0.574&\hspace{0.5cm}3.886&3.073&\hspace{0.5cm}2.637&2.195\\
5/2&(1.162)&0.614&\hspace{0.5cm}3.800&2.976&\hspace{0.5cm}2.553&2.329\\
7/2&(1.263)&0.764&\hspace{0.5cm}3.729&3.111&\hspace{0.5cm}(2.843)&2.601\\
9/2&1.592&1.026&\hspace{0.5cm}(3.836)&3.386&\hspace{0.5cm}(3.336)&(3.027)\\
11/2&$-$&$-$&\hspace{0.5cm}3.696&(4.446)&\hspace{0.5cm}(3.747)&3.953\\
\botrule
\end{tabular}
\end{table*}
\begin{table*}[!ht]
\tabcolsep=5pt
\caption{\label{PbCTfit} Neutron separation energies~\cite{AME2013}, $S_n$, spin cutoff parameters, $\sigma_c$, temperatures, $T$, resonance spacings, $D_0$ and $D_1$, and minimization uncertainty fitted to the CT-JPI model and compared with the values from RIPL-3~\cite{RIPL3} for $^{204-209}$Pb. }
\begin{tabular}{ldddddd}
\toprule
&\multicolumn{1}{c}{$^{204}$Pb}&\multicolumn{1}{c}{$^{205}$Pb}&\multicolumn{1}{c}{$^{206}$Pb}&\multicolumn{1}{c}{$^{207}$Pb}&\multicolumn{1}{c}{$^{208}$Pb}&
\multicolumn{1}{c}{$^{209}$Pb}\\
\colrule
$S_n$ (MeV)&8.3857&6.73166&8.08666&6.73778&7.36781&3.9373\\
$\sigma_c$&3.29&2.73&2.88&2.99&4.83&2.34\\
T $_{CT-JPI}$(MeV) &0.722&0.732&0.736&0.746&0.765&0.795\\
T $_{RIPL-3}$(MeV) &$-$&0.695&$-$&0.752&0.920&0.622\\
$D_0$(CT-JPI)(keV)&0.0510&5.97&0.427&38.4&33.5&396\\
$D_0$(RIPL-3)(keV)&$-$&2.1(2)&$-$&\multicolumn{1}{c}{\quad30(6)}&\multicolumn{1}{c}{\quad30(8)}&\multicolumn{1}{c}{\quad90(15)}\\
$D_1$(CT-JPI)(keV)&0.0104&0.0918&0.0338&3.30&8.03&50\\
$D_1$(RIPL-3)(keV)&$-$&0.69(6)&$-$&5.4(5)&4.8(8)&\multicolumn{1}{c}{\quad30(5)}\\
$\Delta$N&0.144&0.135&0.122&0.133&0.127&0.120\\
\botrule
\end{tabular}
\end{table*}

No spin assignments are known for the measured $^{238}$U resonances however 128 levels are known with ${J=1}$.  The temperature and back shifts for $^{238}$U were fit to the first 12 ${J=1^-}$ level energies and the first 17 ${J=1^+}$ levels giving $T$=0.420 MeV, $E_0 (1^-)$=0.0.764 MeV, and $E_0 (1^+)$=0.943 MeV with $\overline{\Delta N}(J=1)$=0.145.  The $E_0 (J^{\pi})$ back shifts for levels with ${J^{\pi}=0^+,1,2,3,4^+,6^+}$ were then fit, assuming the constant temperature with $\overline{\Delta N}(J^{\pi})$=0.133 averaged over 54 levels.  The average deviation of the fitted spin distribution from the calculated value is 0.6\%.

For $^{239}$U the temperature and ${J=1/2^+}$ back shift were fit to the 8 known ${J^{\pi}=1/2^+,1/2^-,3/2^-}$ level energies and 27 ${J^{\pi}=1/2^+,1/2^-,3/2^-}$ resonance energies to give $T$=0.419 MeV, $E_0 (1/2^+)$=0.665 MeV, $E_0 (1/2^-)$=0.740 MeV, and $E_0 (3/2^-)$=0.759 MeV with $\overline{\Delta N}(J^{\pi})$=0.144.  The $E_0 (1/2^+,1/2^-,3/2^-)$ back shifts for levels with ${J^{\pi}=1/2,3/2,5/2,7/2^+,9/2^+}$ were then fit assuming a constant temperature with $\overline{\Delta N}(J^{\pi})$=0.139 averaged over 56 levels.  The average deviation of the fitted spin distribution from the calculated value is 0.4\%.

\subsection{Lead}

The lead isotopes test of the CT-JPI model near the Z=82, N=126, doubly magic closed shell.  The model has been applied to the isotopes $^{204-209}$Pb where sufficient nuclear structure and resonance data are available to provide reasonable fits.  The fitted $E_0(J^{\pi})$ back shifts are shown in Table~\ref{Pbdata} and the corresponding neutron separation energies, $S_n$, temperatures, $T$, spin cutoff parameters, $\sigma_c$, resonance spacings, $D_0$ and $D_1$, and quality of fit, $\overline{\Delta N}(J^{\pi})$, are shown in Table~\ref{PbCTfit}.  The average fit is $\overline{\Delta N}(J^{\pi})$=0.130(9) in excellent agreement with the expected value from the folded Normal distribution. The temperatures and resonance spacings fitted to the CT-JPI model are compared with the values from RIPL-3~\cite{RIPL3} in Table~\ref{PbCTfit}.  The fitted temperatures vary wildly from RIPL-3 values with $^{208}$Pb 80\% of the RIPL-3 value and $^{205}$Pb 128\% of the RIPL-3 value.  As in the uranium isotopes the fitted $D_0$ and $D_1$ values vary up to of 7.5 times the RIPL-3 values.  The average spin cutoff parameter for $^{204-207,209}$Pb, $\sigma_c$=2.8(3), is far smaller that the value predicted by T. von Egidy \textit{et al}~\cite{Egidy88}, $\sigma_c$=4.6, which is slightly less than the value for doubly magic $^{208}$Pb with $\sigma_c$=4.83.
\begin{figure*}[!ht]
  \centering
    \includegraphics[width=8cm]{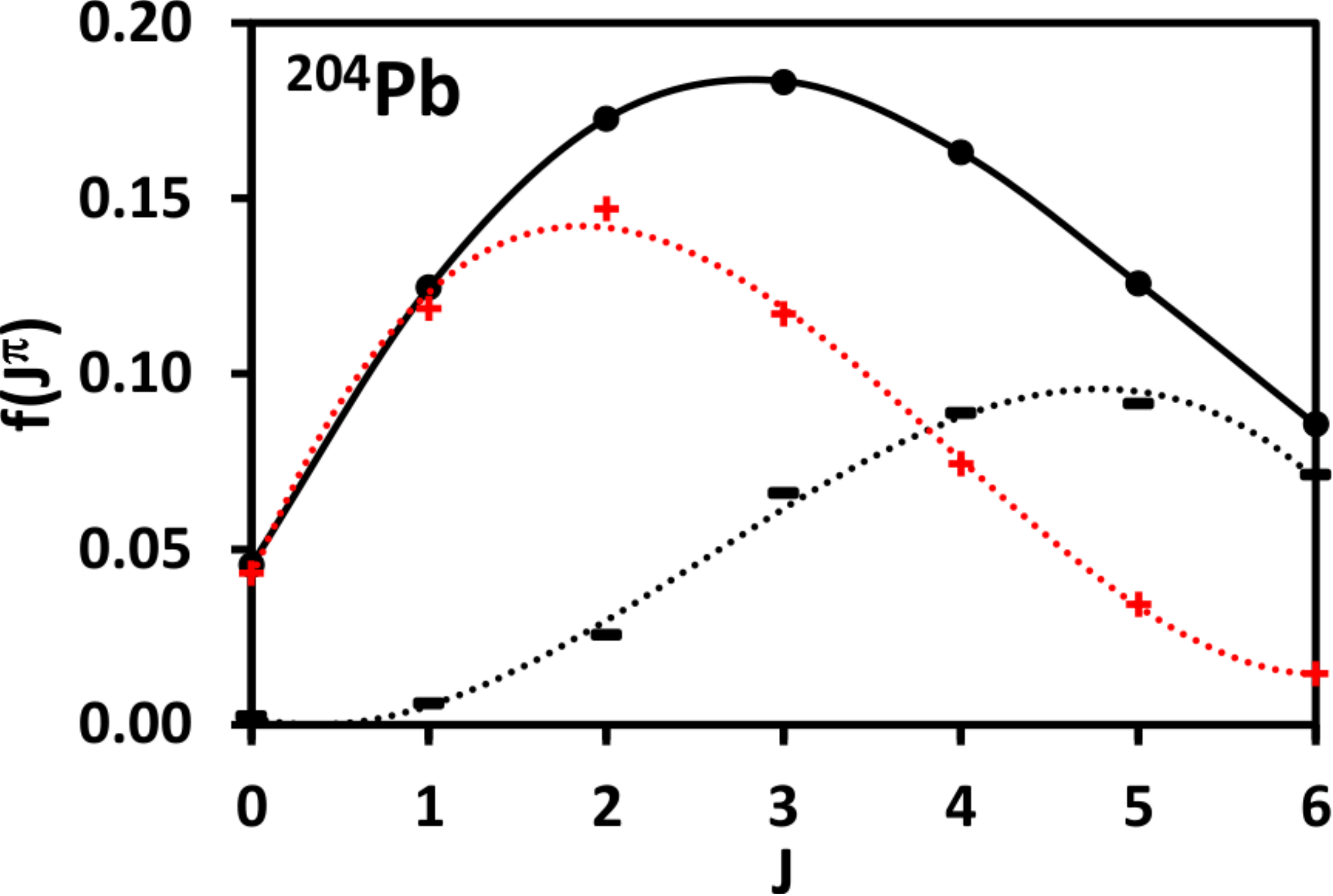}
    \includegraphics[width=8cm]{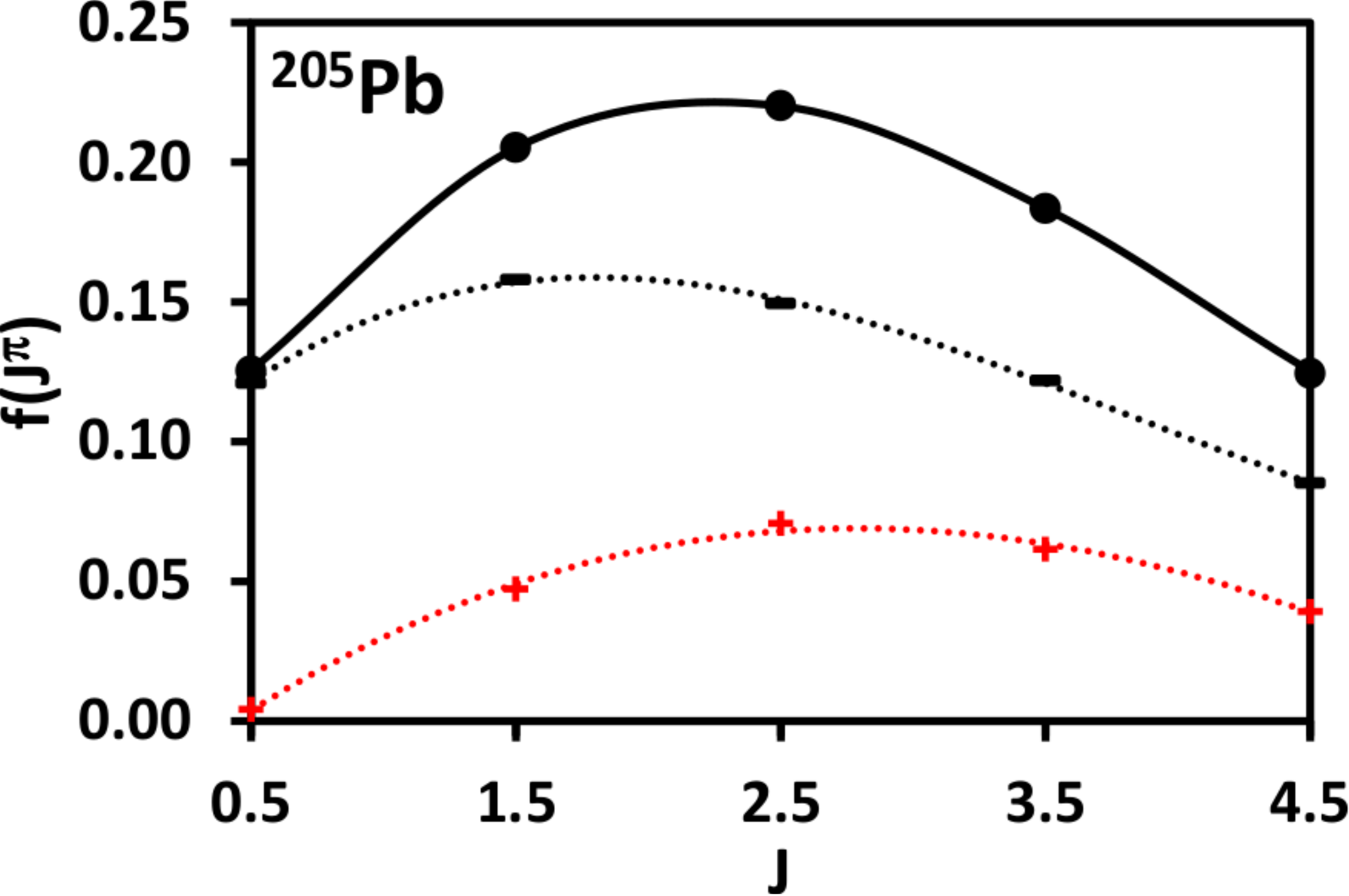}
    \includegraphics[width=8cm]{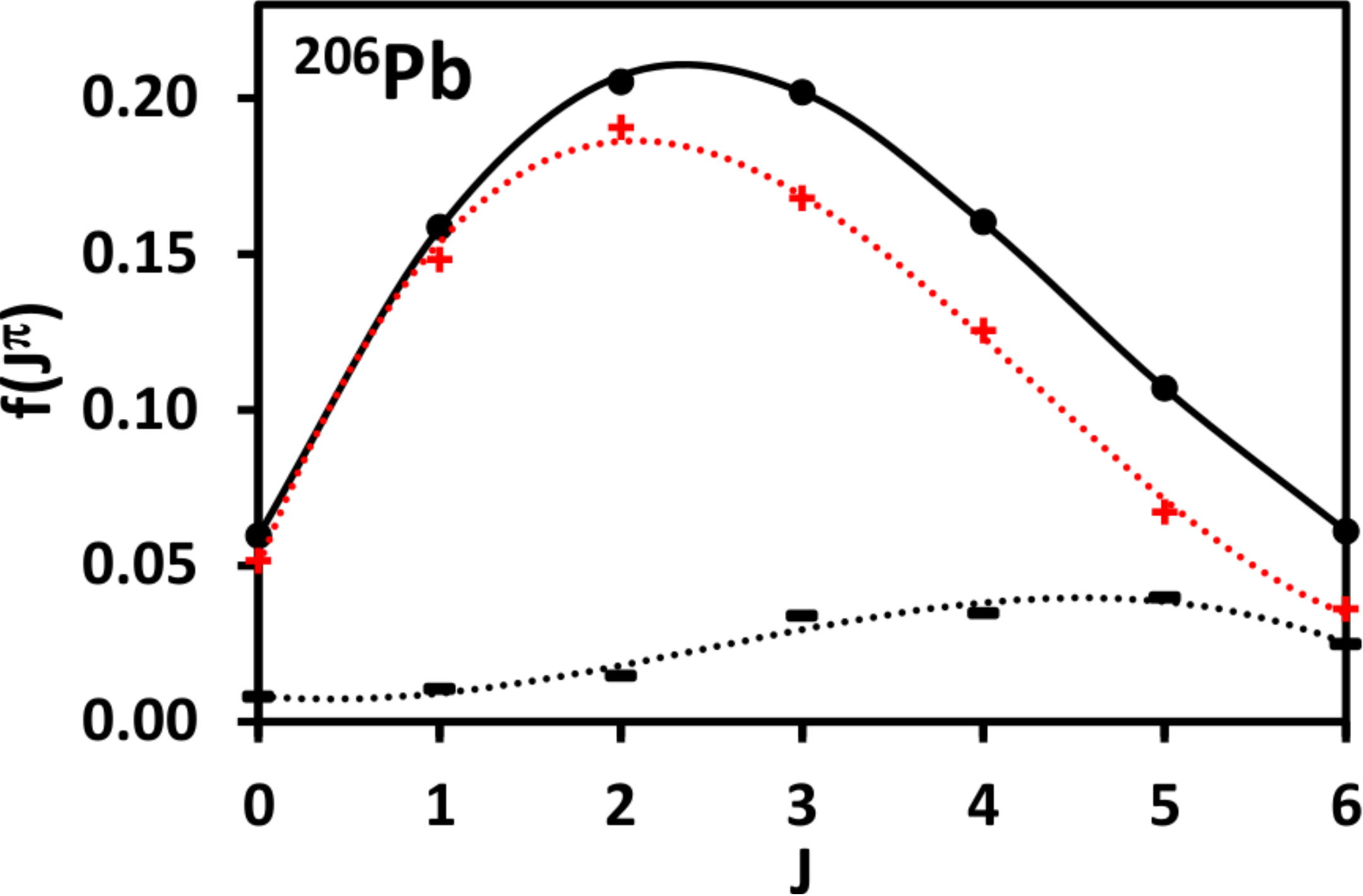}
    \includegraphics[width=8cm]{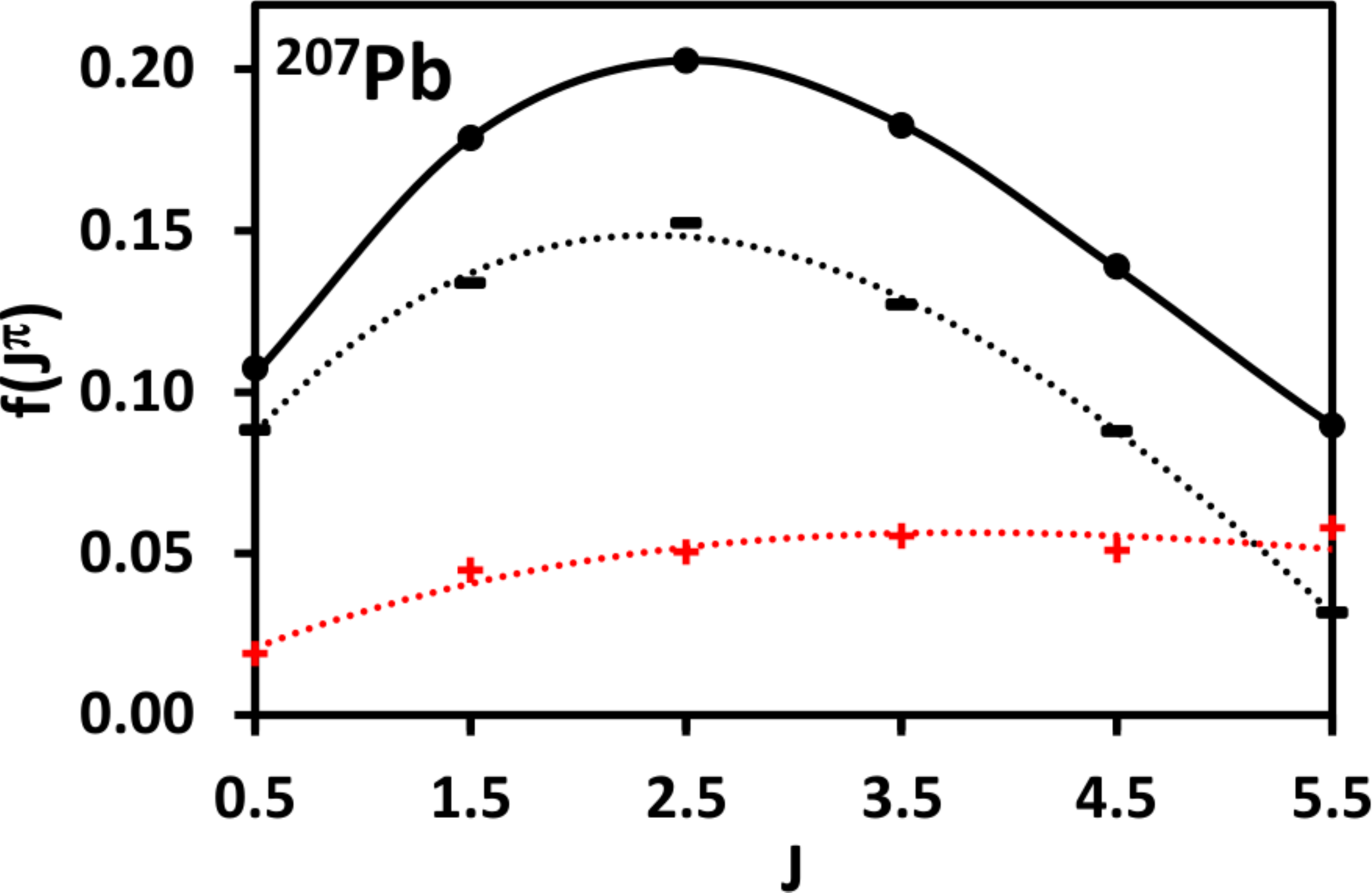}
    \includegraphics[width=8cm]{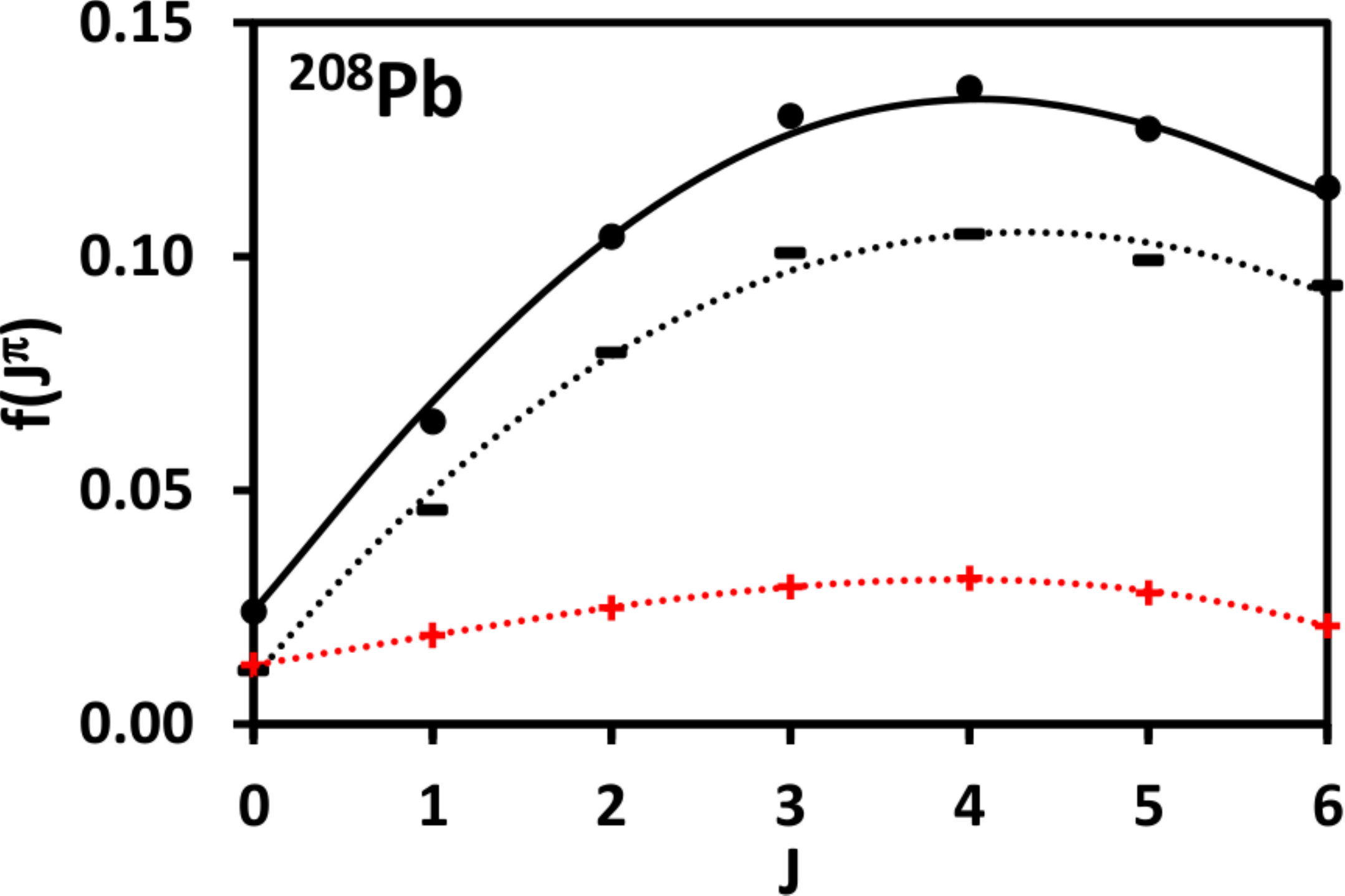}
    \includegraphics[width=8cm]{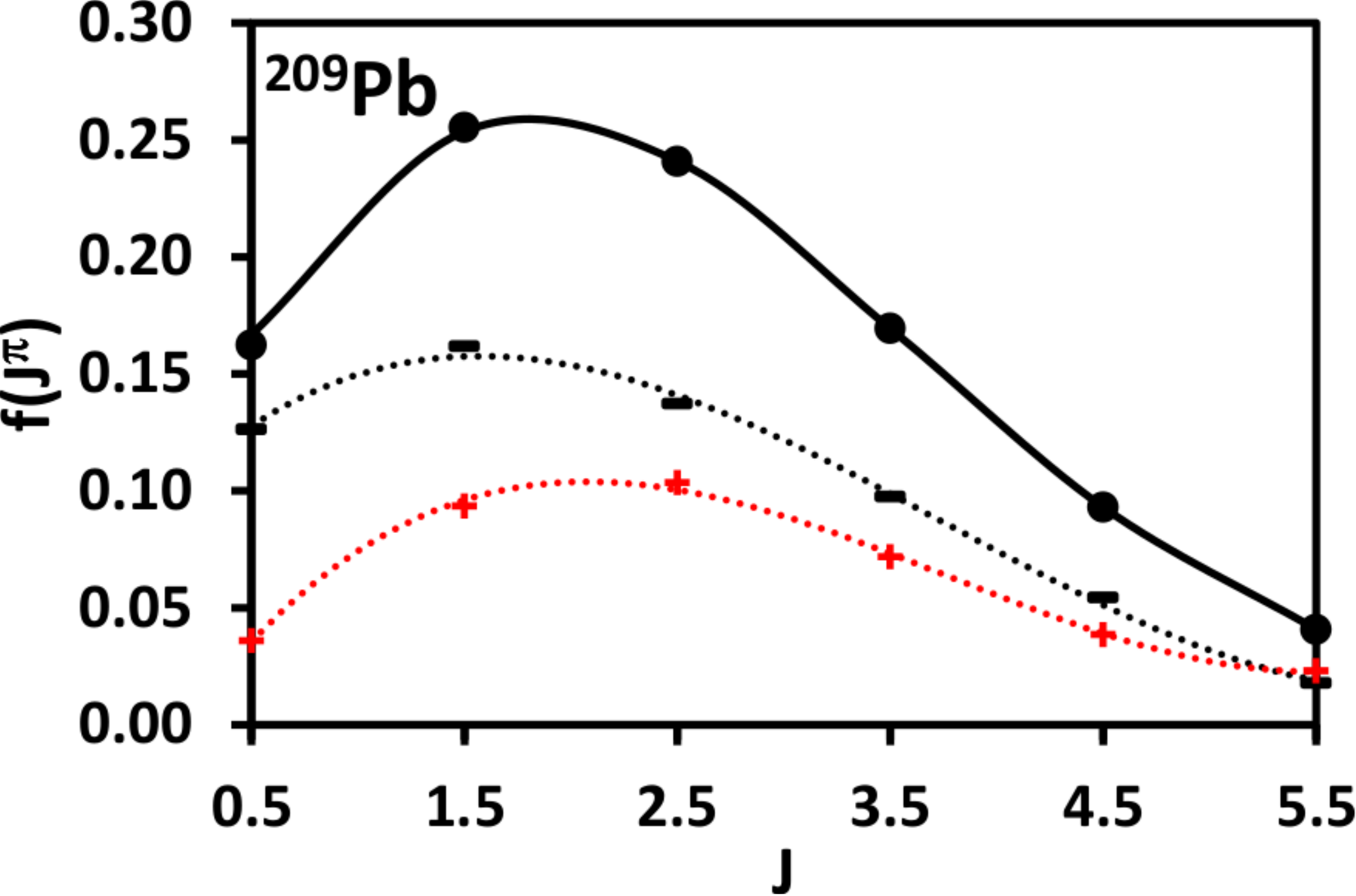}
  \caption{Fit of the CT-JPI model level densities ($\bullet$) to the spin distribution function for $^{204-209}$Pb (solid black lines).  The dotted lines show third order polynomial fits to the positive ($\textbf{\textcolor{red}+}$) and negative (\textbf{-}) parity $J^{\pi}$ fractions.}
  \label{Pbfig}
\end{figure*}

The fitted positive parity, negative parity and total spin fractions are show shown in Fig.~\ref{Pbfig} where they are compared to the spin fractions from the spin distribution function assuming the spin cutoff parameters, $\sigma_c$, shown in Table~\ref{PbCTfit}.  The distribution of both positive and negative parity spins varies smoothly as shown by a polynomial fit in Fig.~\ref{Pbfig}.  This fit is presented only to guide the eye and no fundamental importance should be taken from this.  The fitted and calculated spin distributions differ by $\lesssim$1\%.  The distributions of both positive and negative parity spins are seen to vary smoothly and were fit with a third order polynomial to guide the eye in Fig.~\ref{Pbfig}.

For $^{204}$Pb there are no resonance data so the temperature and back shift were fit to the energies of 8 well established ${J^{\pi}=2^+}$ levels giving $T$=0.722 MeV, $E_0 (2^+)$=1,165 MeV, and $E_0 (3^+)$=0.935 MeV with $\overline{\Delta N}(2^+)$=0.150.  The $E_0 (J^{\pi})$ back shifts for all levels with ${J^{\pi}=0^+,2^+,3^+,4^+,5,6}$ were then fit, assuming the constant temperature, with $\overline{\Delta N}(J^{\pi})$=0.144 averaged over 37 levels.  The average deviation of the fitted spin distribution from the calculated value is 0.08\%.

Both s-wave and p-wave resonance data are available for $^{205}$Pb so the temperature is fit to the ${J^{\pi}=1/2^+,1/2^-,3/2^-}$ level and resonance data giving $T$=0.732 MeV with $\overline{\Delta N}(1/2^+,1/2^-,3/2^-)$=0.138.  The $E_0 (J^{\pi})$ back shifts for all levels with ${J=1/2,3/2^-,5/2^-,7/2^-,9/2}$ were then fit, assuming the constant temperature, with $\overline{\Delta N}(J^{\pi})$=0.135 averaged over 46 levels.  No data were available for J=11/2 levels.  The average deviation of the fitted spin distribution from the calculated value is 0.2\%.
\begin{table*}[!ht]
\tabcolsep=5pt
\caption{\label{Hodata} Back shifts, $E_0(J^{\pi})$, derived from the CT-JPI model, for $^{163-166}$Ho. }
\begin{tabular}{cddddcdddd}
\toprule
&\multicolumn{2}{c}{$\hspace{0.5cm}E_0 (J^{\pi}) ^{164}Ho$}&\multicolumn{2}{c}{$\hspace{0.5cm}E_0 (J^{\pi}) ^{166}Ho$}&&\multicolumn{2}{c}{$\hspace{0.5cm}E_0 (J^{\pi}) ^{163}Ho$}&\multicolumn{2}{c}{$\hspace{0.5cm}E_0 (J^{\pi}) ^{165}Ho$}\\
J&\multicolumn{1}{c}{$\hspace{0.5cm}\pi=+$}&\multicolumn{1}{c}{$\hspace{0.5cm}\pi=-$}&\multicolumn{1}{c}{$\hspace{0.5cm}\pi=+$}&\multicolumn{1}{c}{$\hspace{0.5cm}\pi=-$}&
J&\multicolumn{1}{c}{$\hspace{0.5cm}\pi=+$}&\multicolumn{1}{c}{$\hspace{0.5cm}\pi=-$}&\multicolumn{1}{c}{$\hspace{0.5cm}\pi=+$}&\multicolumn{1}{c}{$\hspace{0.5cm}\pi=-$}\\
\colrule
0&$-$&$-$&(1.756)&(1.044)&\hspace{0.0cm}1/2&0.450&(1.247)&1.013&\hspace{0.5cm}(0.901)\\
1&(0.325)&(0.999)&0.519&(0.943)&\hspace{0.0cm}3/2&0.305&0.525&0.692&\hspace{0.5cm}(0.651)\\
2&(0.186)&(0.660)&0.271&0.726&\hspace{0.0cm}5/2&0.339&0.252&0.674&\hspace{0.5cm}0.547\\
3&0.170&0.537&0.170&0.551&\hspace{0.0cm}7/2&0.431&0.151&0.705&\hspace{0.5cm}0.614\\
4&0.262&(0.512)&0.194&0.414&\hspace{0.0cm}9/2&0.588&0.180&0.820&\hspace{0.5cm}0.825\\
5&0.389&(0.630)&0.264&0.323&\hspace{0.0cm}11/2&0.688&0.338&0.916&\hspace{0.5cm}1.217\\
6&(0.570)&(0.830)&0.376&0.305&\hspace{0.0cm}&&&&\hspace{0.5cm}\\
\botrule
\end{tabular}
\end{table*}

No resonance data are available for  $^{206}$Pb however considerable data are available for ${J^{\pi}=1^-,2^+}$ levels.  The temperature and back shift were fit to the energies of twelve well established ${J^{\pi}=1^-,2^+}$ levels giving $T$=0.736 MeV  with $\overline{\Delta N}(1^-,2^+)$=0.137.   The $E_0 (J^{\pi})$ back shifts for all levels with ${J=0^+,1,2,3^+,4,5^-,6}$ were then fit, assuming the constant temperature, with $\overline{\Delta N}(J^{\pi})$=0.122 averaged over 40 levels.  The average deviation of the fitted spin distribution from the calculated value is 0.2\%.

For $^{207}$Pb there are considerable s-wave, p-wave, and d-wave resonance data available so the temperature is fit to the ${J^{\pi}=1/2^+,1/2^-,3/2^+,3/2^-5/2^+}$ level and resonance data giving $T$=0.732 MeV with ${\overline{\Delta N}(1/2^+,1/2^-,3/2^+,3/2^-5/2^+)=0.143}$. The $E_0 (J^{\pi})$ back shifts for all levels with ${J=1/2,3/2,5/2,7/2,9/2^-,11/2^+}$ were then fit, assuming the constant temperature, with $\overline{\Delta N}(J^{\pi})$=0.133 averaged over 77 levels.  The average deviation of the fitted spin distribution from the calculated value is 0.4\%.

For the doubly magic nucleus $^{208}$Pb there are considerable s-wave, p-wave, and d-wave resonance data available so the temperature is fit to the ${J^{\pi}=0,1,2,3^-}$ resonance and level data giving $T$=0.765 MeV with $\overline{\Delta N}(0,1,2,3^-)$=0.139.  The $E_0 (J^{\pi})$ back shifts for all levels with ${J=0,1,2,3^-,4,5^-,6}$ were then fit, assuming the constant temperature, with $\overline{\Delta N}(J^{\pi})$=0.127 averaged over 143 levels.  The average deviation of the fitted spin distribution from the calculated value is 0.7\%.  The spin cutoff parameter, $\sigma_c$=4.83, is considerably larger than for the other lead isotopes although comparable to the von Egidy \textit{et al}~\cite{Sigmac} prediction, $\sigma_c$(calc)=4.61, suggesting that $\sigma_c$ is strongly dependent on nuclear structure.

A wide range of resonance data with ${J^{\pi}=1/2,3/2,5/2,7/2^-}$ is available for $^{209}$Pb which give temperature, $T$=0.795 MeV, with $\overline{\Delta N}(J^{\pi})$=0.130.  The $E_0 (J^{\pi})$ back shifts for all levels with ${J=1/2,3/2,5/2,7/2^-,11/2^-}$ were then fit, assuming the constant temperature, with $\overline{\Delta N}(J^{\pi})$=0.120 averaged over 128 levels.   The average deviation of the fitted spin distribution from the calculated value is 0.5\%.

\subsection{Holmium}

The holmium isotopes test the CT-JPI model for odd-Z, odd/even-N nuclei.  The model has been applied to the isotopes $^{163-166}$Ho where sufficient nuclear structure and resonance data are available to provide reasonable fits.  The fitted $E_0(J^{\pi})$ back shifts are shown in Table~\ref{Hodata} and the corresponding neutron separations, $S_n$, temperatures, $T$, spin cutoff parameters, $\sigma_c$, resonance spacings, $D_0$ and $D_1$, and quality of fit, $\overline{\Delta N}(J^{\pi})$, are shown in Table~\ref{HoCTfit}.  The average fit is $\overline{\Delta N}(J^{\pi})$=0.112(21) in excellent agreement with the expected value from the folded Normal distribution.  The temperatures and resonance spacings fitted to the CT-JPI model are compared with the values from RIPL-3~\cite{RIPL3} in Table~\ref{DyCTfit}.  The fitted temperature and $D_0$ value for $^{166}$Ho are both $\approx11\%$ higher than the  RIPL-3 values.  The average spin cutoff parameter for $^{163-165}$Ho, $\sigma_c$=3.3(3), is smaller that the value predicted by T. von Egidy \textit{et al}~\cite{Egidy88}, $\sigma_c$=4.3, but comparable to the values found for other nuclei, but the value for $^{166}$Ho, $\sigma_c$=4.82, is 12\% higher than the von Egidy value.

\begin{table}[!ht]
\tabcolsep=0pt
\caption{\label{HoCTfit} Neutron separation energies~\cite{AME2013}, $S_n$, spin cutoff parameters, $\sigma_c$, temperatures, $T$, resonance spacings, $D_0$, and minimization uncertainty fitted to the CT-JPI model and compared with the values from RIPL-3~\cite{RIPL3} for $^{163-166}$Ho. }
\begin{tabular}{ldddd}
\toprule
&\multicolumn{1}{c}{$^{163}$Ho}&\multicolumn{1}{c}{$^{164}$Ho}&\multicolumn{1}{c}{$^{165}$Ho}&\multicolumn{1}{c}{$^{166}$Ho}\\
\colrule
$S_n$ (MeV)&8.408&6.6745&6.2206&6.7483\\
$\sigma_c$&3.60&3.45&2.96&4.82\\
T $_{CT-JPI}$(MeV) &0.535&0.559&0.540&0.571\\
T $_{RIPL-3}$(MeV) &$-$&$-$&$-$&0.512\\
$D_0$(CT-JPI)(eV)&100&4.86&12.5&4.76\\
$D_0$(RIPL-3)(eV)&$-$&$-$&$-$&\multicolumn{1}{c}{\quad4.2(5)}\\
$D_1$(CT-JPI)(eV)&72.7&1.55&6.31&1.50\\
$\Delta$N&0.116&0.085&0.110&0.137\\
\botrule
\end{tabular}
\end{table}

\begin{figure}[!ht]
  \centering
    \includegraphics[width=8cm]{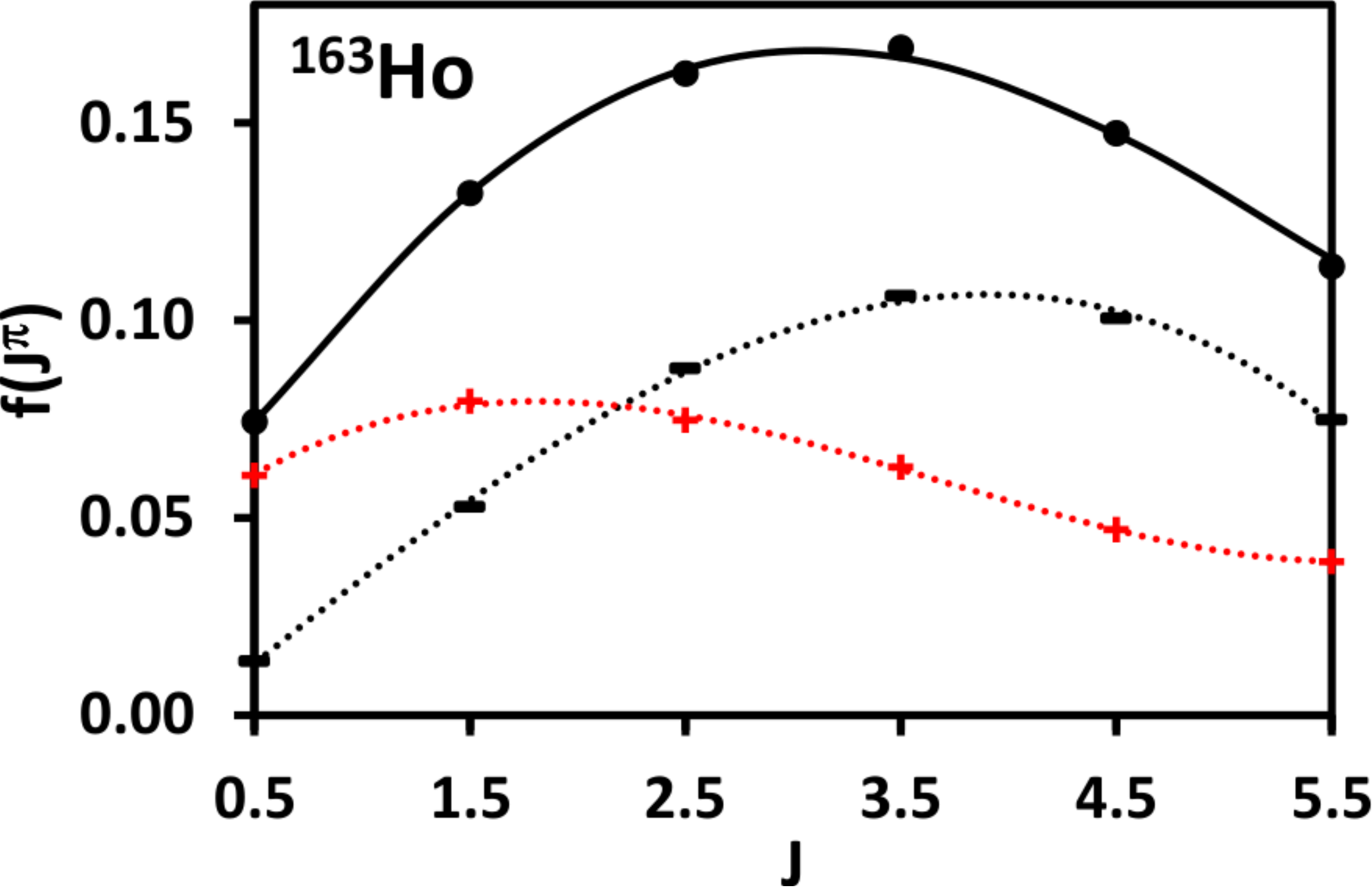}
    \includegraphics[width=8cm]{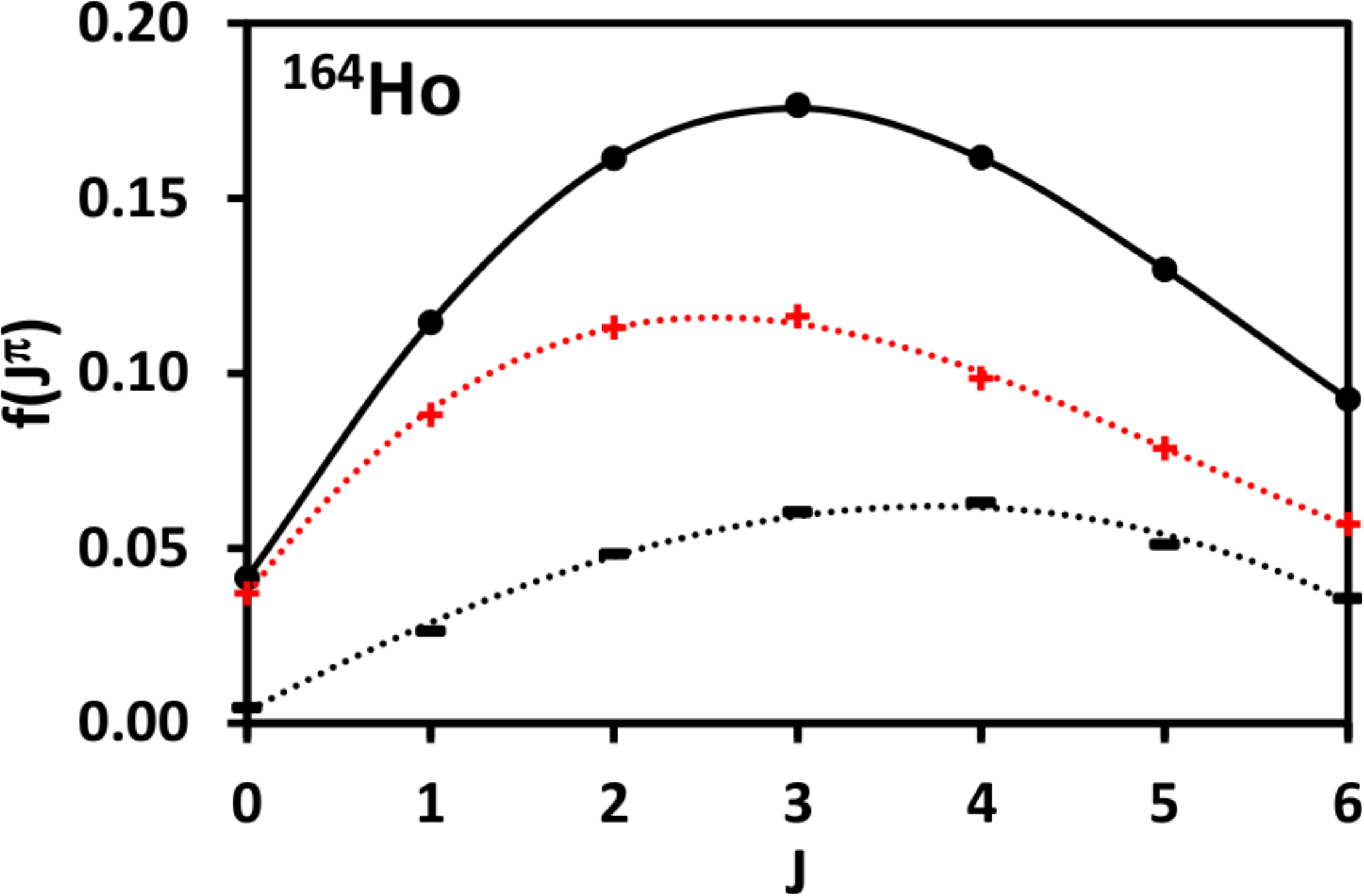}
    \includegraphics[width=8cm]{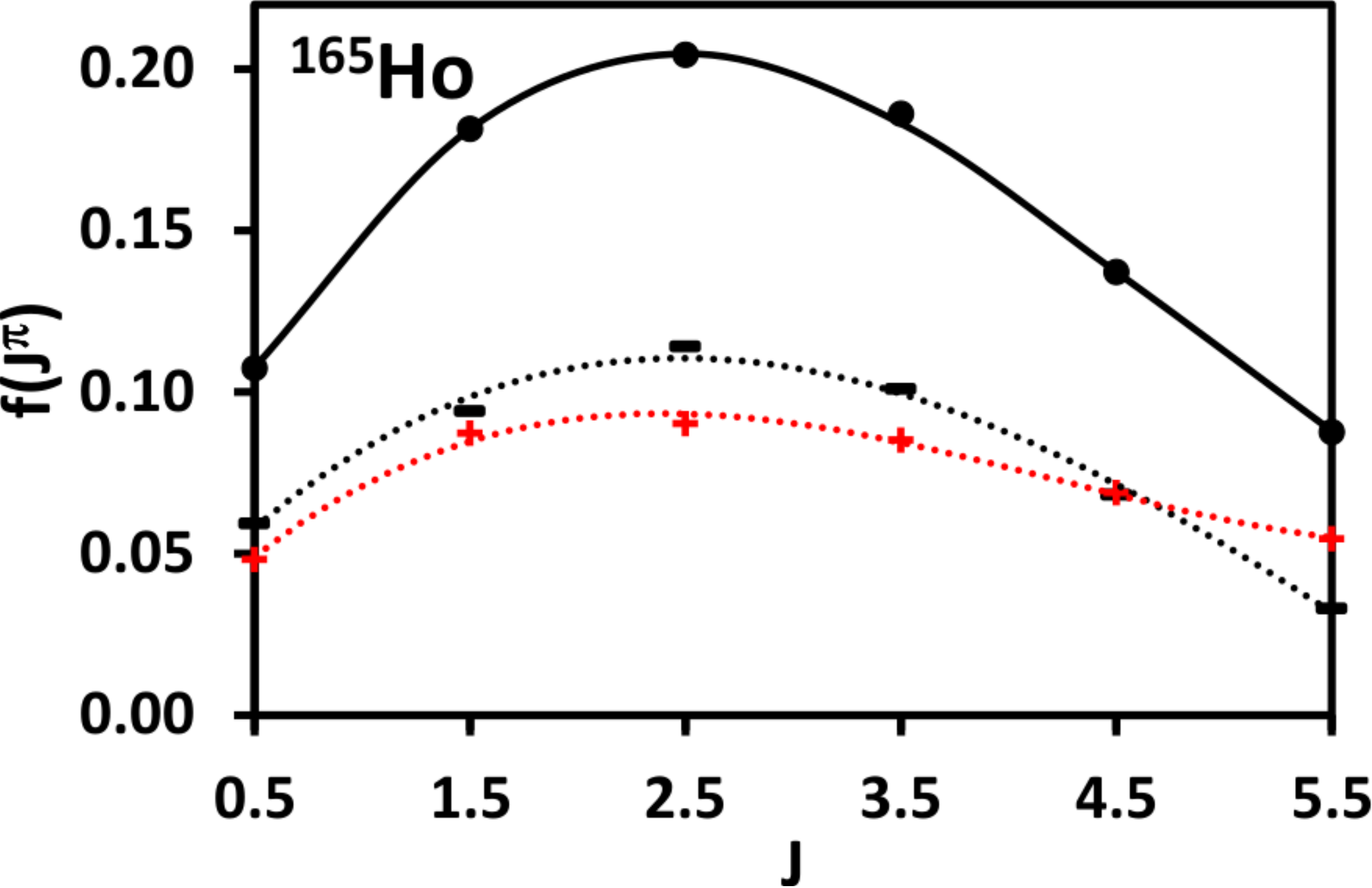}
    \includegraphics[width=8cm]{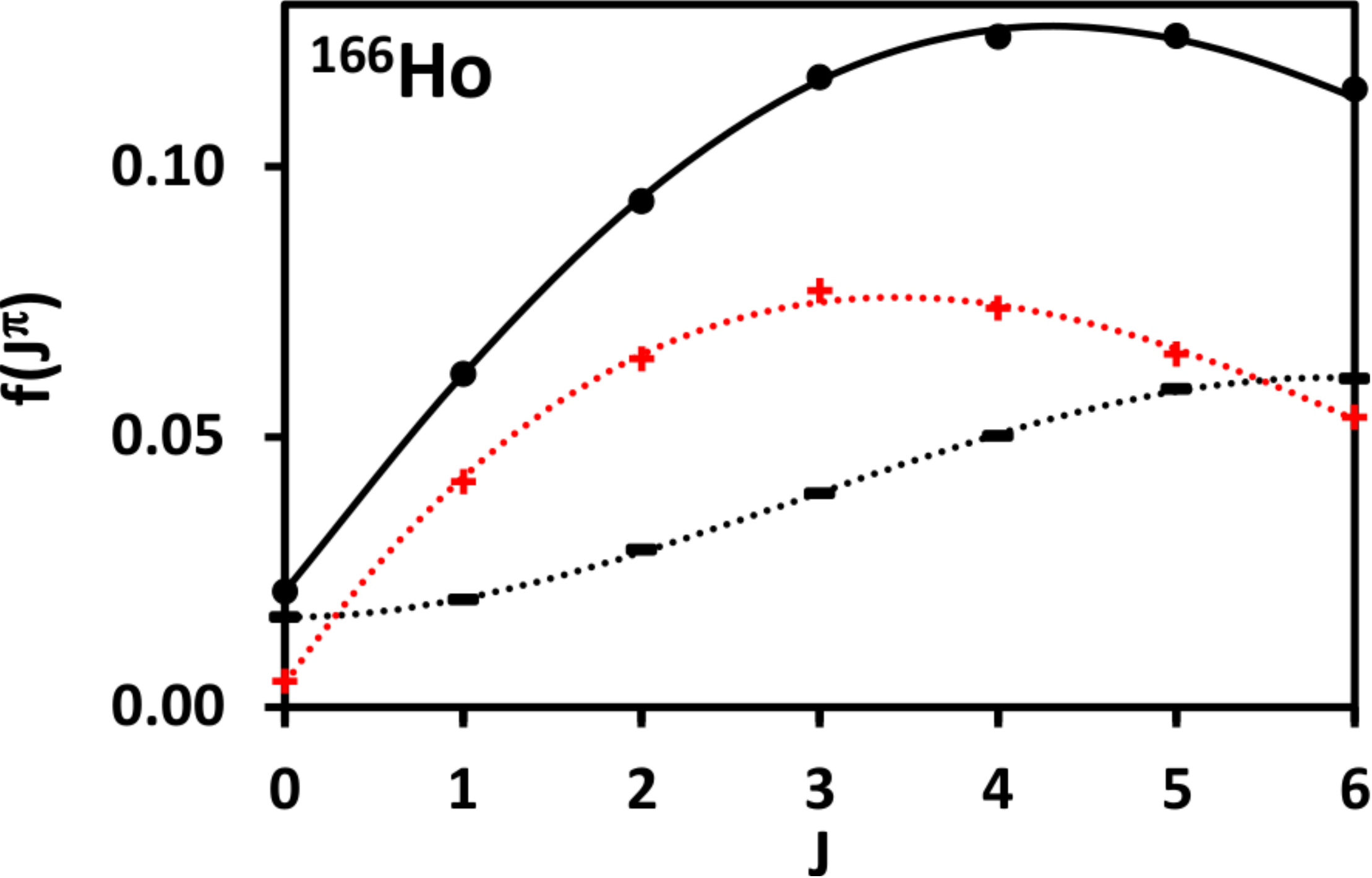}
  \caption{Fit of the CT-JPI model level densities ($\bullet$) to the spin distribution function for $^{164-166}$Ho (solid black lines).  The dotted lines show third order polynomial fits to the positive ($\textbf{\textcolor{red}+}$) and negative (\textbf{-}) parity $J^{\pi}$ fractions.}
  \label{Hofig}
\end{figure}

The fitted positive parity, negative parity and total spin fractions are show shown in Fig.~\ref{Hofig} where they are compared to the spin fractions from the spin distribution function assuming the spin cutoff parameters, $\sigma_c$, shown in Table~\ref{HoCTfit}.  The distribution of both positive and negative parity spins varies smoothly as shown by a polynomial fit in Fig.~\ref{Hofig}.  This fit is presented only to guide the eye and no fundamental importance should be taken from this.  The fitted and calculated spin distributions differ by $<$0.7\%.  The distributions of both positive and negative parity spins are seen to vary smoothly and were fit with a third order polynomial to guide the eye in Fig.~\ref{Hofig}.

There are no resonance data for $^{163}$Ho so the temperature and back shifts were fit to 34 level energies with ${J=1/2^+,3/2,5/2,7/2,9/2,11/2}$ giving $T$=0.535 MeV and $\overline{\Delta N}(J^{\pi})$=0.116.    The average deviation of the fitted spin distribution from the calculated value is 0.7\%.

For $^{164}$Ho there are no resonance data so the temperature and back shifts were fit to 12 level energies with ${J=3,4^+,5^+}$ giving $T$=0.559 MeV and $\overline{\Delta N}(J^{\pi})$=0.085.  The average deviation of the fitted spin distribution from the calculated value is 0.1\%.  Although only a minimal amount of nuclear structure data are available for $^{164}$Ho an excellent fit to the data was obtained.

No resonance data exist for $^{165}$Ho so the temperature and back shifts were fit to 32 level energies with ${J=1/2^+,3/2^+,5/2,7/2,9/2,11/2}$ giving $T$=0.540 MeV and $\overline{\Delta N}(J^{\pi})$=0.110.  The average deviation of the fitted spin distribution from the calculated value is 0.4\%.

Both s-wave resonance data and considerable nuclear structure data exist for $^{166}$Ho.  The temperature and back shifts could be fit to the ${J^{\pi}=3^-,4^-}$ resonances and levels giving $T$=0.571 MeV, $E_0 (3^-)$=0.551 MeV, and $E_0 (4^-)$=0.414 MeV with $\overline{\Delta N}(3^-,4^-)$=0.144 averaged over 23 levels.  The $E_0 (J^{\pi})$ back shifts for levels with ${J=1^+,2,3,4,5,6}$ were fit, assuming a constant temperature, with $\overline{\Delta N}(J^{\pi})$=0.137 averaged over 65 levels.  The average deviation of the fitted spin distribution from the calculated value is 0.6\%.
\begin{table*}[!ht]
\tabcolsep=5pt
\caption{\label{Dydata} Back shifts, $E_0(J^{\pi})$, derived from the CT-JPI model, for $^{160-165}$Dy. }
\begin{tabular}{cdddddd}
\toprule
&\multicolumn{2}{c}{$\hspace{0.5cm}E_0 (J^{\pi}) ^{160}Dy$}&\multicolumn{2}{c}{$\hspace{0.5cm}E_0 (J^{\pi}) ^{162}Dy$}&\multicolumn{2}{c}{$\hspace{0.5cm}E_0 (J^{\pi}) ^{164}Dy$}\\
J&\multicolumn{1}{c}{$\hspace{0.5cm}\pi=+$}&\multicolumn{1}{c}{$\hspace{0.5cm}\pi=-$}&\multicolumn{1}{c}{$\hspace{0.5cm}\pi=+$}&\multicolumn{1}{c}{$\hspace{0.5cm}\pi=-$}&
\multicolumn{1}{c}{$\hspace{0.5cm}\pi=+$}&\multicolumn{1}{c}{$\hspace{0.5cm}\pi=-$}\\
\colrule
0&1.554&(2.780)&\hspace{0.5cm}2.154&(2.023)&\hspace{0.5cm}1.780&(1.727)\\
1&1.168&1.436&\hspace{0.5cm}1.546&1.461&\hspace{0.5cm}0.972&1.367\\
2&0.945&1.341&\hspace{0.5cm}1.337&1.215&\hspace{0.5cm}0.761&1.234\\
3&0.874&1.283&\hspace{0.5cm}1.230&1.197&\hspace{0.5cm}0.745&1.106\\
4&0.902&1.377&\hspace{0.5cm}1.126&1.442&\hspace{0.5cm}0.775&1.156\\
5&0.990&1.527&\hspace{0.5cm}1.225&1.560&\hspace{0.5cm}(0.915)&1.275\\
6&1.298&1.587&\hspace{0.5cm}1.324&(2.027)&\hspace{0.5cm}1.190&1.393\\
\toprule
&\multicolumn{2}{c}{$\hspace{0.5cm}E_0 (J^{\pi}) ^{161}Dy$}&\multicolumn{2}{c}{$\hspace{0.5cm}E_0 (J^{\pi}) ^{163}Dy$}&\multicolumn{2}{c}{$\hspace{0.5cm}E_0 (J^{\pi}) ^{165}Dy$}\\
J&\multicolumn{1}{c}{$\hspace{0.5cm}\pi=+$}&\multicolumn{1}{c}{$\hspace{0.5cm}\pi=-$}&\multicolumn{1}{c}{$\hspace{0.5cm}\pi=+$}&\multicolumn{1}{c}{$\hspace{0.5cm}\pi=-$}&
\multicolumn{1}{c}{$\hspace{0.5cm}\pi=+$}&\multicolumn{1}{c}{$\hspace{0.5cm}\pi=-$}\\
\colrule
1/2&0.890&0.493&\hspace{0.5cm}0.900&0.752&\hspace{0.5cm}0.981&(0.372)\\
3/2&0.420&0.312&\hspace{0.5cm}0.499&0.533&\hspace{0.5cm}0.463&0.157\\
5/2&0.247&0.212&\hspace{0.5cm}0.482&0.445&\hspace{0.5cm}0.273&0.155\\
7/2&0.097&0.180&\hspace{0.5cm}(0.571)&0.490&\hspace{0.5cm}(0.241)&0.299\\
9/2&(0.049)&0.228&\hspace{0.5cm}(0.875)&0.588&\hspace{0.5cm}(0.311)&(0.574)\\
11/2&(-0.006)&0.443&\hspace{0.5cm}(1.183)&0.851&\hspace{0.5cm}(0.502)&(0.921)\\
\botrule
\end{tabular}
\end{table*}

\subsection{Dysprosium}

The dysprosium nuclei test of the CT-JPI model in a region of high deformation.  The model has been applied to the isotopes $^{160-165}$Dy where sufficient nuclear structure and resonance data are available to provide reasonable fits.   The fitted $E_0(J^{\pi})$ back shifts are shown in Table~\ref{Dydata} and the corresponding neutron separation energies, $S_n$, temperatures, $T$, spin cutoff parameters, $\sigma_c$, resonance spacings, $D_0$ and $D_1$, and quality of fit, $\overline{\Delta N}(J^{\pi})$, are shown in Table~\ref{DyCTfit}.  The average fit is $\overline{\Delta N}(J^{\pi})$=0.129(12) in excellent agreement with the expected value from the folded Normal distribution.  The temperatures and resonance spacings fitted to the CT-JPI model are compared with the values from RIPL-3~\cite{RIPL3} in Table~\ref{DyCTfit}.  The fitted temperatures average ${\approx4\%}$ higher than the  RIPL-3 values .  The fitted $D_0$ values are comparable to RIPL-3 values for $^{161-164}$Dy, but the $^{165}$Dy $D_0$ value is 3 times the RIPL-3 value.  The average spin cutoff parameter for $^{160-165}$Dy, $\sigma_c$=3.5(6), is smaller that the value predicted by T. von Egidy \textit{et al}~\cite{Egidy88}, $\sigma_c$=4.3, but comparable to the values found for other nuclei.

\begin{table*}[t]
\tabcolsep=5pt
\caption{\label{DyCTfit} Neutron separation energies~\cite{AME2013}, $S_n$, spin cutoff parameters, $\sigma_c$, temperatures, $T$, resonance spacings, $D_0$, and minimization uncertainty fitted to the CT-JPI model and compared with the values from RIPL-3~\cite{RIPL3} for $^{160-165}$Dy. }
\begin{tabular}{ldddddd}
\toprule
&\multicolumn{1}{c}{$^{160}$Dy}&\multicolumn{1}{c}{$^{161}$Dy}&\multicolumn{1}{c}{$^{162}$Dy}&\multicolumn{1}{c}{$^{163}$Dy}&\multicolumn{1}{c}{$^{164}$Dy}&
\multicolumn{1}{c}{$^{165}$Dy}\\
\colrule
$S_n$ (MeV)&8.5765&6.45439&8.19699&6.27101&7.65811&5.716\\
$\sigma_c$&3.54&4.62&3.50&2.90&3.47&3.02\\
T $_{CT-JPI}$(MeV) &0.575&0.548&0.591&0.590&0.593&0.588\\
T $_{RIPL-3}$(MeV) &$-$&0.540&0.550&0.580&0.564&0.549\\
$D_0$(CT-JPI)(eV)&1.06&21.4&1.47&65.7&5.26&186\\
$D_0$(RIPL-3)(eV)&$-$&\multicolumn{1}{c}{\quad27(5)}&2.4(2)&\multicolumn{1}{c}{\quad62(5)}&6.8(6)&\multicolumn{1}{c}{\quad150(10)}\\
$D_1$(CT-JPI)(eV)&0.310&4.35&1.29&20.9&1.43&27.0\\
$\Delta$N&0.136&0.145&0.144&0.145&0.117&0.117\\
\botrule
\end{tabular}
\end{table*}
\begin{figure*}[!ht]
  \centering
    \includegraphics[width=8cm]{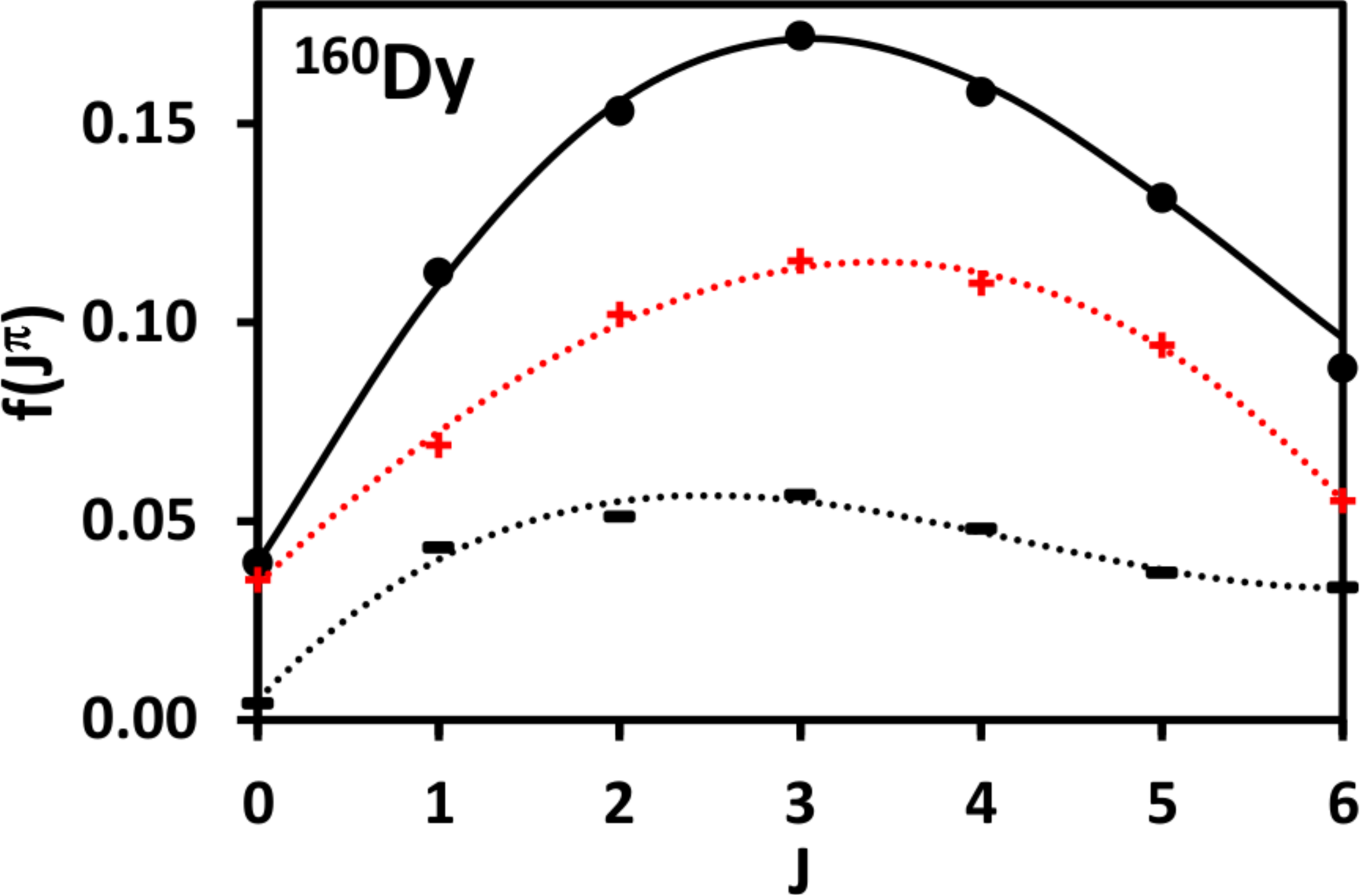}
    \includegraphics[width=8cm]{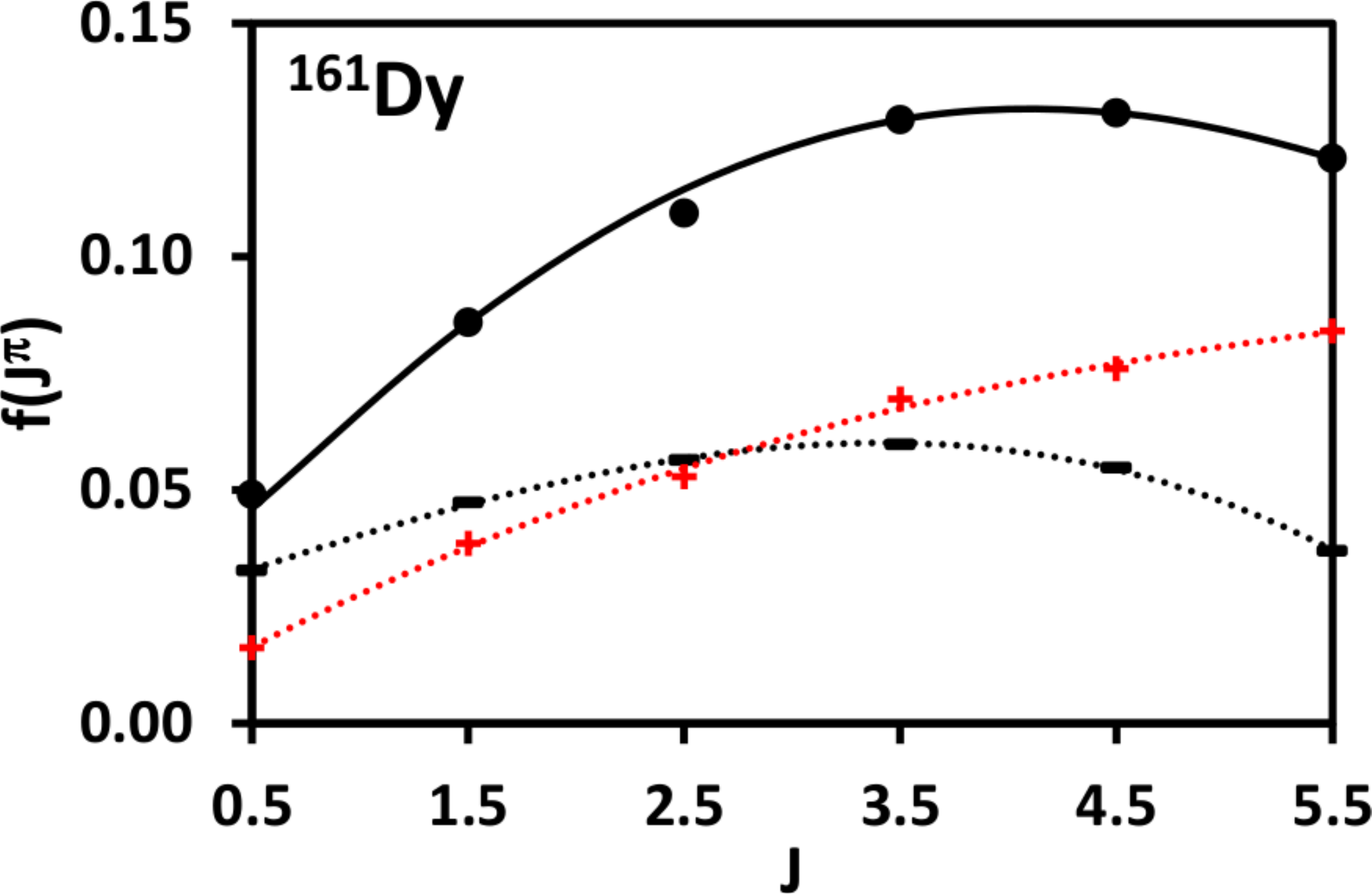}
    \includegraphics[width=8cm]{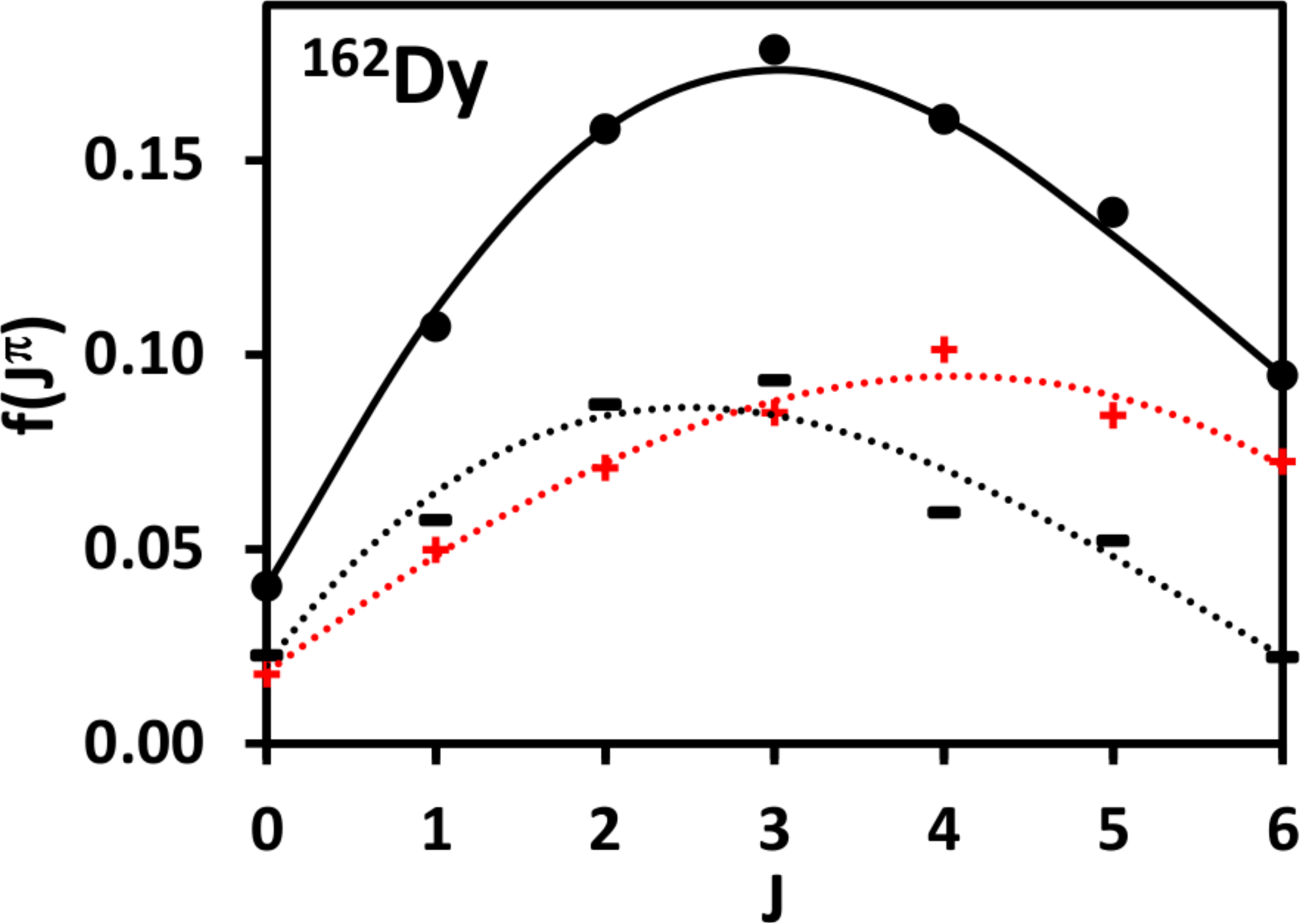}
    \includegraphics[width=8cm]{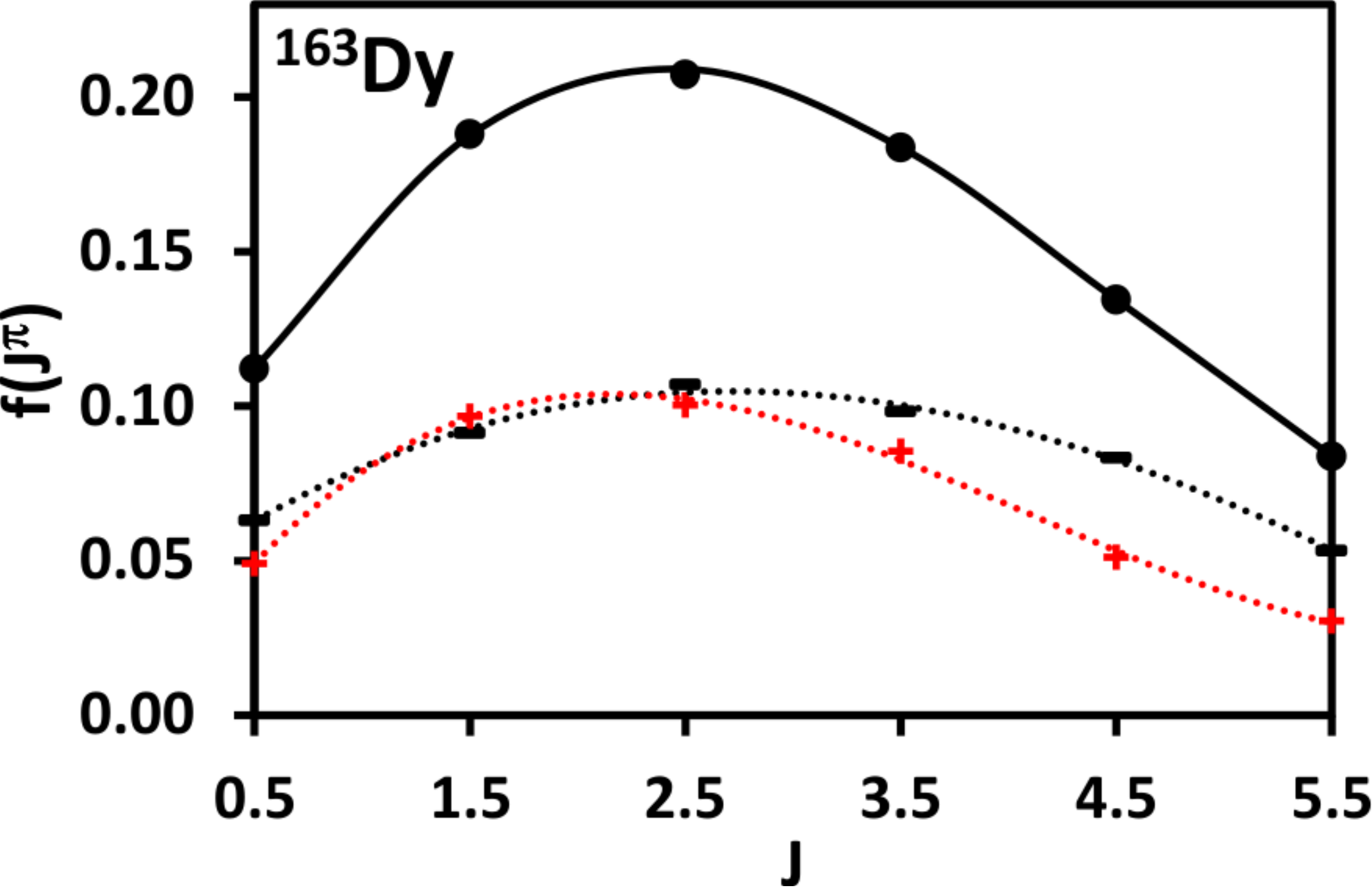}
    \includegraphics[width=8cm]{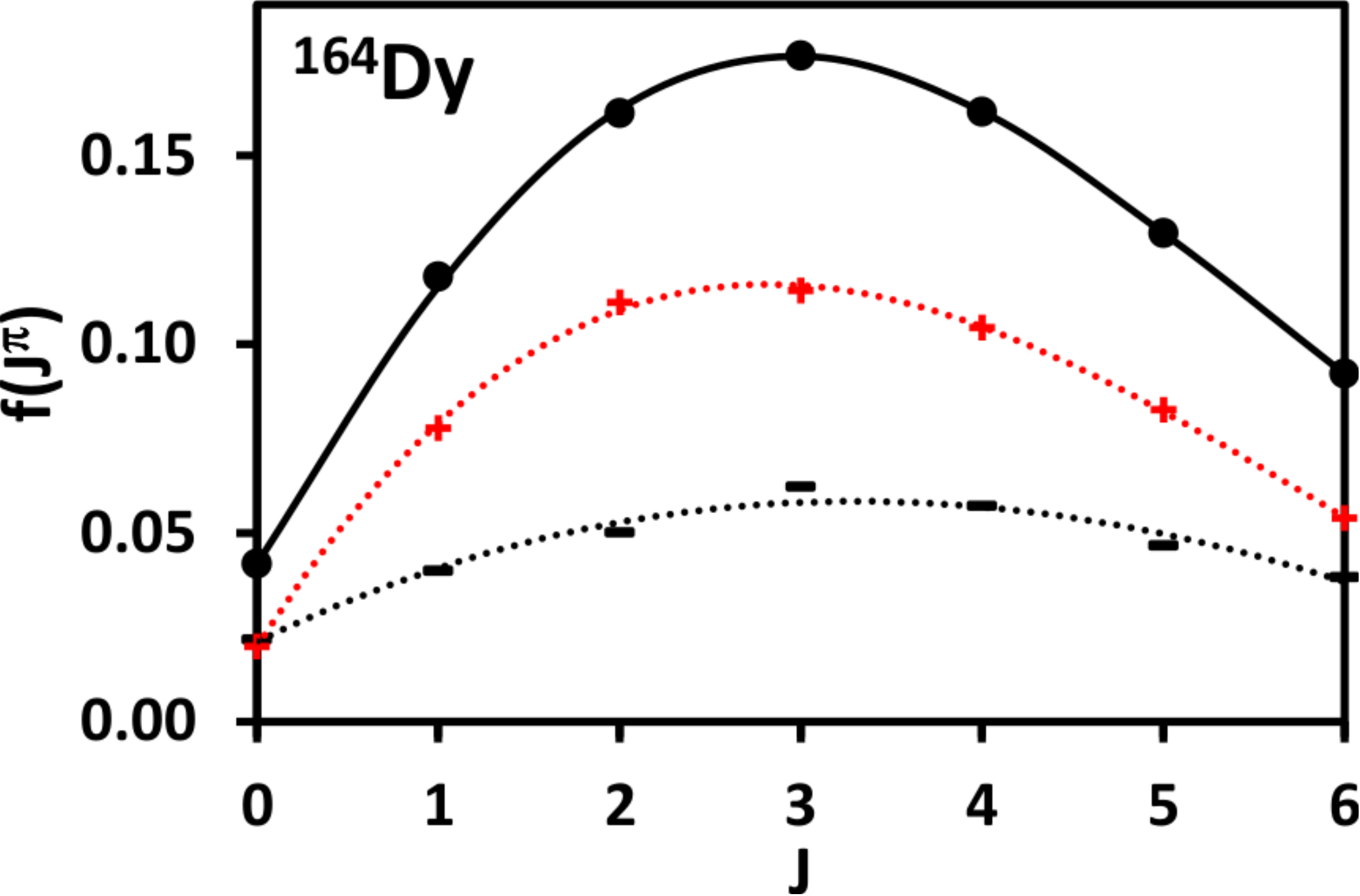}
    \includegraphics[width=8cm]{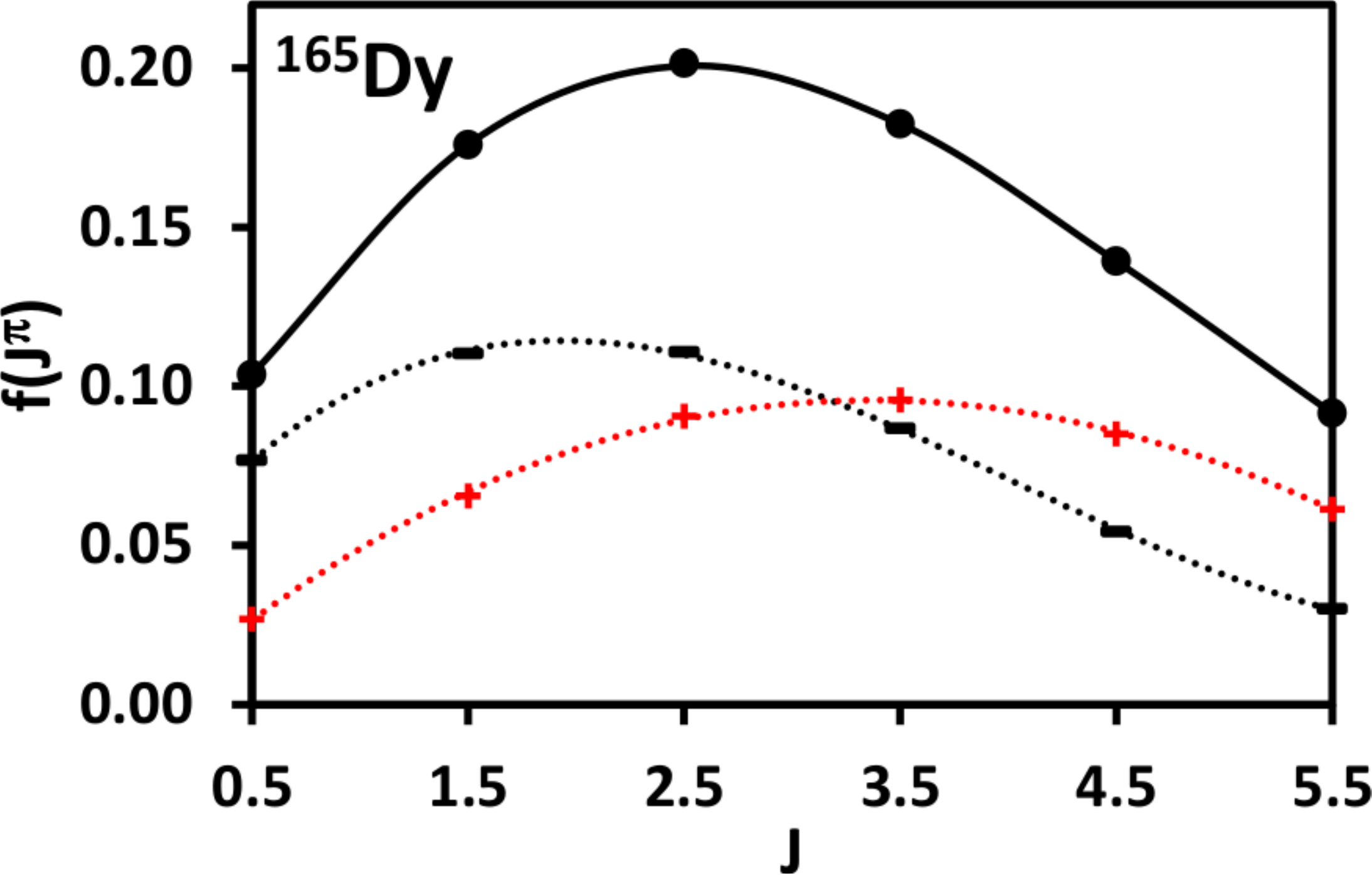}
  \caption{Fit of the CT-JPI model level densities ($\bullet$) to the spin distribution function for $^{159-165}$Dy (solid black lines).  The dotted lines show third order polynomial fits to the positive ($\textbf{\textcolor{red}+}$) and negative (\textbf{-}) parity $J^{\pi}$ fractions.}
  \label{Dyfig}
\end{figure*}

The fitted positive parity, negative parity and total spin fractions are show shown in Fig.~\ref{Dyfig} where they are compared to the spin fractions from the spin distribution function assuming the spin cutoff parameters, $\sigma_c$, shown in Table~\ref{DyCTfit}.  The distribution of both positive and negative parity spins varies smoothly as shown by a polynomial fit in Fig.~\ref{Dyfig}.  This fit is presented only to guide the eye and no fundamental importance should be taken from this.  The fitted and total calculated spin distributions differ by $\lesssim$2\%.  The distributions of both positive and negative parity spins are seen to vary smoothly and were fit with a third order polynomial to guide the eye in Fig.~\ref{Dyfig}.  The fitted total spin distribution agrees remarkably will with the spin distribution function confirming the utility of the CT-JPI model.

For $^{160}$Dy there are no resonance data so the temperature and back shifts were fit to the large quantity of available nuclear structure data giving $T$=0.575 MeV.  The $E_0 (J^{\pi})$ back shifts for all levels with ${J=0^+,2^+,3^+,4^+,5,6}$ were fit assuming a constant temperature with $\overline{\Delta N}(J^{\pi})$=0.136 averaged over 56 levels.  The average deviation of the fitted spin distribution from the calculated value is 2.1\%.

Considerable s-wave resonance data are available for $^{161}$Dy so the temperature and back shift could be fit to 12 ${J^{\pi}=1/2^+}$ levels giving $T$=0.548 MeV and $E_0 (1/2^+)$=0.895 MeV with $\overline{\Delta N}(1/2^+)$=0.143 averaged over 12 levels.  The back shifts for the ${J^{\pi}=1/2,3/2,5/2,7/2,9/2^-,11/2^-}$ levels were then fit to the constant temperature with $\overline{\Delta N}(J^{\pi})$=0.140 fit over 42 levels.  The average deviation of the fitted spin distribution from the calculated value is 1.8\%.

For $^{162}$Dy s-wave resonance and considerable nuclear structure data were available so the temperature and back shifts could be fit to the ${J^{\pi}=2^+,3^+}$ levels and resonances giving $T$=0.591 MeV, $E_0 (2^+)$=1.337 MeV, and $E_0 (3^+)$=1.230 MeV with $\overline{\Delta N}(2^+,3^+)$=0.138 averaged over 23 levels.  The back shifts for the ${J^{\pi}=0^+,1,2,3,4,5,6^+}$ levels were then fit to the constant temperature with $\overline{\Delta N}(J^{\pi})$=0.133 fit over 65 levels.  The average deviation of the fitted spin distribution from the calculated value is 1.6\%.

Both s-wave resonance data and considerable nuclear structure data were available for $^{163}$Dy so the temperature and back shift could be fit to the ${J^{\pi}=1/2^+}$ levels giving $T$=0.590 MeV and $E_0 (1/2^+)$=0.900 MeV with $\overline{\Delta N}(1/2^+)$=0.121 averaged over 13 levels and resonances.  The back shifts for the ${J^{\pi}=1/2,3/2,5/2,7/2^-,9/2^-,11/2^-}$ levels were then fit to the constant temperature with $\overline{\Delta N}(J^{\pi})$=0.135 fit over 47 levels.  The average deviation of the fitted spin distribution from the calculated value is 0.2\%.

For $^{164}$Dy s-wave resonance and considerable nuclear structure data were available so the temperature and back shifts could be fit to the ${J^{\pi}=2^-,3^-}$ levels giving $T$=0.593 MeV, $E_0 (2^-)$=1.234 MeV, and $E_0 (3^-)$=1.105 MeV with $\overline{\Delta N}(2^-,3^-)$=0.138 averaged over 21 levels and resonances.  The back shifts for the ${J^{\pi}=0^+,1,2,3,4,5^-,6^-}$ levels were then fit to the constant temperature with $\overline{\Delta N}(J^{\pi})$=0.111 fit over 46 levels.  The average deviation of the fitted spin distribution from the calculated value is 0.5\%.

Considerable s-wave resonance data were available, but no ${J^{\pi}=1/2^+}$ nuclear structure data are available for $^{165}$Dy, so the temperature and ${J^{\pi}=1/2^+}$ back shift were fit to the first 10 resonances giving $T$=0.588 MeV and $E_0(1/2^+)$=0.987 MeV $\overline{\Delta N}(1/2^+)$=0.131.  Considerable nuclear structure data were available for ${J^{\pi}=3/2,5/2,7/2^-}$ levels to determine their back shifts with $\overline{\Delta N}(J^{\pi})$=0.111 fit for 32 levels.    The average deviation of the fitted spin distribution from the calculated value is 0.07\%.

\subsection{Indium}

\begin{table*}[!ht]
\tabcolsep=5pt
\caption{\label{Indata} Back shifts, $E_0(J^{\pi})$, derived from the CT-JPI model, for $^{114-116}$In. }
\begin{tabular}{cddddcdd}
\toprule
&\multicolumn{2}{c}{$\hspace{0.5cm}E_0 (J^{\pi}) ^{114}In$}&\multicolumn{2}{c}{$\hspace{0.5cm}E_0 (J^{\pi}) ^{116}In$}&&\multicolumn{2}{c}{$\hspace{0.5cm}E_0 (J^{\pi}) ^{115}In$}\\
J&\multicolumn{1}{c}{$\hspace{0.5cm}\pi=+$}&\multicolumn{1}{c}{$\hspace{0.5cm}\pi=-$}&\multicolumn{1}{c}{$\hspace{0.5cm}\pi=+$}&\multicolumn{1}{c}{$\hspace{0.5cm}\pi=-$}&
J&\multicolumn{1}{c}{$\hspace{0.5cm}\pi=+$}&\multicolumn{1}{c}{$\hspace{0.5cm}\pi=-$}\\
\colrule
0&(1.346)&(1.672)&(1.317)&(1.600)&1/2&1.242&(1.631)\\
1&(0.580)&(1.151)&(0.598)&(1.446)&3/2&(0.954)&(1.461)\\
2&(0.380)&(0.848)&(0.380)&(1.140)&5/2&0.955&(1.440)\\
3&0.309&0.805&20.33&0.853&7/2&1.189&(1.648)\\
4&0.366&0.880&410.42&0.584&9/2&1.459&(2.189)\\
5&0.471&(1.093)&0.612&0.464&11/2&(1.826)&(3.306)\\
6&0.590&(1.673)&(0.946)&0.398&&&\\
\botrule
\end{tabular}
\end{table*}
The indium isotopes test the CT-JPI model for odd-Z, even/odd-N nuclei near the Z=50 closed shell.  The model has been applied to the isotopes $^{114-116}$In where sufficient nuclear structure and resonance data are available to provide reasonable fits.   The fitted $E_0(J^{\pi})$ back shifts are shown in Table~\ref{Indata} and the corresponding neutron separation energies, $S_n$, temperatures, $T$, spin cutoff parameters, $\sigma_c$, resonance spacings, $D_0$ and $D_1$, and quality of fit, $\overline{\Delta N}(J^{\pi})$,
are shown in Table~\ref{InCTfit}.  The average fit is $\overline{\Delta N}(J^{\pi})$=0.128(2) in excellent agreement with the expected value from the folded Normal distribution.  The temperatures and resonance spacings fitted to the CT-JPI model are compared with the values from RIPL-3~\cite{RIPL3} in Table~\ref{InCTfit}.  The fitted temperatures are comparable to the  RIPL-3 values and the fitted $D_0$ values vary by up to a factor of 2 larger than the RIPL-3 values.  The spin cutoff parameters vary from $\sigma_c$=2.4-4.5 compared to the value predicted by T. von Egidy \textit{et al}~\cite{Egidy88}, $\sigma_c$=3.9.

\begin{figure*}[!ht]
  \centering
    \includegraphics[width=8cm]{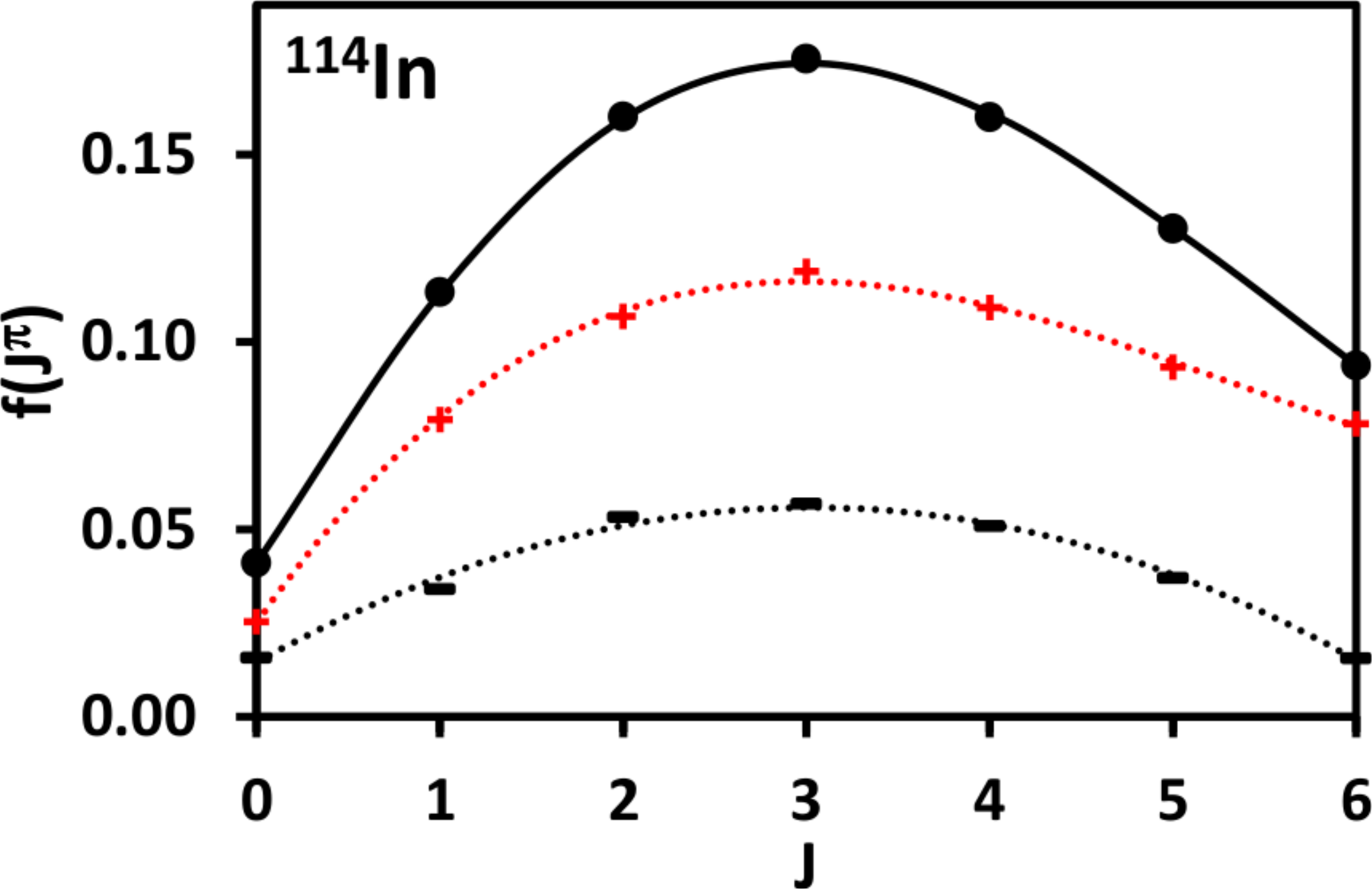}
    \includegraphics[width=8cm]{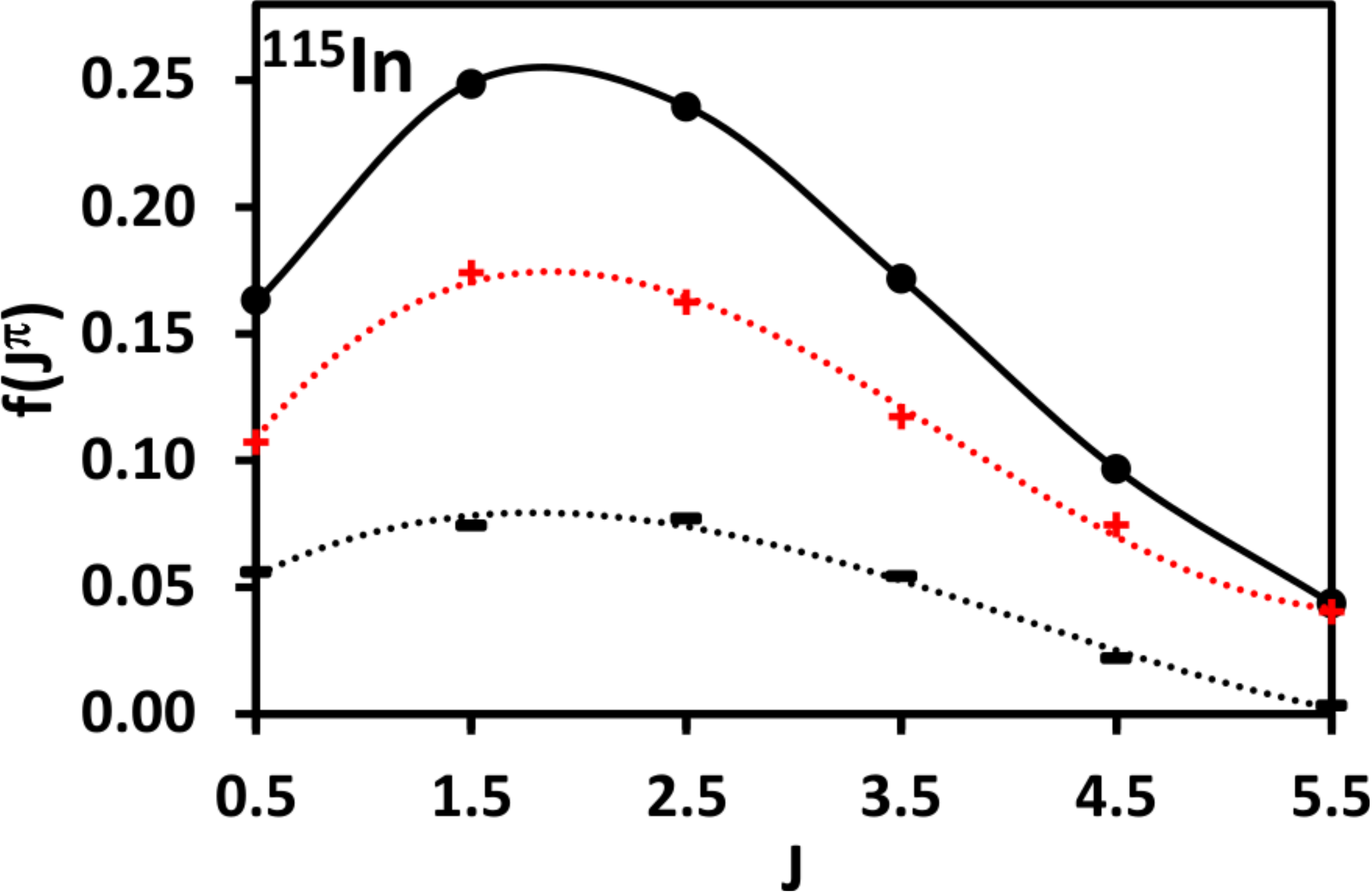}
    \includegraphics[width=8cm]{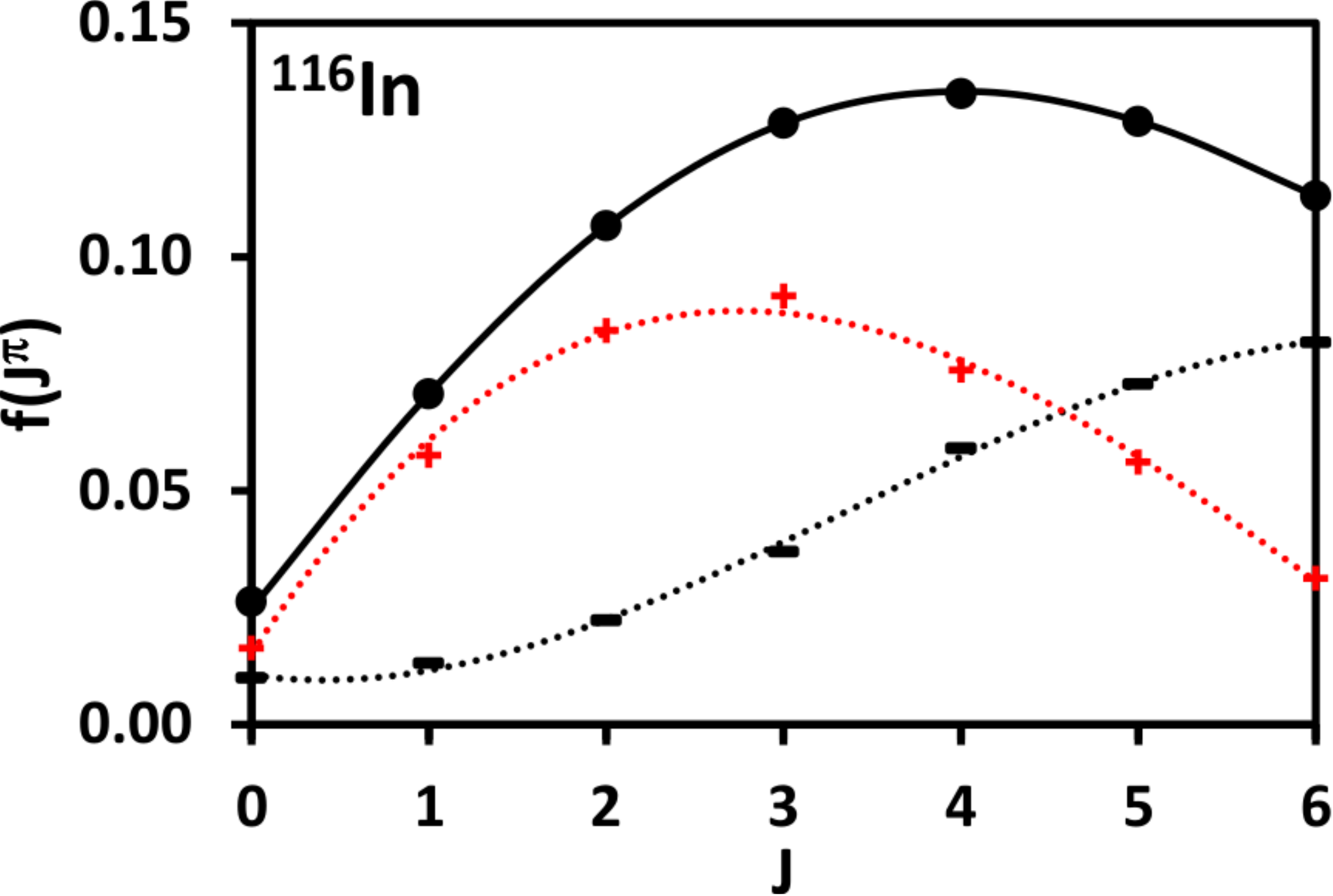}
  \caption{Fit of the CT-JPI model level densities ($\bullet$) to the spin distribution function for $^{114-116}$In (solid black lines).  The dotted lines show third order polynomial fits to the positive ($\textbf{\textcolor{red}+}$) and negative (\textbf{-}) parity $J^{\pi}$ fractions.}
  \label{Infig}
\end{figure*}

The fitted positive parity, negative parity and total spin fractions are shown in Fig.~\ref{Infig} where they are compared to the spin fractions from the spin distribution function assuming the spin cutoff parameters, $\sigma_c$, shown in Table~\ref{InCTfit}.  The distribution of both positive and negative parity spins varies smoothly as shown by a polynomial fit in Fig.~\ref{Infig}.  This fit is presented only to guide the eye and no fundamental importance should be taken from this.  The fitted and calculated spin distributions differ by $<$1\%.  The distributions of both positive and negative parity spins are seen to vary smoothly and were fit with a third order polynomial to guide the eye in Fig.~\ref{Infig}.

For $^{114}$In there are limited s-wave resonance data which is combined with the ${J^{\pi}=4^+,5^+}$ nuclear structure data to determine the temperature $T$=0.672 MeV, and back shifts $E_0 (4^+)$=0.366 MeV and $E_0 (5^+)$=0.471 MeV with $\overline{\Delta N}(4^+,5^+)$=0.130 for 14 levels and resonances.  The $E_0 (J^{\pi})$ back shifts for all levels with ${J=3,4,5^+,6^+}$ were fit, assuming a constant temperature, with $\overline{\Delta N}(J^{\pi})$=0.127 averaged over 27 levels.  The average deviation of the fitted spin distribution from the calculated value is 0.4\%.
\begin{table}[!ht]
\tabcolsep=3pt
\caption{\label{InCTfit} Neutron separation energies~\cite{AME2013}, $S_n$, spin cutoff parameters, $\sigma_c$, temperatures, $T$, resonance spacings, $D_0$, and minimization uncertainty fitted to the CT-JPI model and compared with the values from RIPL-3~\cite{RIPL3} for $^{114-116}$In. }
\begin{tabular}{lddd}
\toprule
&\multicolumn{1}{c}{$^{114}$In}&\multicolumn{1}{c}{$^{115}$In}&\multicolumn{1}{c}{$^{116}$In}\\
\colrule
$S_n$ (MeV)&7.2739&9.0393&6.78472\\
$\sigma_c$&3.48&2.37&4.48\\
T $_{CT-JPI}$(MeV) &0.672&0.596&0.572\\
T $_{RIPL-3}$(MeV) &0.649&$-$&0.579\\
$D_0$(CT-JPI)(eV)&12.5&1.24&4.99\\
$D_0$(RIPL-3)(eV)&\multicolumn{1}{c}{\quad10.7(8)}&$-$&\multicolumn{1}{c}{\quad9.5(5)}\\
$D_1$(CT-JPI)(eV)&15.8&1.02&2.63\\
$D_1$(RIPL-3)(eV)&$-$&$-$&\multicolumn{1}{c}{\quad5.1(4)}\\
$\Delta$N&0.127&0.127&0.130\\
\botrule
\end{tabular}
\end{table}

No resonance data are available for $^{115}$In so the temperature and back shifts with fit to the ${J^{\pi}=5.2^+,7/2^+,9/2^+}$ nuclear structure data giving $T$=0.596 MeV with $\overline{\Delta N}(J^{\pi})$=0.127 averaged over 19 levels.  The average deviation of the fitted spin distribution from the calculated value is 0.1\%.

For $^{116}$In there are considerable s-wave and p-wave resonance data so the temperature was fit to the ${J^{\pi}=3^-,4,5}$ data giving $T$=0.572 MeV with $\overline{\Delta N}(J^{\pi})$=0.131 averaged over 51 levels.  The $E_0 (J^{\pi})$ back shifts for all levels with ${J=3,4,5,6^+}$ were fit assuming a constant temperature with $\overline{\Delta N}(J^{\pi})$=0.130 averaged over 59 levels.  The average deviation of the fitted spin distribution from the calculated value is 0.9\%.

\subsection{Germanium}

The germanium isotopes test the CT-JPI model for low-Z nuclei away from closed shells.  The model has been applied to the isotopes $^{68-76}$Ge where sufficient nuclear structure and resonance data are available to provide reasonable fits.   The fitted $E_0(J^{\pi})$ back shifts are shown in Table~\ref{Gedata} and the corresponding neutron separation energies, $S_n$, temperatures, $T$, spin cutoff parameters, $\sigma_c$, resonance spacings, $D_0$ and $D_1$, and quality of fit, $\overline{\Delta N}(J^{\pi})$, are shown in Table~\ref{GeCTfit}.  The average fit is $\overline{\Delta N}(J^{\pi})$=0.132(10) in excellent agreement with the expected value from the folded Normal distribution.  The temperatures and resonance spacings fitted to the CT-JPI model are compared with the values from RIPL-3~\cite{RIPL3} in Table~\ref{GeCTfit}.  The fitted temperatures range from 4-11\% higher than the  RIPL-3 values and the fitted $D_0$ values vary by up to a factor of 2 larger than the RIPL-3 values.  The spin cutoff parameters average $\sigma_c$=2.6(2) compared to the value predicted by T. von Egidy \textit{et al}~\cite{Egidy88}, $\sigma_c$=3.4.

\begin{table*}[!ht]
\tabcolsep=2pt
\caption{\label{Gedata} Back shifts, $E_0(J^{\pi})$, derived from the CT-JPI model, for $^{68-76}$Ge. }
\begin{tabular}{ccccccccccc}
\toprule
&\multicolumn{2}{c}{$\hspace{0.5cm}E_0 (J^{\pi}) ^{68}Ge$}&\multicolumn{2}{c}{$\hspace{0.5cm}E_0 (J^{\pi}) ^{70}Ge$}&\multicolumn{2}{c}{$\hspace{0.5cm}E_0 (J^{\pi}) ^{72}Ge$}&\multicolumn{2}{c}{$\hspace{0.5cm}E_0 (J^{\pi}) ^{74}Ge$}&\multicolumn{2}{c}{$\hspace{0.5cm}E_0 (J^{\pi}) ^{76}Ge$}\\
J&\multicolumn{1}{c}{$\hspace{0.5cm}\pi=+$}&\multicolumn{1}{c}{$\hspace{0.5cm}\pi=-$}&\multicolumn{1}{c}{$\hspace{0.5cm}\pi=+$}&\multicolumn{1}{c}{$\hspace{0.5cm}\pi=-$}&
\multicolumn{1}{c}{$\hspace{0.5cm}\pi=+$}&\multicolumn{1}{c}{$\hspace{0.5cm}\pi=-$}&\multicolumn{1}{c}{$\hspace{0.5cm}\pi=+$}&\multicolumn{1}{c}{$\hspace{0.5cm}\pi=-$}
&\multicolumn{1}{c}{$\hspace{0.5cm}\pi=+$}&\multicolumn{1}{c}{$\hspace{0.5cm}\pi=-$}\\
\colrule
0&2.108&(3.511)&\hspace{0.5cm}2.988&(4.125)&\hspace{0.5cm}2.897&(3.008)&\hspace{0.5cm}2.202&(4.189)&\hspace{0.5cm}2.148&(8.091)\\
1&(1.273)&(2.490)&\hspace{0.5cm}2.253&2.751&\hspace{0.5cm}1.794&2.470&\hspace{0.5cm}1.464&2.366&\hspace{0.5cm}(1.294)&(3.185)\\
2&1.117&(2.367)&\hspace{0.5cm}2.250&2.358&\hspace{0.5cm}1.542&2.339&\hspace{0.5cm}1.249&(2.094)&\hspace{0.5cm}1.071&(2.867)\\
3&(1.350)&2.405&\hspace{0.5cm}(2.384)&2.495&\hspace{0.5cm}(1.670)&2.665&\hspace{0.5cm}(1.286)&2.117&\hspace{0.5cm}(1.219)&2.734\\
4&1.749&(2.957)&\hspace{0.5cm}2.943&(2.732)&\hspace{0.5cm}1.967&(3.440)&\hspace{0.5cm}1.520&(2.339)&\hspace{0.5cm}1.546&(3.149)\\
5&(2.357)&(3.663)&\hspace{0.5cm}(3.777)&3.187&\hspace{0.5cm}(2.587)&3.967&\hspace{0.5cm}1.950&2.658&\hspace{0.5cm}(2.150)&3.490\\
6&3.096&(5.434)&\hspace{0.5cm}(4.952)&(3.823)&\hspace{0.5cm}3.667&(4.109)&\hspace{0.5cm}2.593&(3.016)&\hspace{0.5cm}(2.965)&(3.999)\\
\toprule
&\multicolumn{2}{c}{$\hspace{0.5cm}E_0 (J^{\pi}) ^{69}Ge$}&\multicolumn{2}{c}{$\hspace{0.5cm}E_0 (J^{\pi}) ^{71}Ge$}&\multicolumn{2}{c}{$\hspace{0.5cm}E_0 (J^{\pi}) ^{73}Ge$}&\multicolumn{2}{c}{$\hspace{0.5cm}E_0 (J^{\pi}) ^{75}Ge$}&\\
J&\multicolumn{1}{c}{$\hspace{0.5cm}\pi=+$}&\multicolumn{1}{c}{$\hspace{0.5cm}\pi=-$}&\multicolumn{1}{c}{$\hspace{0.5cm}\pi=+$}&\multicolumn{1}{c}{$\hspace{0.5cm}\pi=-$}&
\multicolumn{1}{c}{$\hspace{0.5cm}\pi=+$}&\multicolumn{1}{c}{$\hspace{0.5cm}\pi=-$}&\multicolumn{1}{c}{$\hspace{0.5cm}\pi=+$}&\multicolumn{1}{c}{$\hspace{0.5cm}\pi=-$}&\\
\colrule
1/2&1.446&0.515&\hspace{0.5cm}1.308&0.718&\hspace{0.5cm}1.031&(0.662)&\hspace{0.5cm}1.455&(0.984)&\hspace{0.5cm}&\\
3/2&(1.255)&0.515&\hspace{0.5cm}0.656&0.479&\hspace{0.5cm}(0.399)&0.332&\hspace{0.5cm}0.882&0.541&\hspace{0.5cm}&\\
5/2&1.380&0.892&\hspace{0.5cm}0.511&0.552&\hspace{0.5cm}0.303&(0.318)&\hspace{0.5cm}0.698&0.545&\hspace{0.5cm}&\\
7/2&1.533&1.477&\hspace{0.5cm}(0.627)&0.846&\hspace{0.5cm}(0.494)&(0.512)&\hspace{0.5cm}(0.854)&(0.756)&\hspace{0.5cm}&\\
9/2&1.992&2.431&\hspace{0.5cm}1.037&(1.249)&\hspace{0.5cm}0.945&(0.840)&\hspace{0.5cm}1.206&(1.160)&\hspace{0.5cm}&\\
11/2&2.539&69&\hspace{0.5cm}(1.745)&(1.700)&\hspace{0.5cm}(1.392)&(1.516)&\hspace{0.5cm}(1.678)&(1.798)&\hspace{0.5cm}&\\
\botrule
\end{tabular}
\end{table*}
\begin{table*}[!ht]
\tabcolsep=-1pt
\caption{\label{GeCTfit} Neutron separation energies~\cite{AME2013}, $S_n$, spin cutoff parameters, $\sigma_c$, temperatures, $T$, resonance spacings, $D_0$, and minimization uncertainty fitted to the CT-JPI model and compared with the values from RIPL-3~\cite{RIPL3} for $^{68-76}$Ge. }
\begin{tabular}{lddddddddd}
\toprule
&\multicolumn{1}{c}{$^{68}$Ge}&\multicolumn{1}{c}{$^{69}$Ge}&\multicolumn{1}{c}{$^{70}$Ge}&\multicolumn{1}{c}{$^{71}$Ge}&\multicolumn{1}{c}{$^{72}$Ge}&
\multicolumn{1}{c}{$^{73}$Ge}&\multicolumn{1}{c}{$^{74}$Ge}&\multicolumn{1}{c}{$^{75}$Ge}&\multicolumn{1}{c}{$^{76}$Ge}\\
\colrule
$S_n$ (MeV)&12.392&8.1932&11.5325&7.41594&10.7508&6.78294&10.19624&6.50584&9.42724\\
$\sigma_c$&2.37&2.37&2.48&2.63&2.45&2.68&2.84&2.74&2.64\\
T $_{CT-JPI}$(MeV) &0.948&0.931&0.992&0.953&0.962&0.959&0.943&1.004&1.064\\
T $_{RIPL-3}$(MeV) &$-$&$-$&$-$&0.873&$-$&0.924&0.860&0.907&$-$\\
$D_0$(CT-JPI)(eV)&12.8&\multicolumn{1}{c}{\quad661}&51.0&\multicolumn{1}{c}{\quad1575}&\multicolumn{1}{c}{\quad112}&\multicolumn{1}{c}{\quad2387}&58.3&\multicolumn{1}{c}{\quad6559}&\multicolumn{1}{c}{\quad3018}\\
$D_0$(RIPL-3)(eV)&$-$&$-$&$-$&\multicolumn{1}{c}{\quad1170(23)}&$-$&\multicolumn{1}{c}{\quad1500(30)}&\multicolumn{1}{c}{\quad62(15)}&\multicolumn{1}{c}{\quad3000(1000)}&$-$\\
$D_1$(CT-JPI)(eV)&2.16&\multicolumn{1}{c}{\quad154}&25.4&\multicolumn{1}{c}{\quad371}&33.4&\multicolumn{1}{c}{\quad674}&65.5&\multicolumn{1}{c}{\quad1606}&\multicolumn{1}{c}{\quad190}\\
$\Delta$N&0.139&0.125&0.136&0.122&0.116&0.122&0.109&0.122&0.117\\
\botrule
\end{tabular}
\end{table*}

\begin{figure*}[!ht]
  \centering
    \includegraphics[width=7cm]{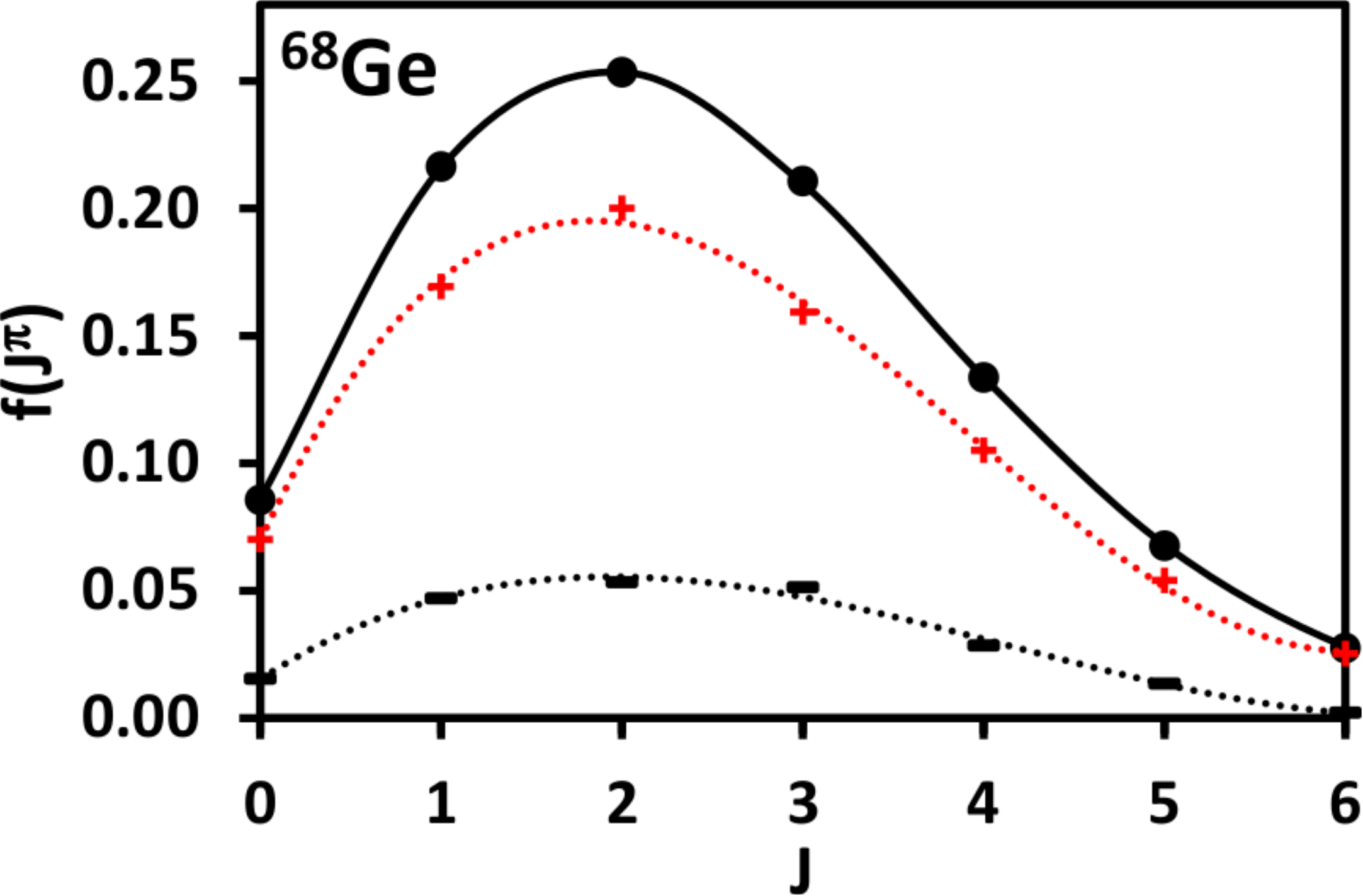}
    \includegraphics[width=7cm]{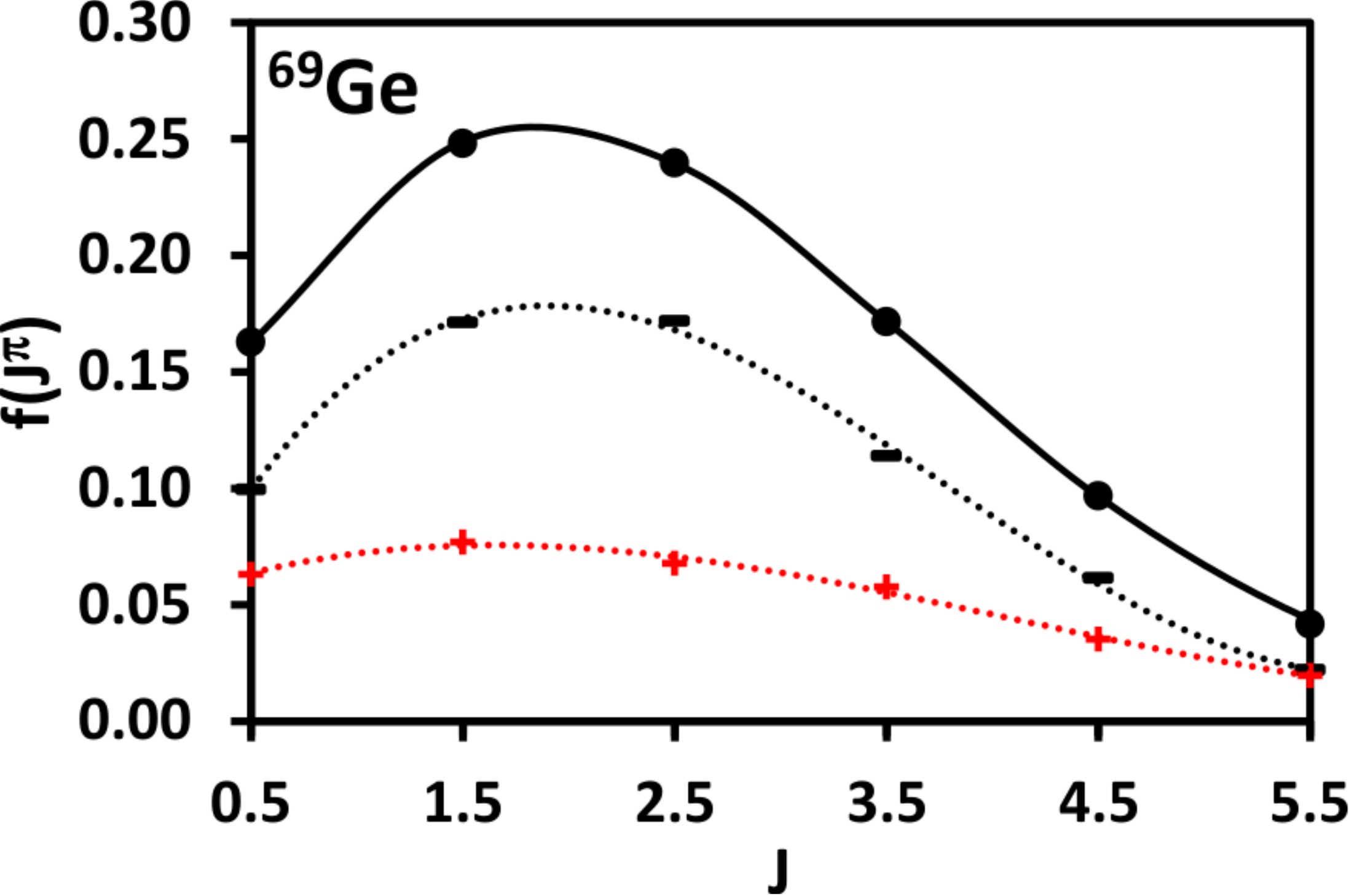}
    \includegraphics[width=7cm]{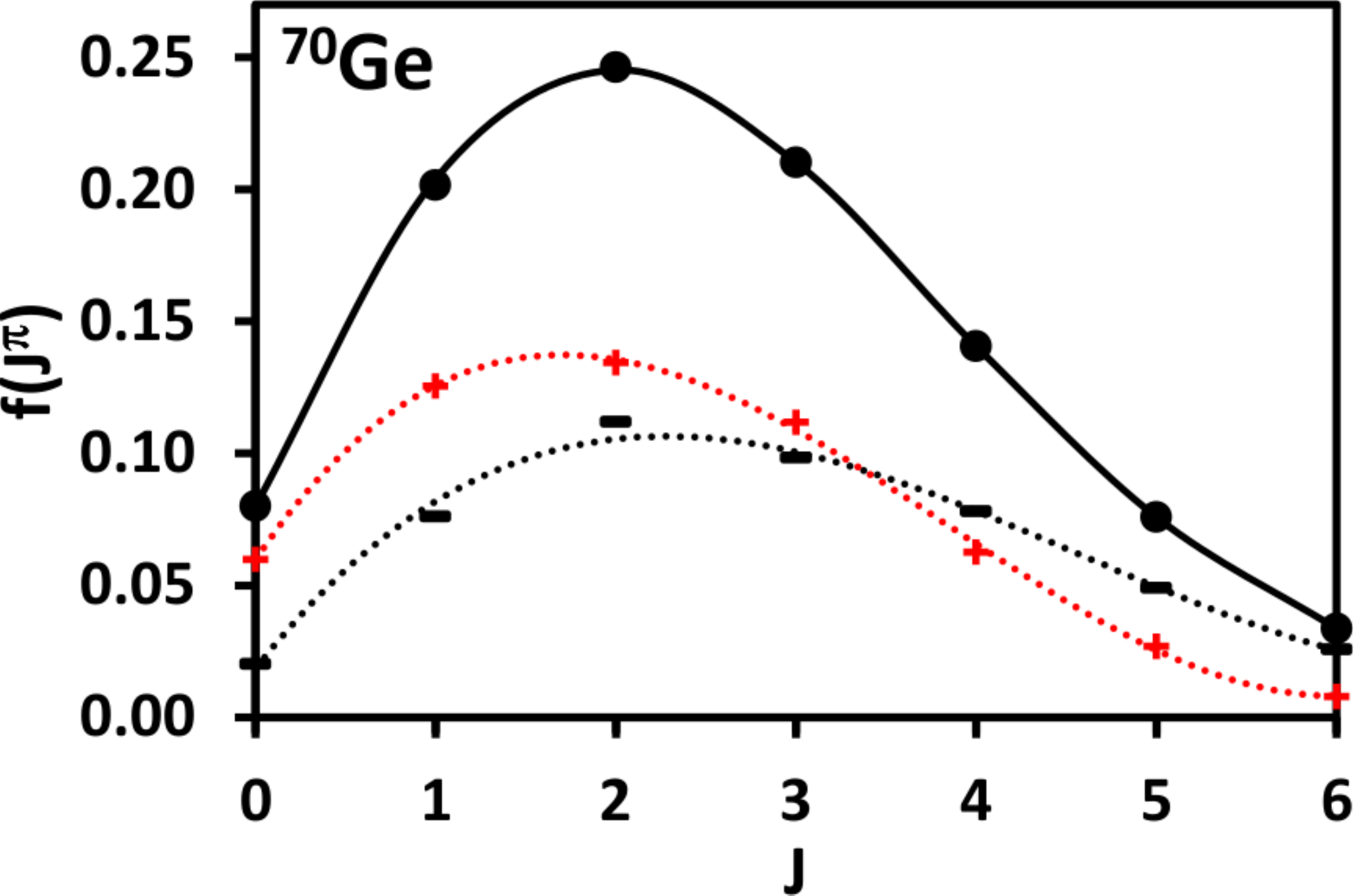}
    \includegraphics[width=7cm]{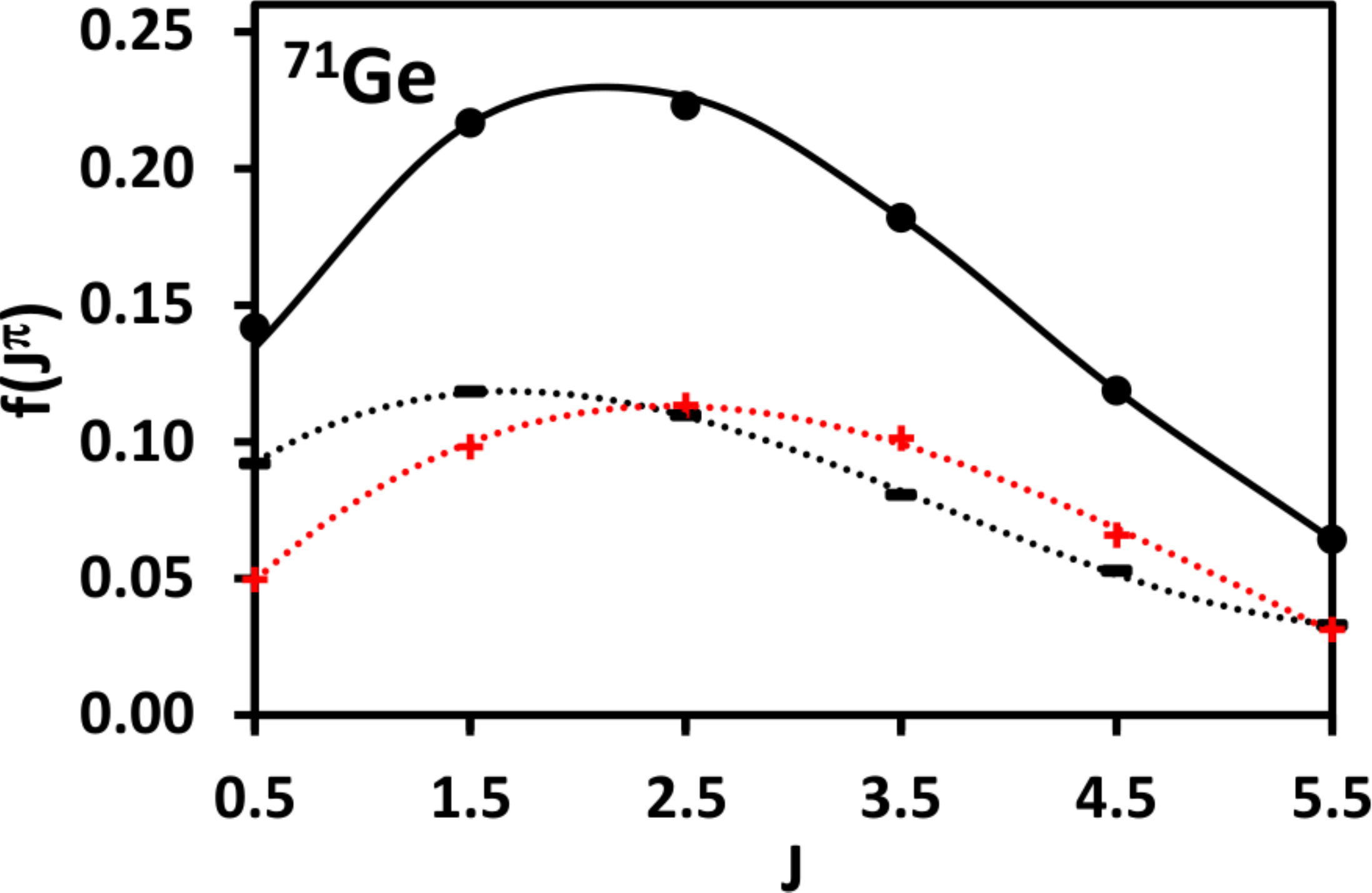}
    \includegraphics[width=7cm]{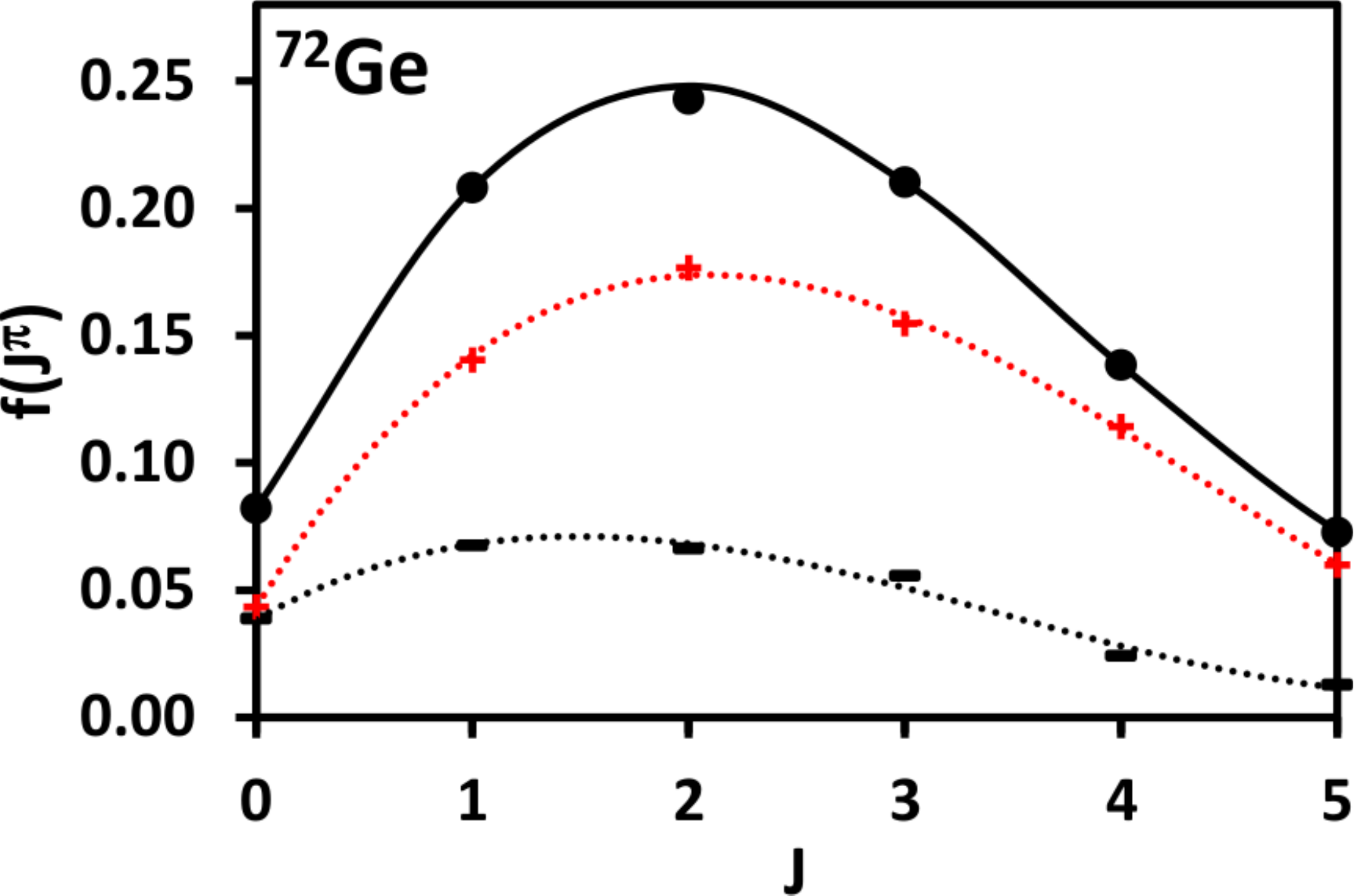}
    \includegraphics[width=7cm]{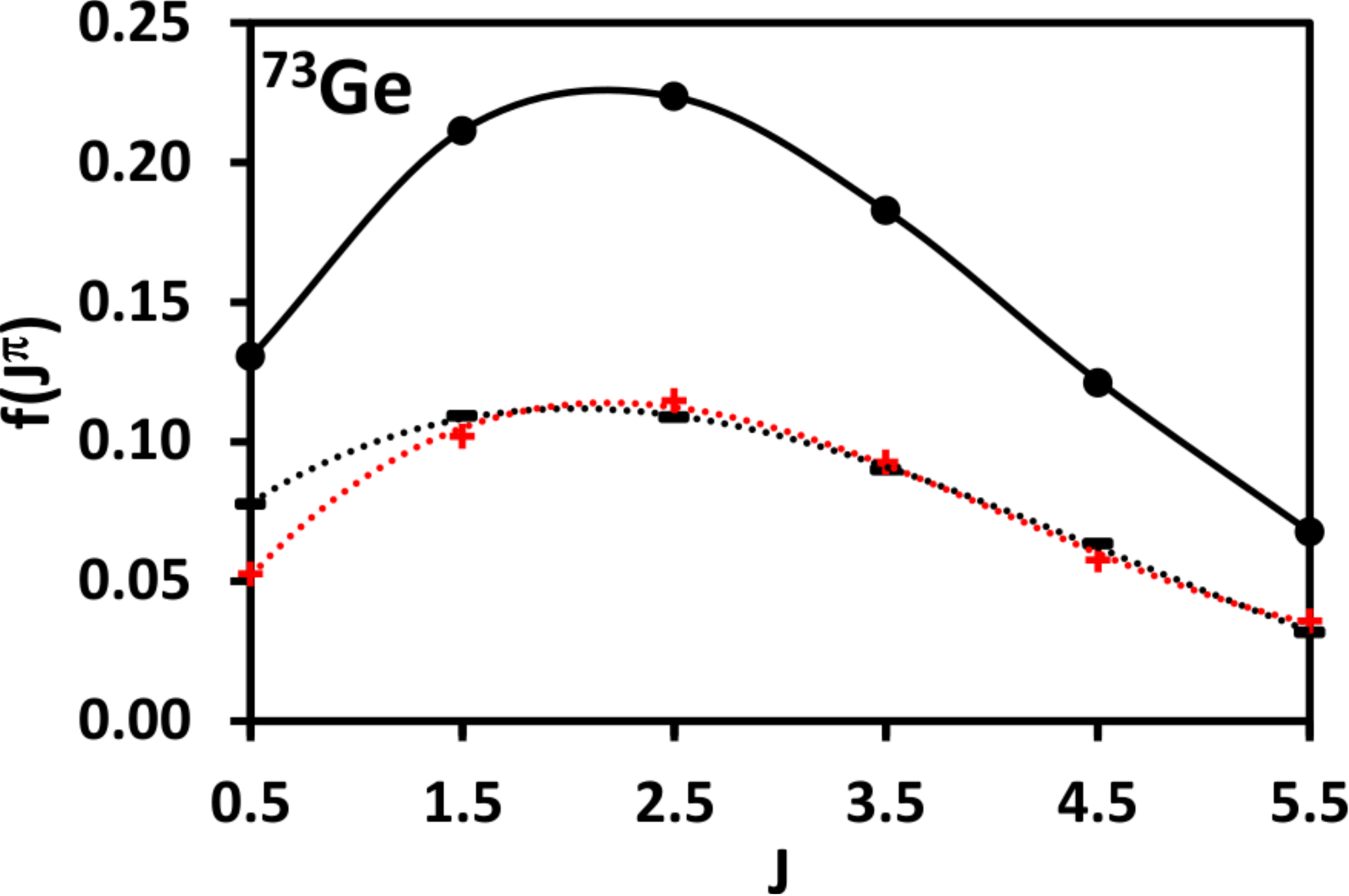}
    \includegraphics[width=7cm]{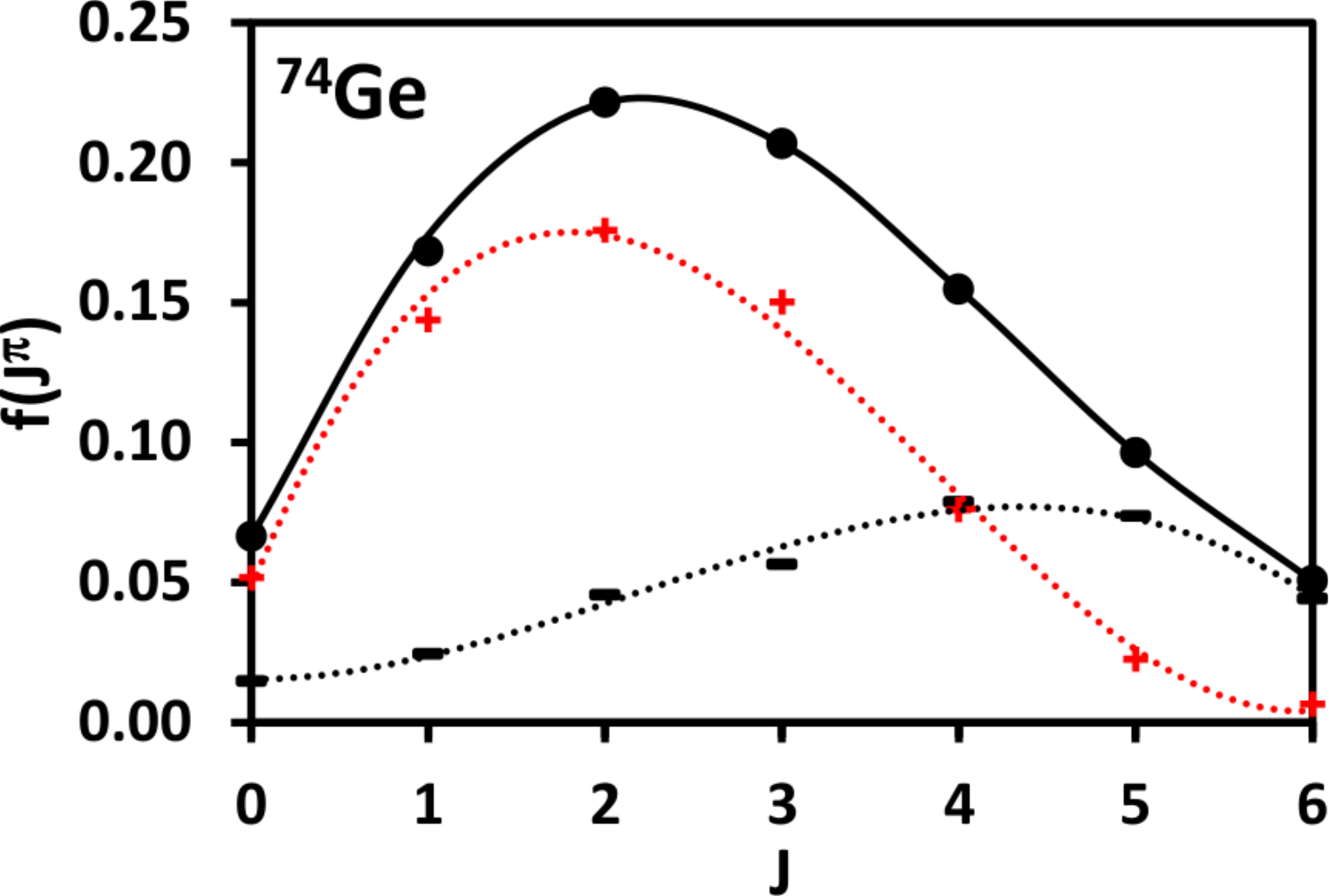}
    \includegraphics[width=7cm]{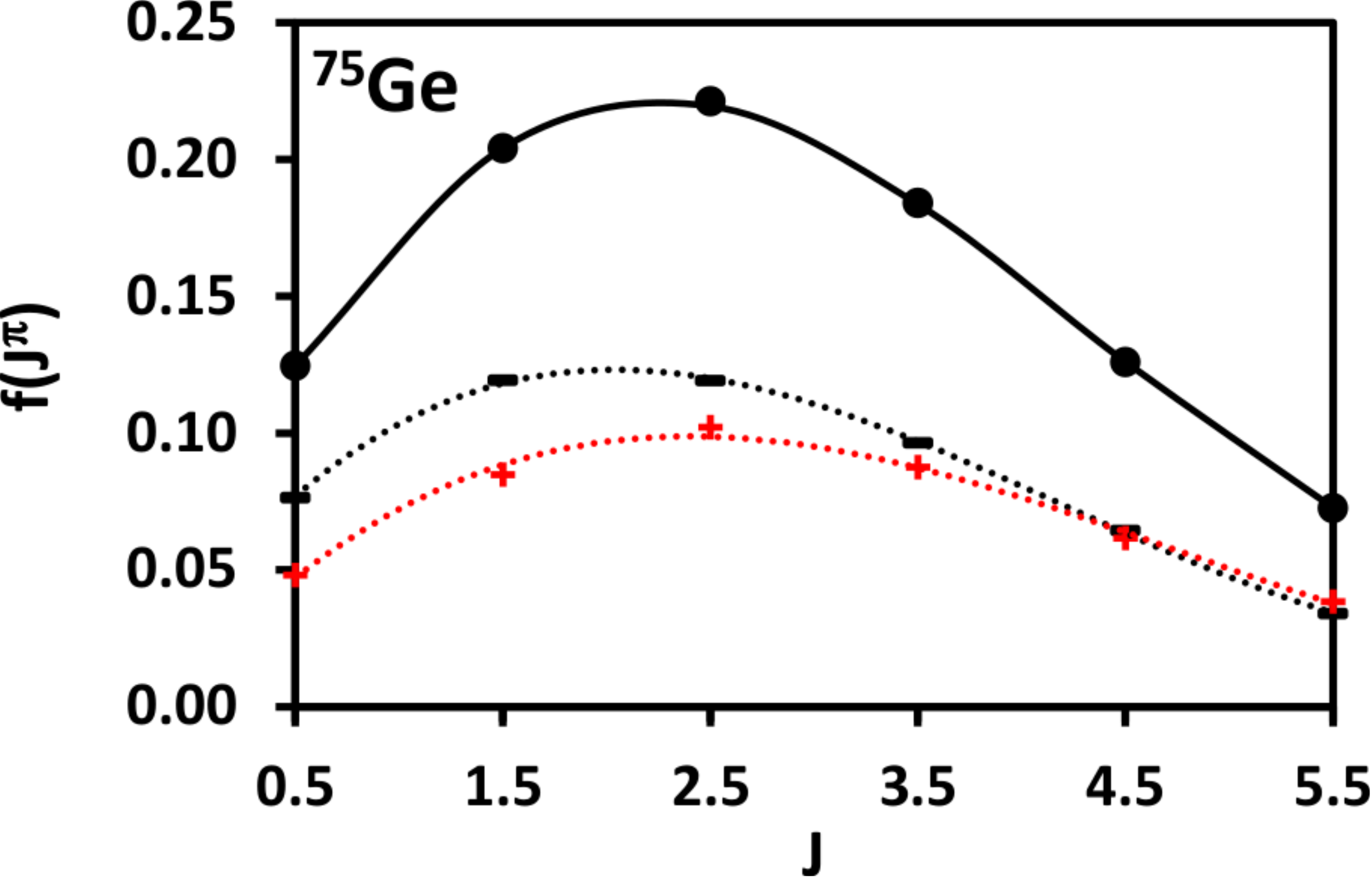}
    \includegraphics[width=7cm]{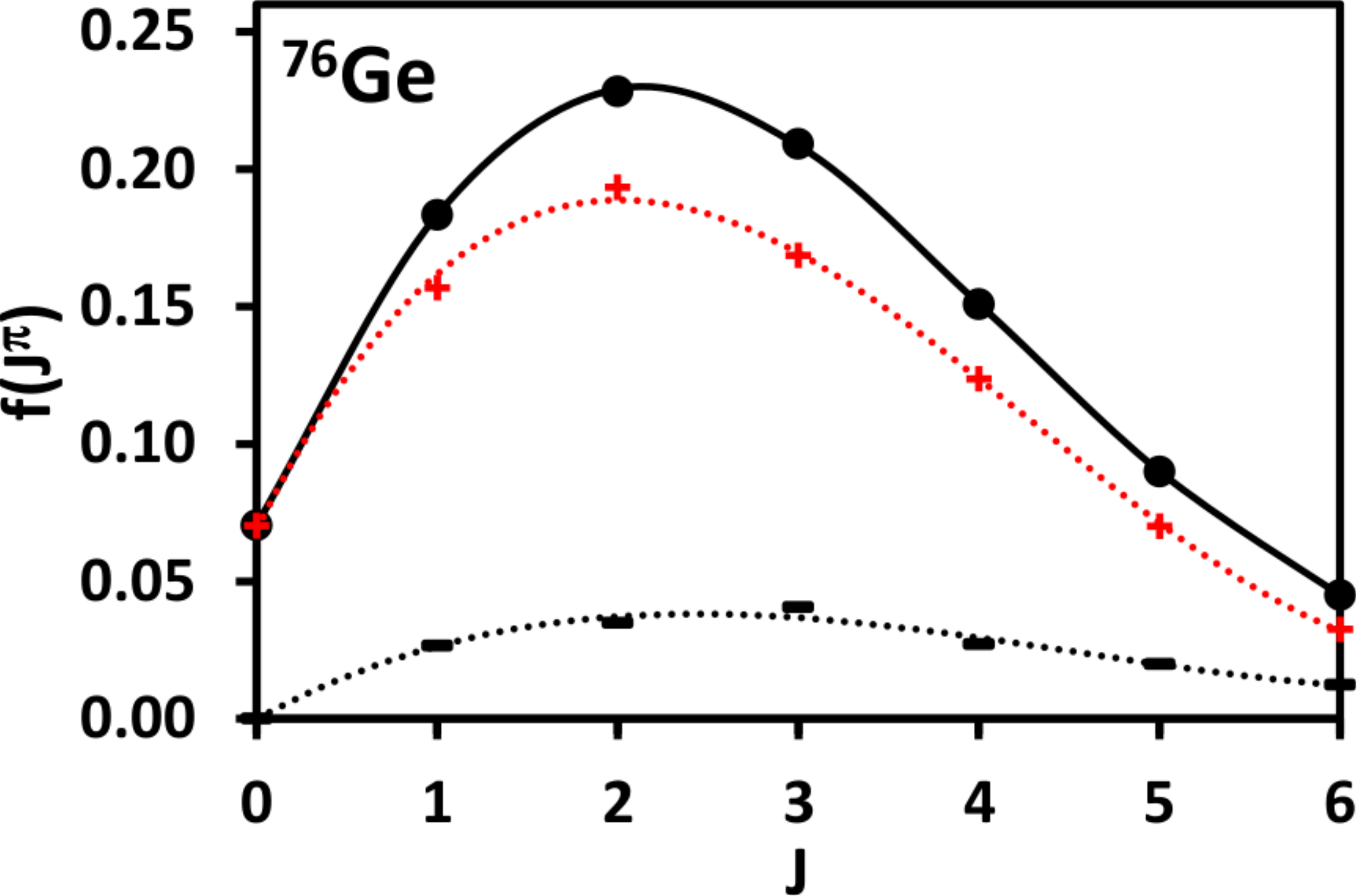}
  \caption{Fit of the CT-JPI model level densities ($\bullet$) to the spin distribution function for $^{68-76}$Ge (solid black lines).  The dotted lines show third order polynomial fits to the positive ($\textbf{\textcolor{red}+}$) and negative (\textbf{-}) parity $J^{\pi}$ fractions.}
  \label{Gefig}
\end{figure*}

The fitted positive parity, negative parity and total spin fractions are show shown in Fig.~\ref{Gefig} where they are compared to the spin fractions from the spin distribution function assuming the spin cutoff parameters, $\sigma_c$, shown in Table~\ref{GeCTfit}.  The distribution of both positive and negative parity spins varies smoothly as shown by a polynomial fit in Fig.~\ref{Gefig}.  This fit is presented only to guide the eye and no fundamental importance should be taken from this.  The fitted and calculated spin distributions differ by $\lesssim$1\%.  The distributions of both positive and negative parity spins are seen to vary smoothly and were fit with a third order polynomial to guide the eye in Fig.~\ref{Gefig}.

For $^{68}$Ge there are no resonance data so the temperature and back shift were fit to 7 ${J^{\pi}=2^+}$ level energies giving $T$=0.948 MeV and a back shift $E_0 (2^+)$=1.117 MeV with $\overline{\Delta N}(2^+)$=0.135.  The $E_0 (J^{\pi})$ back shifts for all levels with ${J=0^+,2^+,3^-,4^+,6^+}$ were fit assuming a constant temperature with $\overline{\Delta N}(J^{\pi})$=0.139 averaged over 24 levels.  The average deviation of the fitted spin distribution from the calculated value is 0.09\%.

There are no resonance data for $^{69}$Ge so the temperature and back shifts were fit to the extensive nuclear structure data fir levels with ${J^{\pi}=1/2,3/2^-,5/2,7/2,9/2,11/2}$ giving $T$=0.931 MeV with $\overline{\Delta N}(J^{\pi})$=0.125 averaged over 39 levels.  The average deviation of the fitted spin distribution from the calculated value is 0.9\%.

For $^{70}$Ge there are no resonance data but considerable nuclear structure data with ${J^{\pi}=0^+,1,2,3^-,4^+,5^-}$ giving a fitted temperature $T$=0.992 MeV with $\overline{\Delta N}(J^{\pi})$=0.136 averaged over 38 levels.  The average deviation of the fitted spin distribution from the calculated value is 0.3\%.

The temperature for $^{71}$Ge was fit to 15 s-wave resonance and ${J^{\pi}=1/2^+}$ level energy data to give $T$=0.953 MeV and $E_0 (1/2^+)$=1.308 MeV with $\overline{\Delta N}(1/2^+)$=0.122.  The back shifts for levels with ${J^{\pi}=1/2,3/2,5/2,7/2^-,9/2^+}$ were then fit giving $\overline{\Delta N}(J^{\pi})$=0.122 averaged over 36 levels and resonance energies..  The average deviation of the fitted spin distribution from the calculated value is 1\%.

No resonance data are available for $^{72}$Ge so the temperature and back shifts were fit to the extensive nuclear structure data with ${J^{\pi}=0^+,1,2,3^-,4^+,5^-,6^+}$ giving $T$=0.962 MeV with $\overline{\Delta N}(J^{\pi})$=0.116 averaged over 32 level energies.  The average deviation of the fitted spin distribution from the calculated value is 0.8\%.

The temperature for $^{73}$Ge was fit to 13 s-wave resonance and ${J^{\pi}=1/2^+}$ level energy data to give $T$=0.959 MeV and $E_0 (1/2^+)$=1.031 MeV with $\overline{\Delta N}(1/2^+)$=0.123.  The back shifts for levels with ${J^{\pi}=1/2^+,3/2^-,5/2^+,9/2^+}$ were then fit giving $\overline{\Delta N}(J^{\pi})$=0.122 averaged over 26 level and resonance energies.  The average deviation of the fitted spin distribution from the calculated value is 0.1\%.

Limited s-wave resonance data are available for $^{74}$Ge which can be combined with ${J^{\pi}=4^+,5^+}$ level energies to give $T$=0.943 MeV, $E_0 (4^+)$=1.520 MeV, and $E_0 (4^+)$=1.950 MeV with $\overline{\Delta N}(4^+,5^+)$=0.115 for 12 resonance and level energies.  The back shifts for levels with ${J^{\pi}=0^+,1,2^+,3^-,4^+,5,6+}$ were then fit giving $\overline{\Delta N}(J^{\pi})$=0.109 averaged over 32 level and resonance energies.  The average deviation of the fitted spin distribution from the calculated value is 0.3\%.

No resonance data exist for $^{75}$Ge so the temperature and back shift were fit to the ${J^{\pi}=1/2^+}$ level giving $T$=1.004 MeV and $E_0 (1/2^+)$=1.455 MeV with $\overline{\Delta N}(1/2^+)$=0.108 for 9 level energies.  The back shifts for levels with ${J^{\pi}=1/2^+,3/2,5/2,9/2^+}$ were then fit giving $\overline{\Delta N}(J^{\pi})$=0.122 averaged over 34 levels.  The average deviation of the fitted spin distribution from the calculated value is 0.1\%.

No resonance data are available for $^{76}$Ge so the temperature and back shifts were fit to the nuclear structure data with ${J^{\pi}=0^+,2^+,3^-,4^+,5^-}$ giving $T$=1.064 MeV with $\overline{\Delta N}(J^{\pi})$=0.117 averaged over 27 levels.  The average deviation of the fitted spin distribution from the calculated value is 0.02\%.

\subsection{Calcium}

\begin{table*}[!ht]
\tabcolsep=2pt
\caption{\label{Cadata} Back shifts, $E_0(J^{\pi})$, derived from the CT-JPI model, for $^{40-46,48}$Ca. }
\begin{tabular}{ccccccccccc}
\toprule
&\multicolumn{2}{c}{$\hspace{0.5cm}E_0 (J^{\pi}) ^{40}Ca$}&\multicolumn{2}{c}{$\hspace{0.5cm}E_0 (J^{\pi}) ^{42}Ca$}&\multicolumn{2}{c}{$\hspace{0.5cm}E_0 (J^{\pi}) ^{44}Ca$}&\multicolumn{2}{c}{$\hspace{0.5cm}E_0 (J^{\pi}) ^{46}Ca$}&\multicolumn{2}{c}{$\hspace{0.5cm}E_0 (J^{\pi}) ^{48}Ca$}\\
J&\multicolumn{1}{c}{$\hspace{0.5cm}\pi=+$}&\multicolumn{1}{c}{$\hspace{0.5cm}\pi=-$}&\multicolumn{1}{c}{$\hspace{0.5cm}\pi=+$}&\multicolumn{1}{c}{$\hspace{0.5cm}\pi=-$}&
\multicolumn{1}{c}{$\hspace{0.5cm}\pi=+$}&\multicolumn{1}{c}{$\hspace{0.5cm}\pi=-$}&\multicolumn{1}{c}{$\hspace{0.5cm}\pi=+$}&\multicolumn{1}{c}{$\hspace{0.5cm}\pi=-$}
&\multicolumn{1}{c}{$\hspace{0.5cm}\pi=+$}&\multicolumn{1}{c}{$\hspace{0.5cm}\pi=-$}\\
\colrule
0&7.702&11.112&\hspace{0.5cm}4.993&(6.068)&\hspace{0.5cm}3.727&(10.181)&\hspace{0.5cm}4.719&(4.750)&\hspace{0.5cm}6.300&(7.162)\\
1&6.286&8.497&\hspace{0.5cm}(3.084)&5.505&\hspace{0.5cm}2.674&4.637&\hspace{0.5cm}(3.347)&(3.601)&\hspace{0.5cm}(5.537)&5.495\\
2&5.742&7.454&\hspace{0.5cm}2.685&(4.602)&\hspace{0.5cm}2.351&(3.851)&\hspace{0.5cm}3.023&(3.138)&\hspace{0.5cm}5.179&5.285\\
3&5.689&6.619&\hspace{0.5cm}(2.594)&4.577&\hspace{0.5cm}(2.399)&3.500&\hspace{0.5cm}(2.951)&3.141&\hspace{0.5cm}(5.271)&5.353\\
4&5.898&6.108&\hspace{0.5cm}2.990&4.334&\hspace{0.5cm}(2.565)&3.716&\hspace{0.5cm}3.158&(3.338)&\hspace{0.5cm}5.443&(5.906)\\
5&(6.067)&6.078&\hspace{0.5cm}(3.376)&4.827&\hspace{0.5cm}(3.025)&3.914&\hspace{0.5cm}(3.534)&3.760&\hspace{0.5cm}(5.846)&6.744\\
6&6.300&(6.239)&\hspace{0.5cm}(4.014)&(5.483)&\hspace{0.5cm}(3.453)&(4.737)&\hspace{0.5cm}(4.224)&(4.215)&\hspace{0.5cm}(6.800)&(7.170)\\
\toprule
&\multicolumn{2}{c}{$\hspace{0.5cm}E_0 (J^{\pi}) ^{41}Ca$}&\multicolumn{2}{c}{$\hspace{0.5cm}E_0 (J^{\pi}) ^{43}Ca$}&\multicolumn{2}{c}{$\hspace{0.5cm}E_0 (J^{\pi}) ^{45}Ca$}&&\\
J&\multicolumn{1}{c}{$\hspace{0.5cm}\pi=+$}&\multicolumn{1}{c}{$\hspace{0.5cm}\pi=-$}&\multicolumn{1}{c}{$\hspace{0.5cm}\pi=+$}&\multicolumn{1}{c}{$\hspace{0.5cm}\pi=-$}&\multicolumn{1}{c}{$\hspace{0.5cm}\pi=+$}
&\multicolumn{1}{c}{$\hspace{0.5cm}\pi=-$}&\\
\colrule
1/2&3.501&3.829&\hspace{0.5cm}2.754&2.725&\hspace{0.5cm}2.430&2.499&\hspace{0.5cm}&&\hspace{0.5cm}&\\
3/2&3.368&2.676&\hspace{0.5cm}(2.006)&2.065&\hspace{0.5cm}(1.792)&1.903&\hspace{0.5cm}&&\hspace{0.5cm}&\\
5/2&3.572&2.470&\hspace{0.5cm}1.860&(2.077)&\hspace{0.5cm}(2.172)&(1.476)&\hspace{0.5cm}&&\hspace{0.5cm}&\\
7/2&3.974&2.869&\hspace{0.5cm}(2.231)&2.256&\hspace{0.5cm}$-$&$-$&\hspace{0.5cm}&&\hspace{0.5cm}&\\
9/2&4.056&(4.072)&\hspace{0.5cm}(2.944)&(2.683)&\hspace{0.5cm}$-$&$-$&\hspace{0.5cm}&&\hspace{0.5cm}&\\
11/2&(4.564)&(5.938)&\hspace{0.5cm}$-$&$-$&\hspace{0.5cm}$-$&$-$&\hspace{0.5cm}&&\hspace{0.5cm}&\\
\botrule
\end{tabular}
\end{table*}

\begin{table*}[!h]
\tabcolsep=-3pt
\caption{\label{CaCTfit} Neutron separation energies~\cite{AME2013}, $S_n$, spin cutoff parameters, $\sigma_c$, temperatures, $T$, resonance spacings, $D_0$, and minimization uncertainty fitted to the CT-JPI model and compared with the values from RIPL-3~\cite{RIPL3} for $^{40-46,48}$Ca. }
\begin{tabular}{ldddddddd}
\toprule
&\multicolumn{1}{c}{$^{40}$Ca}&\multicolumn{1}{c}{$^{41}$Ca}&\multicolumn{1}{c}{$^{42}$Ca}&\multicolumn{1}{c}{$^{43}$Ca}&\multicolumn{1}{c}{$^{44}$Ca}&
\multicolumn{1}{c}{$^{45}$Ca}&\multicolumn{1}{c}{$^{46}$Ca}&\multicolumn{1}{c}{$^{48}$Ca}\\
\colrule
$S_n$ (MeV)&15.635&8.36282&11.48067&7.93289&11.13116&7.41481&10.3976&9.9526\\
$\sigma_c$&4.50&2.53&3.23&2.69&3.21&2.66&3.12&2.72\\
T $_{CT-JPI}$(MeV) &1.537&1.542&1.569&1.420&1.328&1.285&1.272&1.181\\
T $_{RIPL-3}$(MeV) &$-$&1.374&$-$&1.367&1.395&1.225&$-$&$-$\\
$D_0$(CT-JPI)(keV)&1.44&66.0&8.89&37.0&1.32&26.6&2.28&14.8\\
$D_0$(RIPL-3)(keV)&$-$&\multicolumn{1}{c}{\quad45(4)}&$-$&\multicolumn{1}{c}{\quad20(5)}&\multicolumn{1}{c}{\quad1.8(3)}&\multicolumn{1}{c}{\quad24.1(32)}&$-$&$-$\\
$D_1$(CT-JPI)(keV)&1.84&26.2&1.64&14.0&0.073&10.8&1.06&6.31\\
$D_1$(RIPL-3)(keV)&$-$&\multicolumn{1}{c}{\quad16(1)}&$-$&$-$&$-$&$-$&$-$&$-$\\
$\Delta$N&0.125&0.137&0.122&0.128&0.129&0.120&0.112&0.089\\
\botrule
\end{tabular}
\end{table*}

The calcium isotopes test the CT-JPI model for low-Z nuclei near the doubly magic N=Z=20 and N=28, Z=20 closed shells.  The model has been applied to the isotopes $^{40-46,48}$Ca where sufficient nuclear structure and resonance data are available to provide reasonable fits.   The fitted $E_0(J^{\pi})$ back shifts are shown in Table~\ref{Cadata} and the corresponding neutron separation energies, $S_n$, temperatures, $T$, spin cutoff parameters, $\sigma_c$, resonance spacings, $D_0$ and $D_1$, and quality of fit, $\overline{\Delta N}(J^{\pi})$, are shown in Table~\ref{CaCTfit}.  The average fit is $\overline{\Delta N}(J^{\pi})$=0.125(15) in excellent agreement with the expected value from the folded Normal distribution.  The temperatures and resonance spacings fitted to the CT-JPI model are compared with the values from RIPL-3~\cite{RIPL3} in Table~\ref{CaCTfit}.  The fitted temperatures vary by $<$12\% from the RIPL-3 values and the fitted $D_0$ and $D_1$ values vary by up to 85\% from the RIPL-3 values.  The average fitted spin cutoff parameter for $^{41-46,48}$Ca is $\sigma_c$=2.9(3) and the spin cutoff parameter for $^{40}$Ca is $\sigma_c$=4.50 compared to the value predicted by T. von Egidy \textit{et al}~\cite{Egidy88}, $\sigma_c$=3.

\begin{figure*}[p]
  \centering
    \includegraphics[width=8cm]{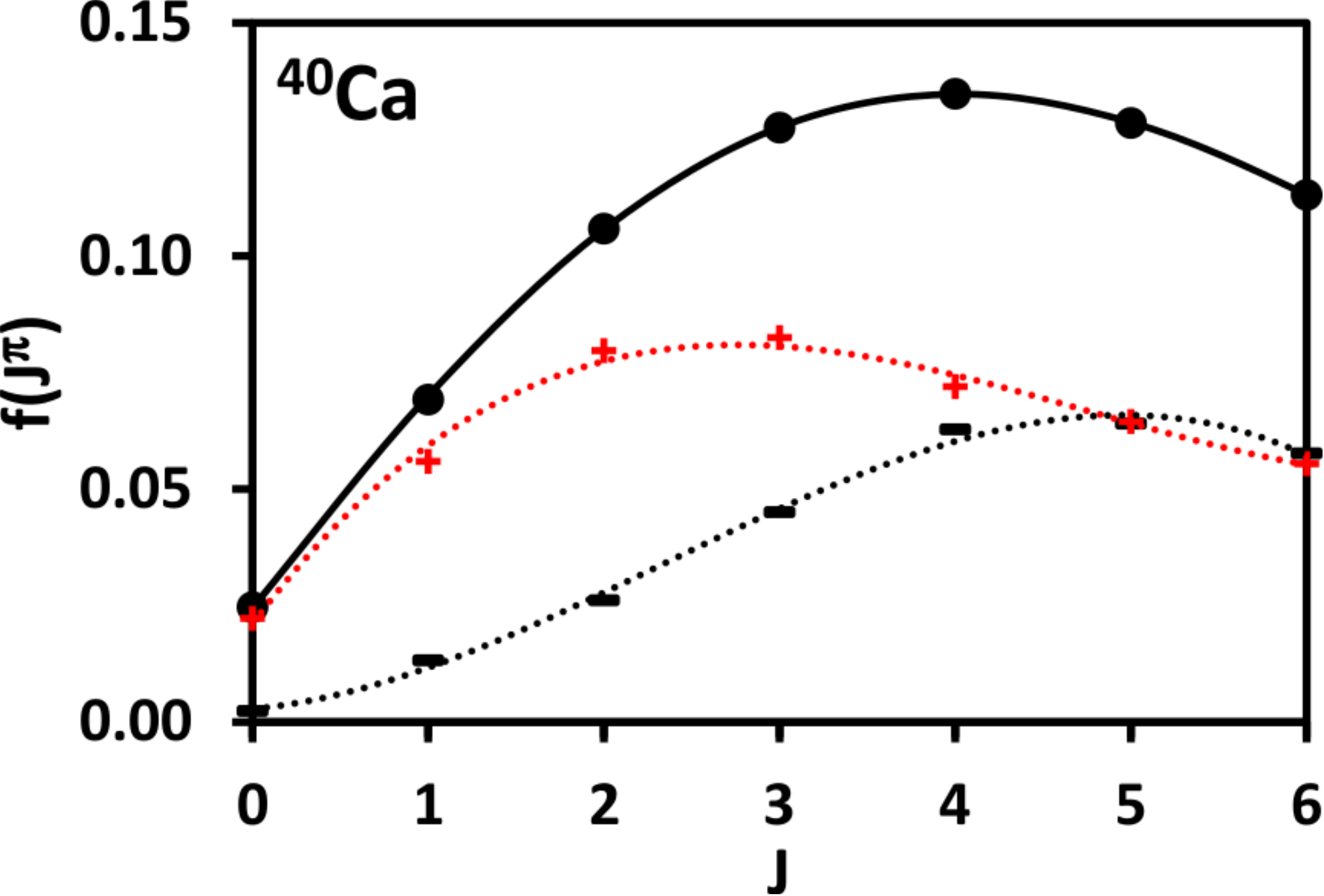}
    \includegraphics[width=8cm]{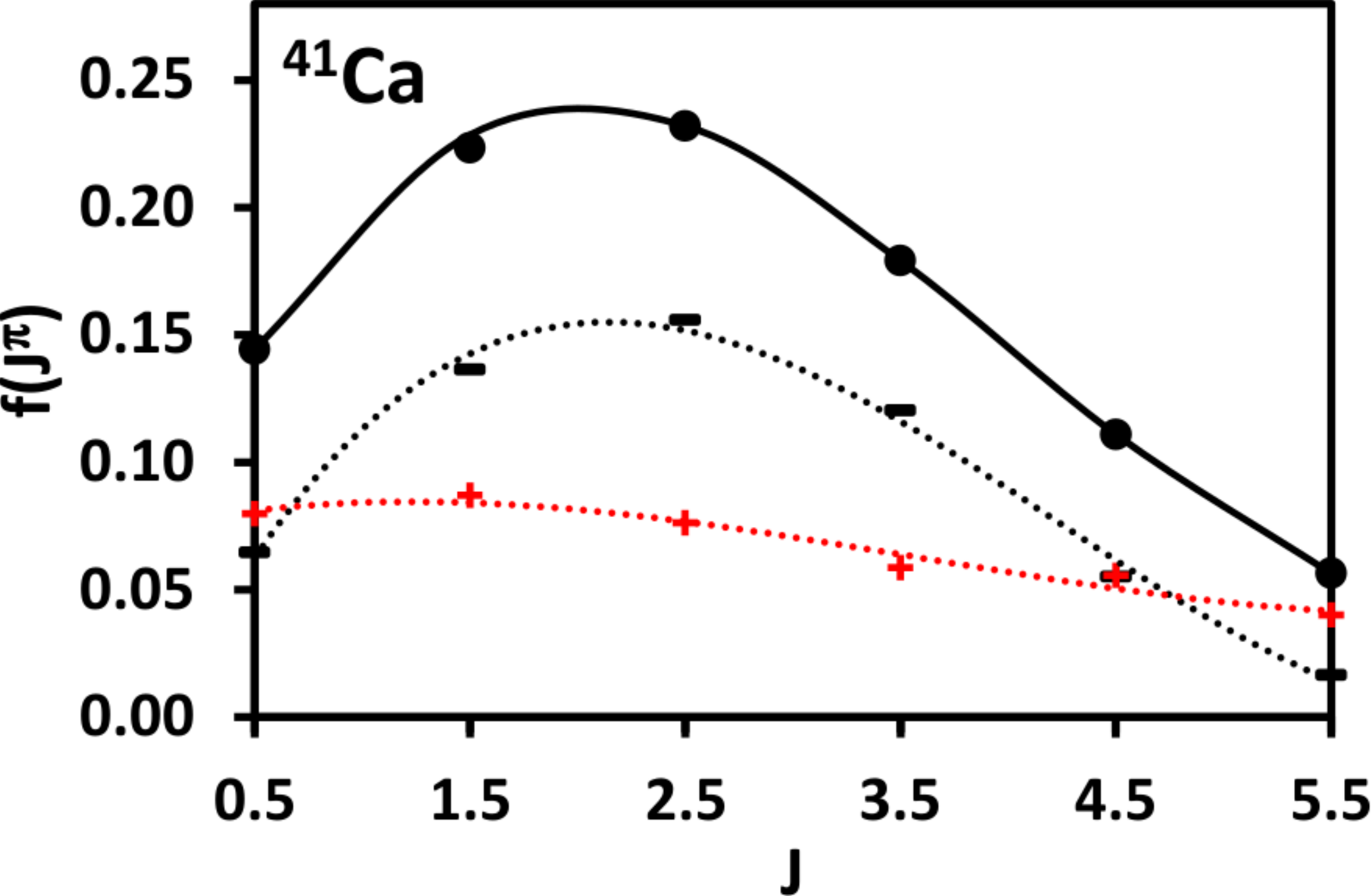}
    \includegraphics[width=8cm]{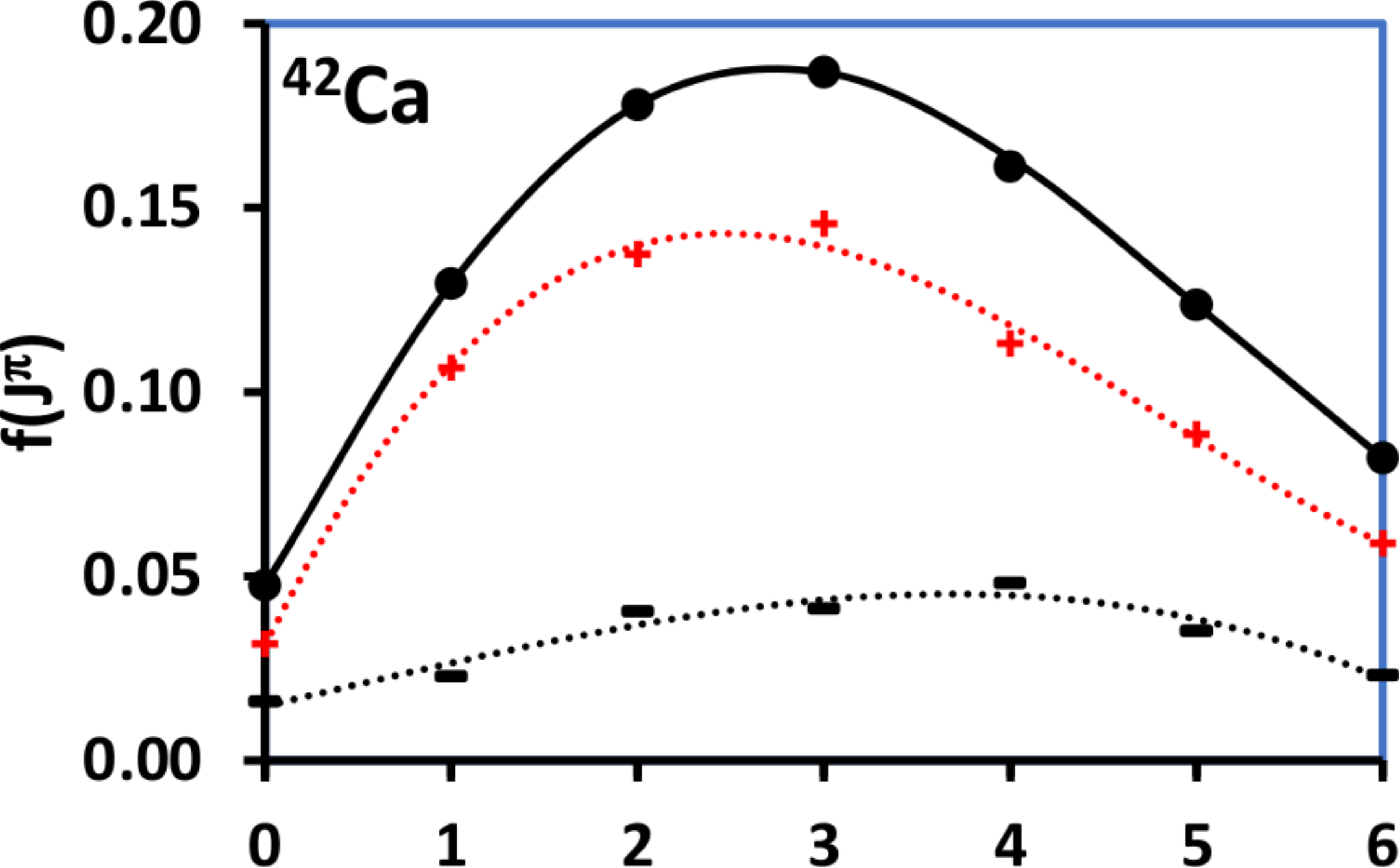}
    \includegraphics[width=8cm]{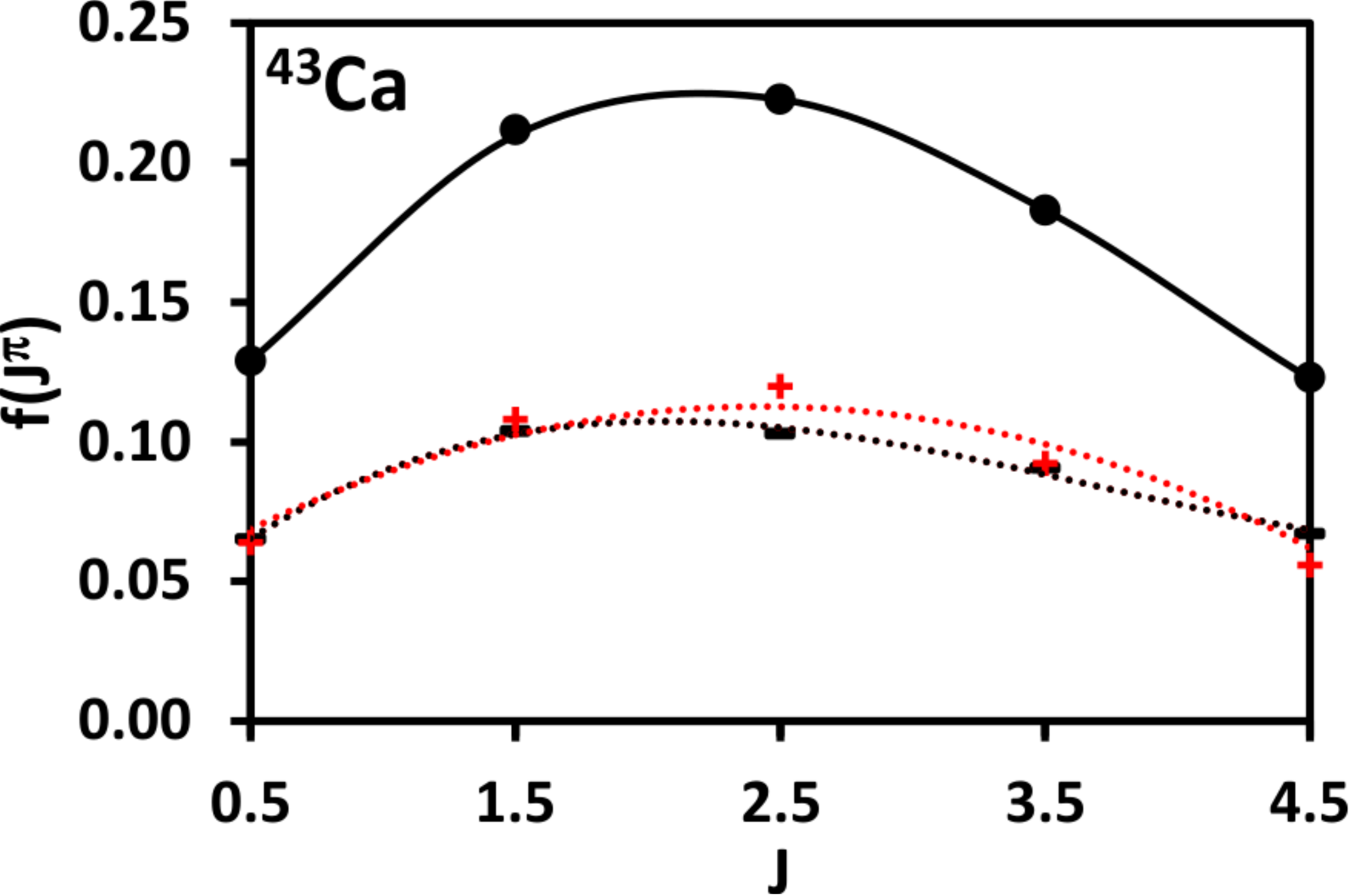}
    \includegraphics[width=8cm]{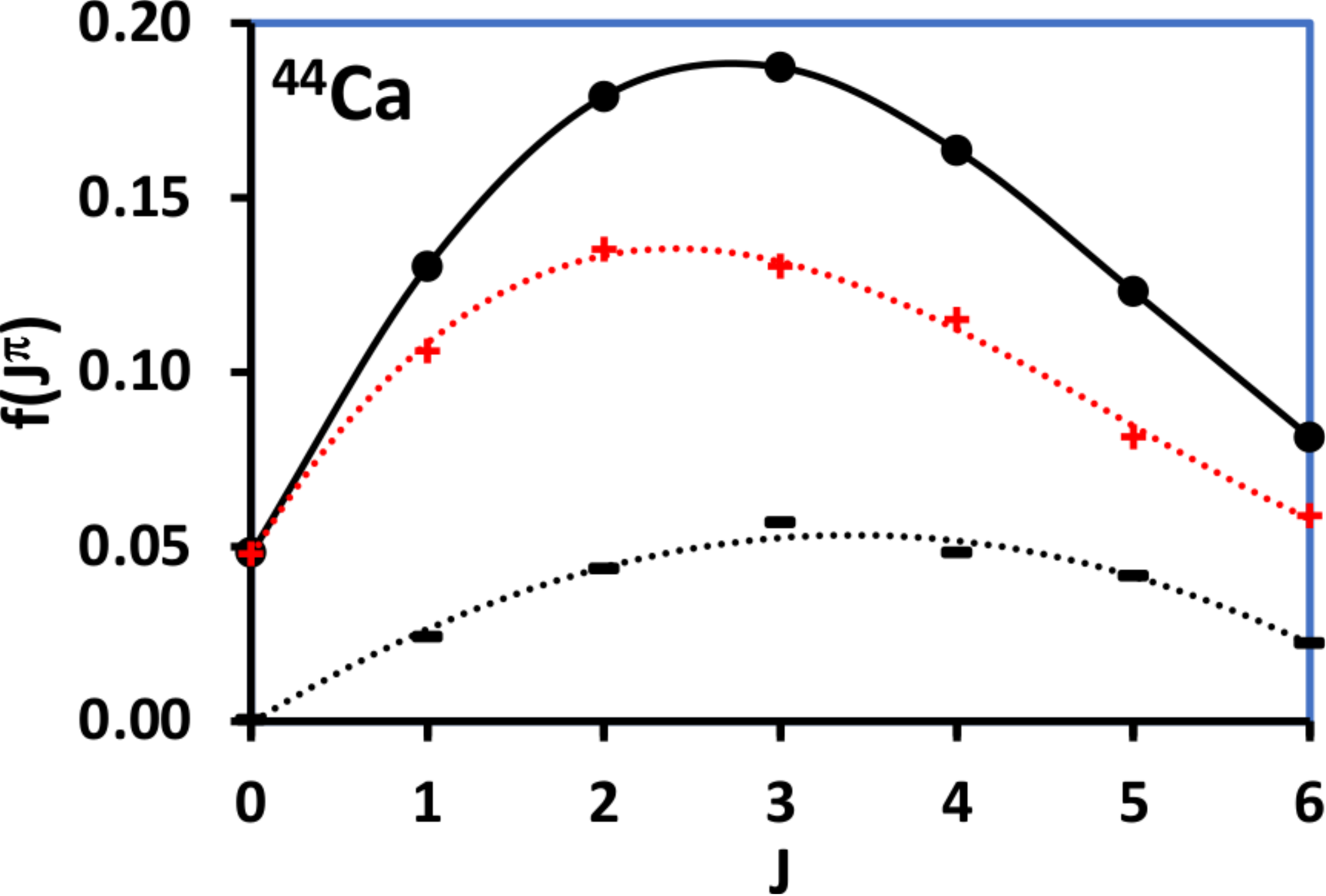}
    \includegraphics[width=8cm]{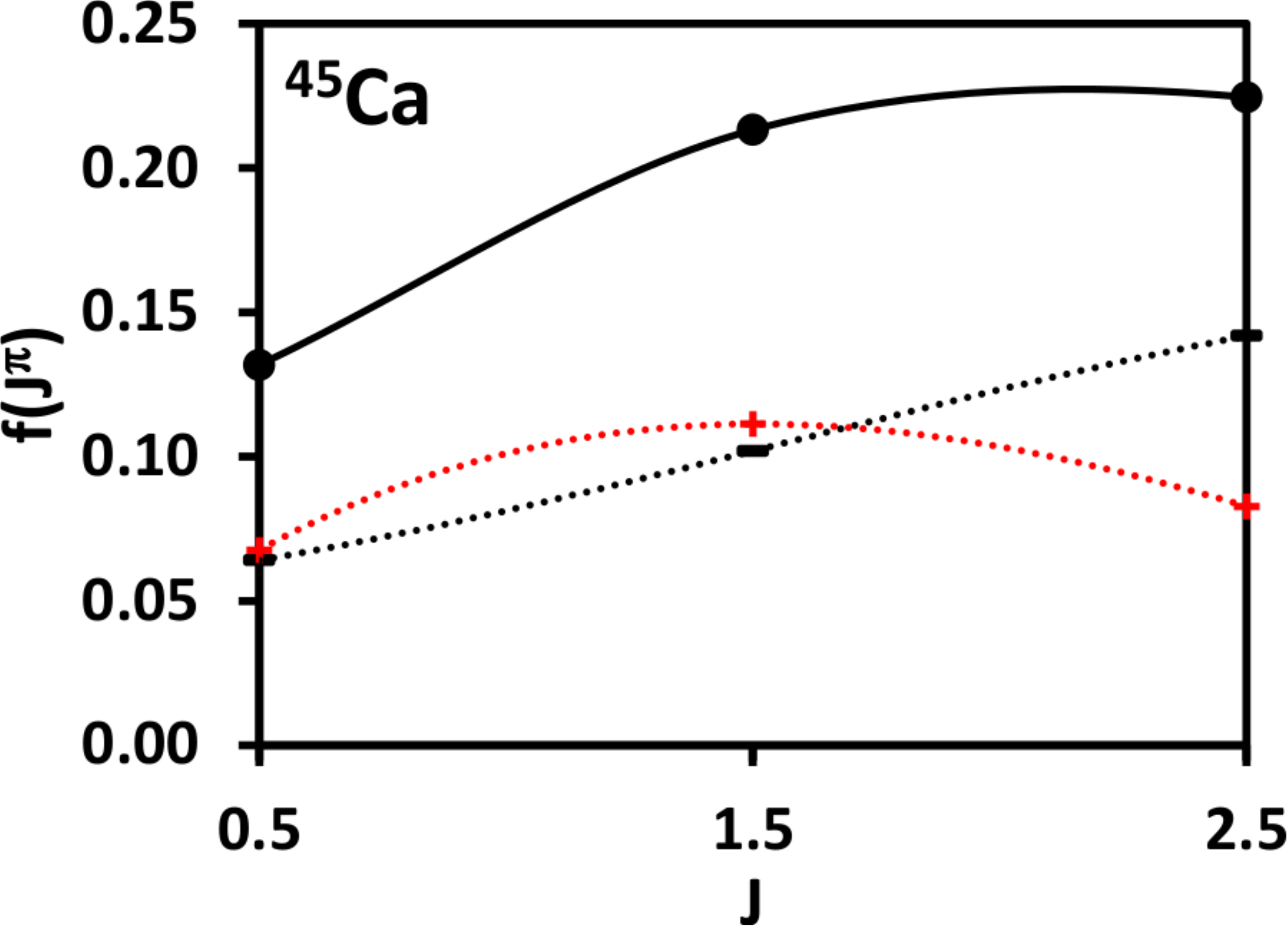}
    \includegraphics[width=8cm]{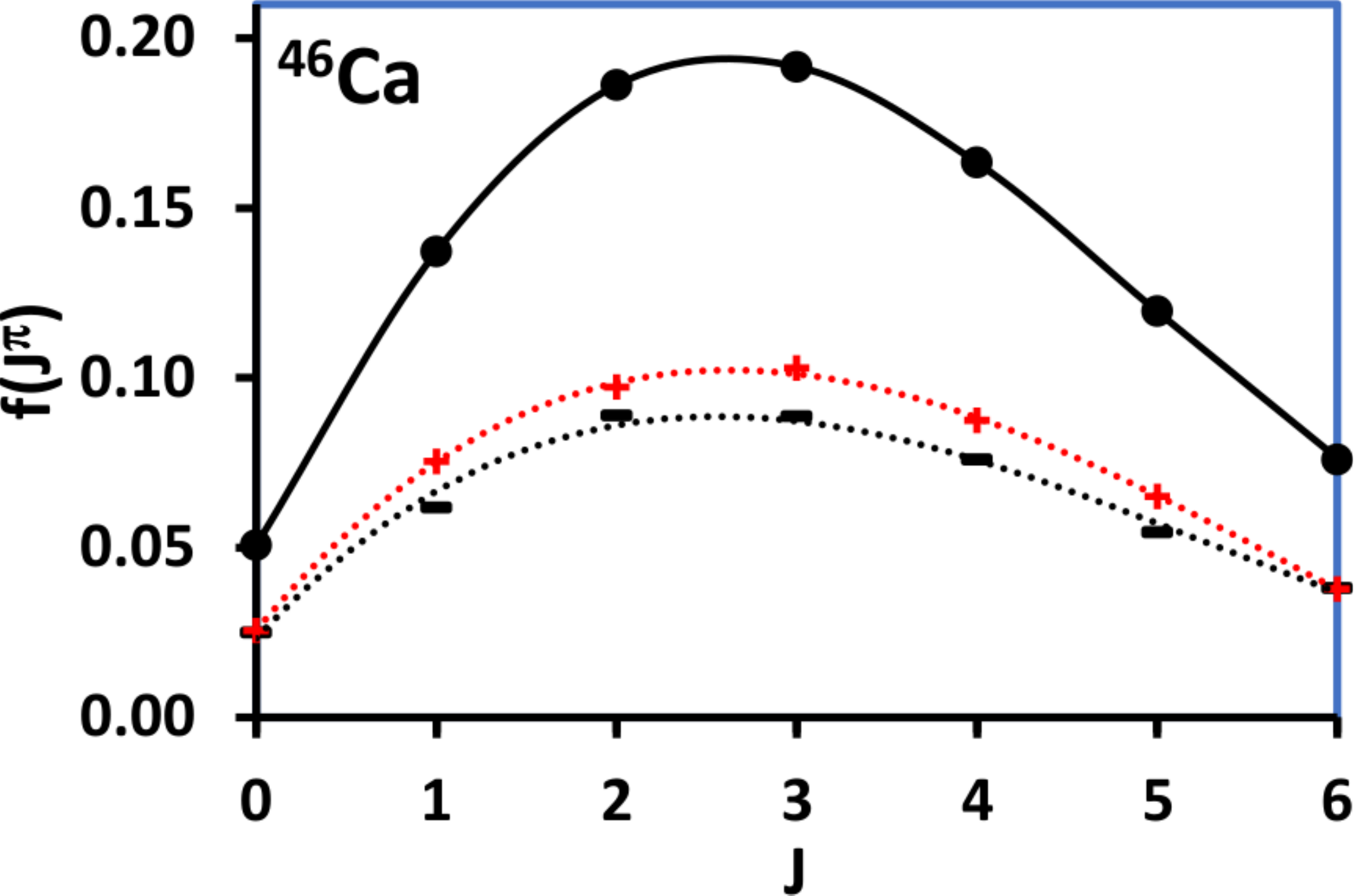}
    \includegraphics[width=8cm]{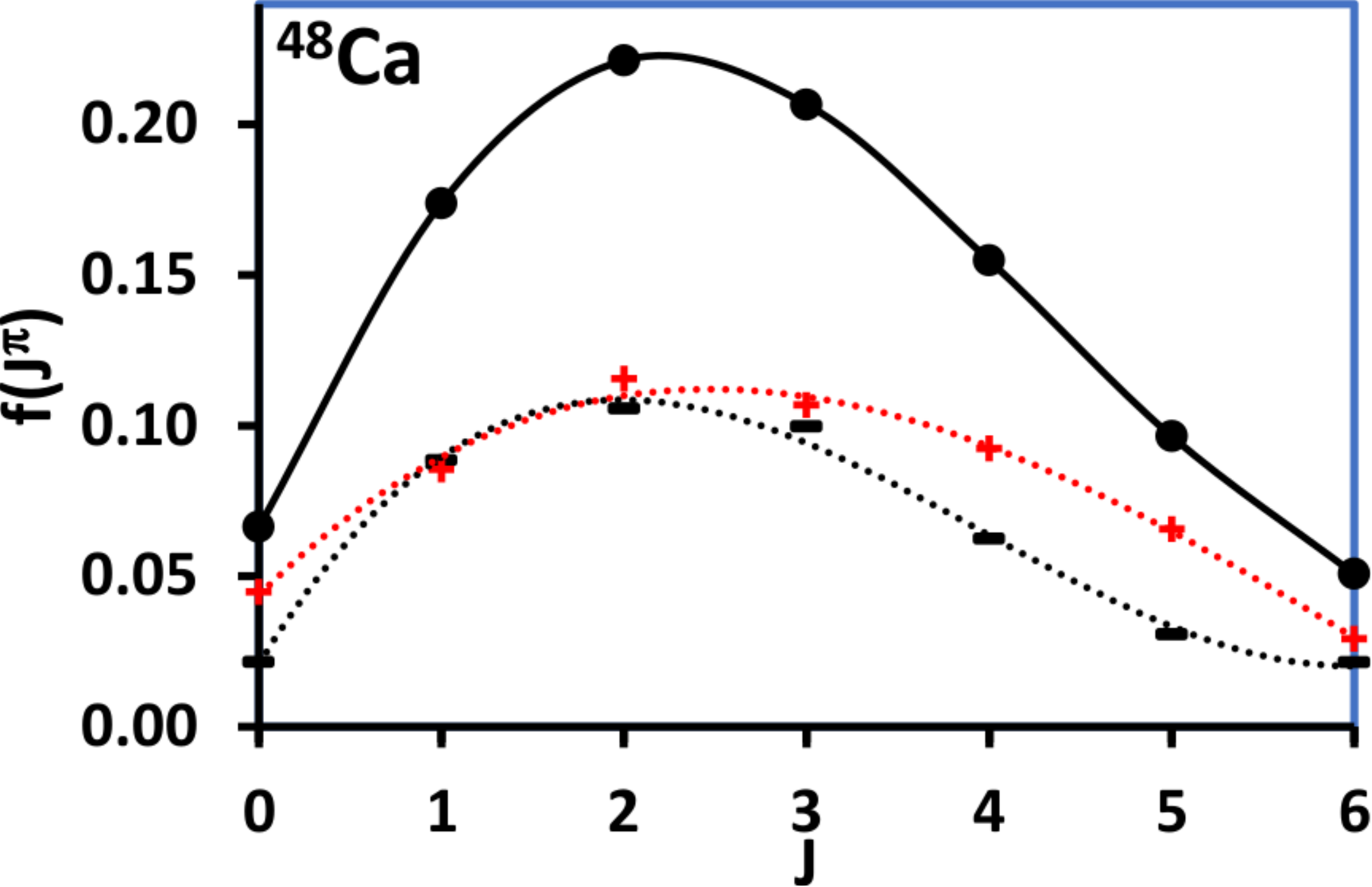}
  \caption{Fit of the CT-JPI model level densities ($\bullet$) to the spin distribution function for $^{40-46,48}$Ca (solid black lines).  The dotted lines show third order polynomial fits to the positive ($\textbf{\textcolor{red}+}$) and negative (\textbf{-}) parity $J^{\pi}$ fractions.}
  \label{Cafig}
\end{figure*}

The fitted positive parity, negative parity and total spin fractions are show shown in Fig.~\ref{Cafig} where they are compared to the spin fractions from the spin distribution function assuming the spin cutoff parameters, $\sigma_c$, shown in Table~\ref{CaCTfit}.  The distribution of both positive and negative parity spins varies smoothly as shown by a polynomial fit in Fig.~\ref{Cafig}.  This fit is presented only to guide the eye and no fundamental importance should be taken from this.  The fitted and calculated spin distributions differ by $\lesssim$1\%.  The distributions of both positive and negative parity spins are seen to vary smoothly and were fit with a third order polynomial to guide the eye in Fig.~\ref{Cafig}.

For $^{40}$Ca there are no resonance data so the temperature and back shifts were fit to the extensive available nuclear level energy data for levels with ${J^{\pi}=0,1,2,3,4,5^-,6^+}$ giving a temperature $T$=1.537 MeV with $\overline{\Delta N}(J^{\pi})$=0.125 averaged over 75 levels.  The average deviation of the fitted spin distribution from the calculated value is 0.3\%.

The temperature for $^{41}$Ca was fit to the substantial s-wave, p-wave, and d-wave resonance data giving $T$=1.542 MeV with $\overline{\Delta N}(1/2,3/2,5/2^+)$=0.134 for 78 level and resonance energies.  The back shifts were fit to the extensive nuclear structure data for levels with ${J^{\pi}=1/2,3/2,5/2,7/2,9/2^+,11/2^+}$ and $\overline{\Delta N}(J^{\pi})$=0.134 averaged over 104 levels.  The average deviation of the fitted spin distribution from the calculated value is 0.4\%.

There are no resonance data for $^{42}$Ca so the temperature and back shifts were fit to the extensive nuclear structure data for levels with ${J^{\pi}=0^+,1^-,2^+,3^-,4,5^+}$ giving $T$=1.569 MeV with $\overline{\Delta N}(J^{\pi})$=0.122 averaged over 50 levels.  The average deviation of the fitted spin distribution from the calculated value is 0.2\%.

Both s-wave and p-wave resonance data are available for $^{43}$Ca which can be fit to give a temperature $T$=1.420 MeV with $\overline{\Delta N}(1/2,3/2^-)$=0.131 averaged over 33 level energies.  Back shifts can be fit to levels with ${J^{\pi}=1/2,3/2^-,5/2^+,7/2^-}$, giving $\overline{\Delta N}(J^{\pi})$=0.128 for 37 level energies and no values were determined, by extrapolation, beyond J=9/2.   The temperature is notably lower than for the lighter calcium isotopes.  The average deviation of the fitted spin distribution from the calculated value is 0.2\%.

Limited s-wave and p-wave resonance energy data are available for $^{44}$Ca which combined with extensive ${J^{\pi}=2^+,3^-,4^+}$ level energy data gives $T$=1.328 MeV with $\overline{\Delta N}(2^+,3^-,4^-)$=0.143, averaged over 24 level and resonance energies.  The back shifts for levels with ${J^{\pi}=0^+,1,2^+,3^-,4^-,5^-}$ can be calculated, assuming the constant temperature, with $\overline{\Delta N}(J^{\pi})$=0.129 averaged over 44 level and resonance energies.  This temperature is lower than for $^{43}$Ca.  The average deviation of the fitted spin distribution from the calculated value is 0.2\%.

Extensive s-wave and p-wave resonance data are available for $^{45}$Ca but little nuclear structure date is available for levels with ${J>3/2}$.  Nevertheless, temperature is well determined by fitting data for 28 levels and resonances with ${J^{\pi}=1/2,3/2-}$ giving $T$=1.285 MeV with $\overline{\Delta N}(1/2,3/2^-)$=0.120, lower than the lighter calcium isotopes.  Calculations could only be extrapolated to ${J=5/2}$ levels.  The average deviation of the fitted spin distribution from the calculated value is 0.01\%.

No resonance data are available for $^{46}$Ca so the temperature was fit to levels with ${J^{\pi}=0^+,2^+,3^-,4^+,5^-}$ giving $T$=1.272 MeV, lower than the lighter calcium isotopes, with $\overline{\Delta N}(J^{\pi})$=0.112 averaged over 20 levels.  The average deviation of the fitted spin distribution from the calculated value is 0.08\%.

There are no resonance data for $^{48}$Ca so the temperature was fit to levels with ${J^{\pi}=0^+,1^-,2,3^-,4^+,5^-}$ giving $T$=1.181 MeV, lower than all other calcium isotopes, with $\overline{\Delta N}(J^{\pi})$=0.089 averaged over 29 levels.  The average deviation of the fitted spin distribution from the calculated value is 0.002\%.  The fit to $^{48}$Ca is better than for any other nuclide analyzed in this work.

\subsection{Nitrogen}

The nitrogen isotopes test the CT-JPI model for low-Z nuclei with well established level schemes.  The model has been applied to the isotopes $^{13-15}$N where sufficient nuclear structure and resonance data are available to provide reasonable fits.   The fitted $E_0(J^{\pi})$ back shifts are shown in Table~\ref{Ndata} and the corresponding neutron separation energies, $S_n$, temperatures, $T$, spin cutoff parameters, $\sigma_c$, resonance spacings, $D_0$ and $D_1$, and quality of fit, $\overline{\Delta N}(J^{\pi})$, are shown in Table~\ref{NCTfit}.  The average fit is $\overline{\Delta N}(J^{\pi})$=0.112(8) in excellent agreement with the expected value from the folded Normal distribution. The average fitted spin cutoff parameter for $^{13-15}$N is $\sigma_c$=2.8(5) which van be compared to the von Egidy \textit{et al} value, $\sigma_c$=2.1.  No RIPL-3 data exist for comparison with then nitrogen isotope data.

\begin{table*}[!ht]
\tabcolsep=3pt
\caption{\label{Ndata} Back shifts, $E_0(J^{\pi})$, derived from the CT-JPI model, for $^{13-15}$N. }
\begin{tabular}{lcccccccccccc}
\toprule
&\multicolumn{4}{c}{$E_0 (J^{\pi}) ^{13}$N}&\multicolumn{4}{c}{\hspace{1cm}$E_0 (J^{\pi}) ^{14}$N}&\multicolumn{4}{c}{\hspace{1cm}$E_0 (J^{\pi}) ^{15}$N}\\
\colrule
&1/2$^+$&8.853&1/2$^-$&10.429&\hspace{1cm}0$^+$&(10.499)&0$^-$&(11.131)&\hspace{1cm}1/2$^+$&10.020&1/2$^-$&(11.166)\\
&3/2$^+$&7.944&3/2$^-$&9.449&\hspace{1cm}1$^+$&8.332&1$^-$&8.863&\hspace{1cm}3/2$^+$&9.374&3/2$^-$&9.482\\
&5/2$^+$&(8.058)&5/2$^-$&9.291&\hspace{1cm}2$^+$&7.557&2$^-$&8.200&\hspace{1cm}5/2$^+$&9.060&5/2$^-$&9.734\\
&7/2$^+$&8.650&7/2$^-$&(10.000)&\hspace{1cm}3$^+$&7.1414&3$^-$&8.252&\hspace{1cm}7/2$^+$&(9.400)&7/2$^-$&(10.280)\\
&9/2$^+$&(9.732)&9/2$^-$&(11.200)&\hspace{1cm}4$^+$&7.348&4$^-$&8.566&\hspace{1cm}9/2$^+$&10.172&9/2$^-$&(11.346)\\
&11/2$^+$&$-$&11/2$^-$&$-$&\hspace{1cm}5$^+$&(7.898)&5$^-$&8.970&\hspace{1cm}11/2$^+$&(11.352)&11/2$^-$&(12.394)\\
\botrule
\end{tabular}
\end{table*}

\begin{table}[!ht]
\tabcolsep=3pt
\caption{\label{NCTfit} Back shifts, $E_0(J^{\pi})$, derived from the CT-JPI model, for $^{13-15}$N. }
\begin{tabular}{lddd}
\toprule
&\multicolumn{1}{c}{$^{13}$N}&\multicolumn{1}{c}{$^{14}$N}&\multicolumn{1}{c}{$^{15}$N}\\
\colrule
$S_n$ (MeV)&20.0639&10.5534&10.8333\\
$\sigma_c$&2.46&3.41&2.67\\
T $_{CT-JPI}$(MeV) &2.160&2.182&2.69\\
$D_0$(CT-JPI)(keV)&4.77&96.8&2.70\\
$D_1$(CT-JPI)(keV)&5.85&36.7&2.71\\
$\Delta$N&0.105&0.11&0.12\\
\botrule
\end{tabular}
\end{table}

\begin{figure}[!ht]
  \centering
    \includegraphics[width=8cm]{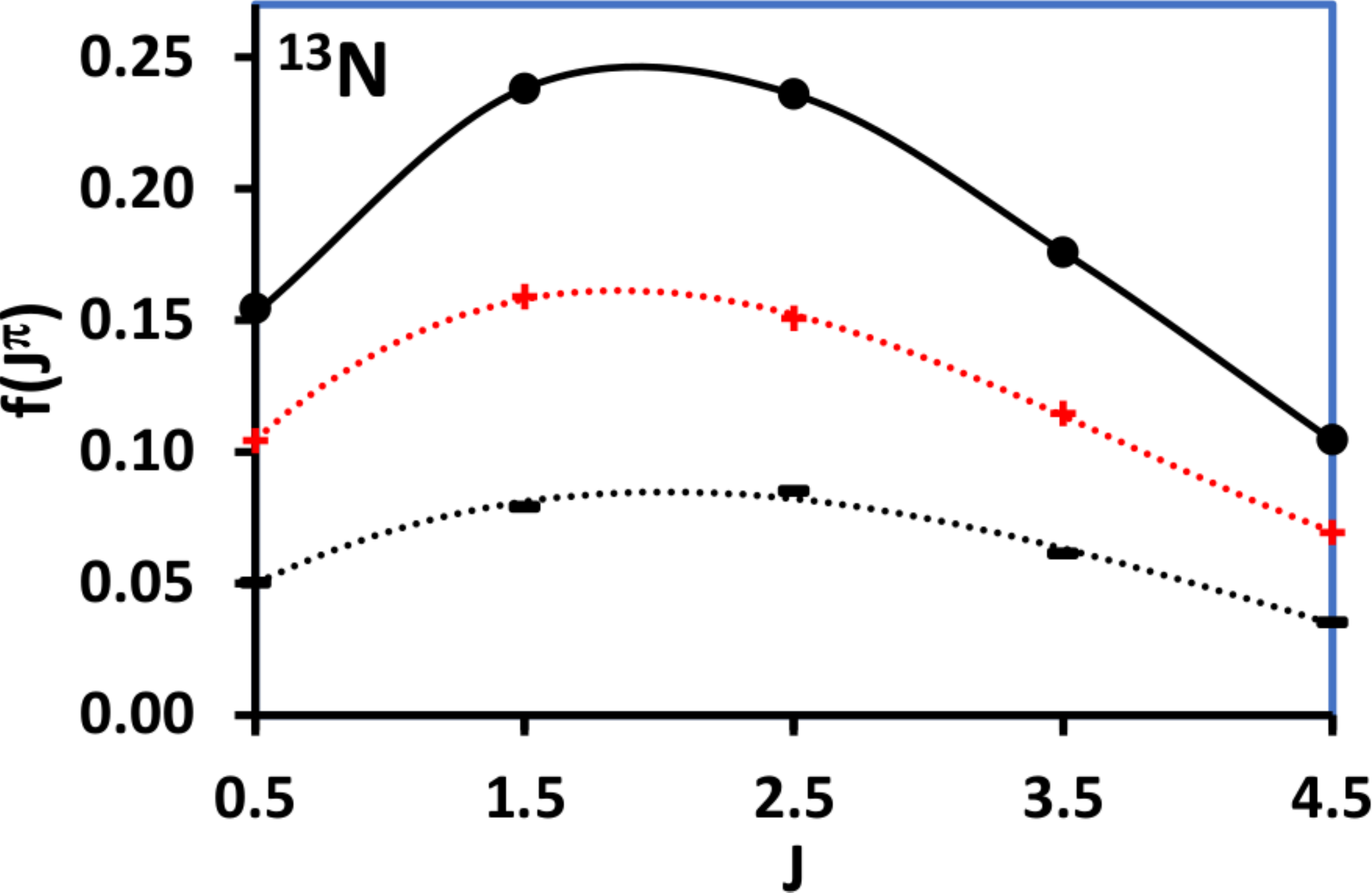}
    \includegraphics[width=8cm]{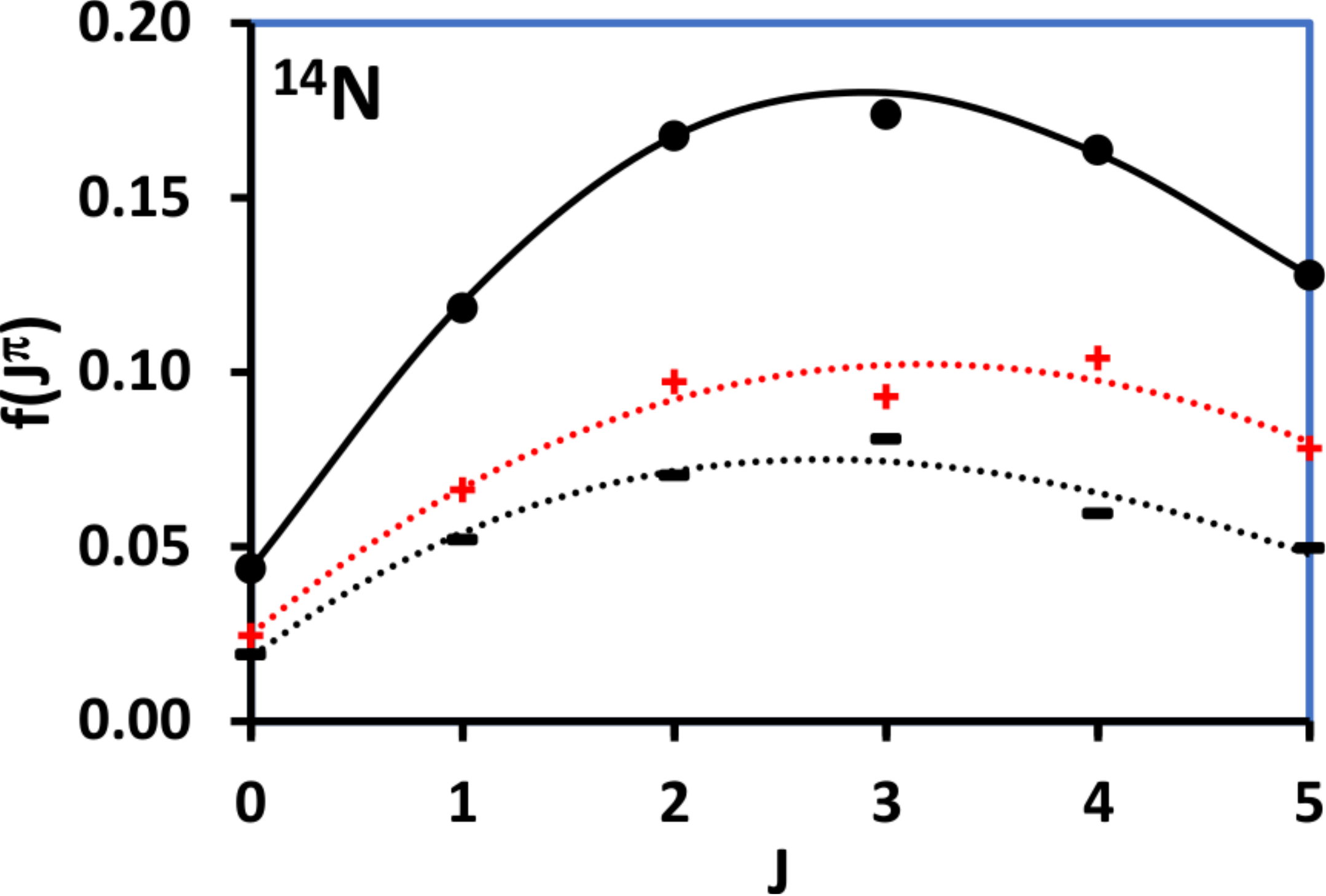}
    \includegraphics[width=8cm]{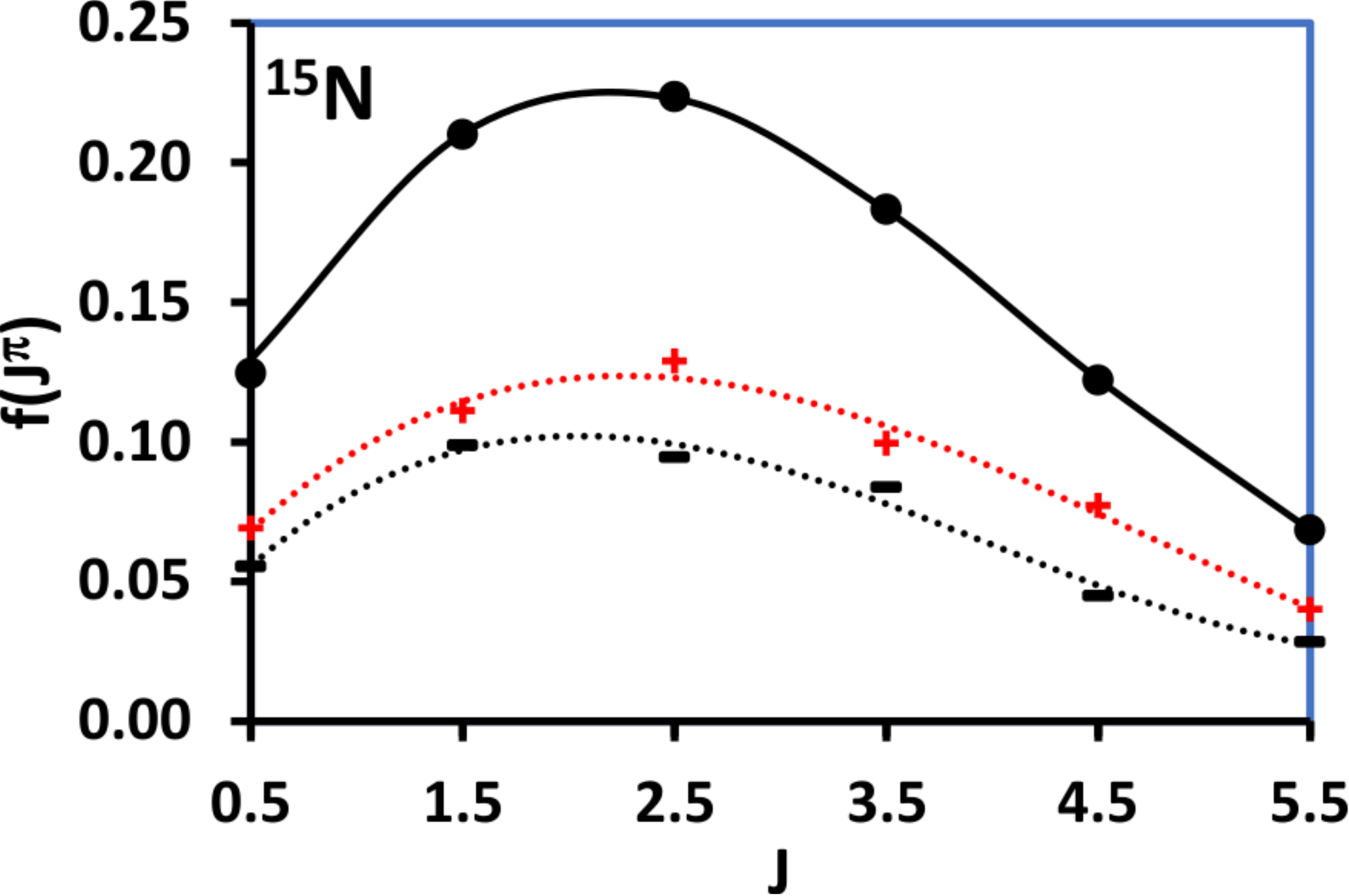}
  \caption{Fit of the CT-JPI model level densities ($\bullet$) to the spin distribution function for $^{40-46,48}$Ca (solid black lines).  The dotted lines show third order polynomial fits to the positive ($\textbf{\textcolor{red}+}$) and negative (\textbf{-}) parity $J^{\pi}$ fractions.}
  \label{Nfig}
\end{figure}

The positive parity, negative parity and total spin distributions are show shown in Fig.~\ref{Nfig} where they are compared to the spin distribution model assuming the spin cutoff parameters, $\sigma_c$, shown in Table~\ref{NCTfit}.  The distribution of both positive and negative parity spins varies smoothly as shown by a polynomial fit in Fig.~\ref{Ndata}.  This fit is presented only to guide the eye and no fundamental importance should be taken from this.  The fitted total spin distribution agrees remarkably will with the spin distribution function confirming the utility of the CT-JPI model.

There are no resonance data for the nitrogen isotopes there so the temperatures and back shifts were fit to the extensive available nuclear structure data available for each isotope.  For $^{13}$N the temperature and back shift were fit to 10 ${J^{\pi}=3/2^+}$ levels giving $T$=2.160 MeV and $E_0 (3/2^+)$=7.944 MeV with $\overline{\Delta N}(3/2^+)$=0.118.  The back shifts were fit to 27 levels with ${J^{\pi}=1/2^+,3/2,5/2^-,7/2^+}$ giving $\overline{\Delta N}(J^{\pi})$=0.105.  The average deviation of the fitted spin distribution from the calculated value is 0.3\%.

For $^{14}$N the temperature and back shifts were fit to 40 level energies with ${J=1,2,3,4,5}$ giving $T$=2.182 MeV with $\overline{\Delta N}(J^{\pi})$=0.110.  The average deviation of the fitted spin distribution from the calculated value is 0.5\%.

The temperature and back shift for $^{15}$N was fit to 11 ${J^{\pi}=3/2^+}$ levels giving $T$=2.166 MeV and $E_0 (3/2^+)$=9.374 MeV with $\overline{\Delta N}(3/2^+)$=0.112.  The back shifts were fit to 34 levels with ${J^{\pi}=1/2^-,3/2,5/2,9/2^+}$ giving $\overline{\Delta N}(J^{\pi})$=0.120.  The average deviation of the fitted spin distribution from the calculated value is 0.5\%.

\section{Conclusions}

The CT model has been shown to be the consequence the level spacing distribution which compresses the random distribution of levels into an exponential distribution.  Although level spacings are commonly thought to follow either a Wigner or Poisson distribution, neither distribution constrains the exponential so that the level sequence number $N(E_0)$=1.00.  Remarkably, a folded Normal distribution constrains the sequence numbers properly and predicts the CT model formulation.  The folded Normal distribution is very similar to the Poisson distribution except that the Poisson distribution is discrete, only applying the spacing to higher lying levels, while the folded Normal distribution is continuous, applying to the spacing of all levels.

The standard CT model fails because it attempts to fit the total level density.  The total level density is punctuated by the appearance of new sequences of levels with various spins and parities at their Yrast energies.  This forces the single back shift parameter, $E_0$, to become negative and nonphysical.  Each sequence of levels with the same spin and parity face no such disruption in their sequence, except possibly near the shell gaps, so they can follow the CT model, each with a back shift $E_0 (J^{\pi})$ that is near the Yrast energy.

The standard CT model also fails because it attempts to constrain the exponential fit to the level density at the neutron separation energy using the spin distribution function~\cite{Eric60} using resonance data.  The spin distribution function contains no parity information while the resonance data contain parity but no total spin information.  This requires the arbitrary assumption that levels of both parities have the same level densities which has no basis in fact.  The CT-JPI model no problem with the comparison with the spin distribution function because the level densities for both parities are determined by the model at the neutron separation energy.  The strongest evidence in support of the CT-JPI model is that using experimental nuclear structure and resonance data the best fits to the temperature and back shifts coincide with the best fit to the spin distribution by varying the spin cutoff parameter.

An iterative procedure has been prescribed to determine the CT-JPI model temperature and back shift parameters.  This procedure is limited by the quality of nuclear structure data that is available.  Levels with established $J^{\pi}$ values from ENSDF were used wherever possible.  Mistaken $J^{\pi}$ assignments in ENSDF likely exist and will deteriorate the quality of the fits.  Levels with large energy uncertainties were avoided in this work, although a weighted fit to the level energies should improve the fit.  It was found that the best fits were obtained using only the first few levels of a sequence.  Although the fitted parameters remained constant when extending the fit to more levels, the quality of fit declined.  This is likely due to the decreasing level spacing at high energies and the compounding of systematic level energy uncertainties in the experimental level schemes.  The CT-JPI model has been successfully applied to 46 nuclei with atomic numbers ranging from Z=7-92.  In all cases the statistical fit of level energies to the CT-JPI is consistent with the uncertainties expected from the folded Normal distribution.  The spin distributions at the neutron separation energy all agree with the spin distribution function predictions to within $\leq$1\%.

\begin{figure*}[t]
  \centering
    \includegraphics[width=8cm]{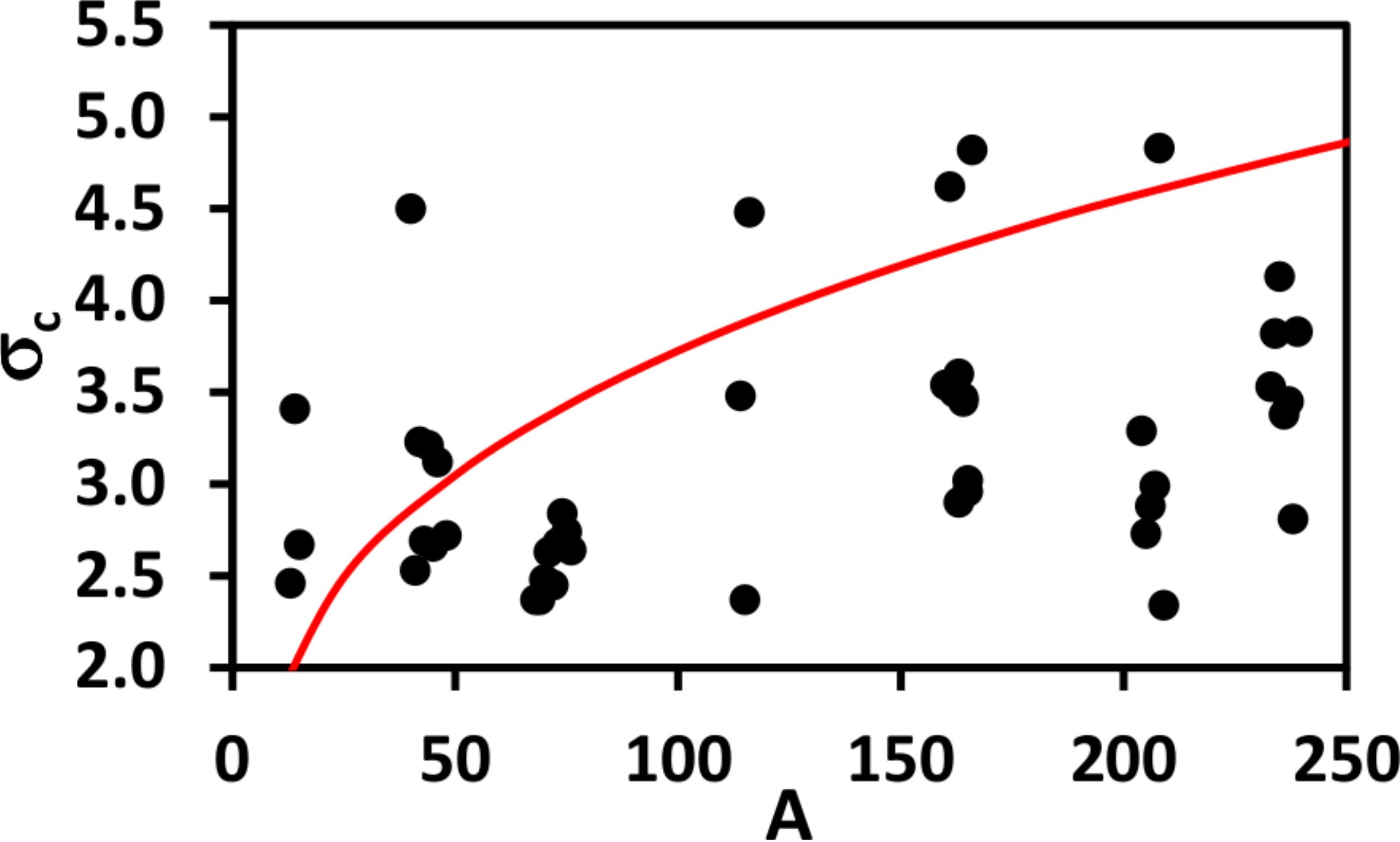}
  \caption{Distribution of fitted spin cutoff parametrs, $\sigma_c$.}, compared with the von Egidy \textit{et al}~\cite{Egidy88} prediction $\sigma_c = 0.98A^{0.29}$ (red curve).
  \label{Sigmac}
\end{figure*}

The excellent agreement with the spin distribution function gives great confidence in the determination of the spin cutoff parameter, $\sigma_c$.  The $\sigma_c$ values determined in this work are compared with the formulation of von Egidy \textit{et al} ~\cite{Egidy88} in Fig.~\ref{Sigmac}.  In each mass region the $\sigma_c$ values vary widely in complete disagreement with calculation.

The CT-JPI model provides complete level density information as a function of spin and parity up to at least the neutron separation energy.  The fitted temperatures and resonance spacings are generally very different from those provide by RIPL-3~\cite{RIPL3}.  That work has required the availability of resonance data while the CT-JPI model analysis only requires good nuclear structure data and is applicable to many more nuclei.  The CT-JPI model also provides an incentive to measure more and higher quality nuclear structure data.  It also provides a challenge to the calculation of level schemes far from stability where the prediction of level energies can be constrained by the model to give more meaningful results.

\section{Acknowledgements}

This work was supported by funding from the University of California retirement system.  Although no federal funding was provided I acknowledge the U.S. Department of Energy, Office of Nuclear Science, Nuclear Data Program for their long history of supporting my research that made this work possible.


\bibliographystyle{apsrev}

\end{document}